\documentclass[12pt, preprint]{aastex}
\usepackage{natbib}
\usepackage{url}
\usepackage{lscape}
\usepackage{float}
\usepackage{amsmath}
\usepackage{graphicx}
\usepackage{hyperref}
\begin{document}
\title{Constraints on the Early Terrestrial Surface UV Environment Relevant to Prebiotic Chemistry}
\author{Sukrit Ranjan\altaffilmark{1,2}, Dimitar D. Sasselov\altaffilmark{1}}

\altaffiltext{1}{Harvard-Smithsonian Center for Astrophysics, Cambridge, MA 02138, USA}
\altaffiltext{2}{60 Garden Street, Mail Stop 10, Cambridge, MA 02138, USA; sranjan@cfa.harvard.edu; 617-495-5676}

\date{\today}

\begin{abstract}
The UV environment is a key boundary condition to abiogenesis. However, considerable uncertainty exists as to planetary conditions and hence surface UV at abiogenesis. Here, we present two-stream multi-layer clear-sky calculations of the UV surface radiance on Earth at 3.9 Ga to constrain the UV surface fluence as a function of albedo, solar zenith angle (SZA), and atmospheric composition. 

Variation in albedo and latitude (through SZA) can affect maximum photoreaction rates by a factor of $>10.4$; for the same atmosphere, photoreactions can proceed an order of magnitude faster at the equator of a snowball Earth than at the poles of a warmer world. Hence, surface conditions are important considerations when computing prebiotic UV fluences.

For climatically reasonable levels of CO$_2$, fluence shortward of 189 nm is screened out, meaning that prebiotic chemistry is robustly shielded from variations in UV fluence due to solar flares or variability. Strong shielding from CO$_2$ also means that the UV surface fluence is insensitive to plausible levels of CH$_4$, O$_2$, and O$_3$.  At scattering wavelengths, UV fluence drops off comparatively slowly with increasing CO$_2$ levels. However, if SO$_2$ and/or H$_2$S can build up to the $\geq1-100$ ppm level as hypothesized by some workers, then they can dramatically suppress surface fluence and hence prebiotic photoprocesses.

H$_2$O is a robust UV shield for $\lambda<198$ nm. This means that regardless of the levels of other atmospheric gases, fluence $\lesssim198$ nm is only available for cold, dry atmospheres, meaning sources with emission $\lesssim198$ (e.g. ArF eximer lasers) can only be used in simulations of cold environments with low abundance of volcanogenic gases. On the other hand, fluence at 254 nm is unshielded by H$_2$O and is available across a broad range of $N_{CO_{2}}$, meaning that mercury lamps are suitable for initial studies regardless of the uncertainty in primordial H$_2$O and CO$_2$ levels

\end{abstract}


\maketitle

\maketitle

\section{Introduction}
Ultraviolet (UV) light plays a key role in prebiotic chemistry (chemistry relevant to the origin of life). UV photons are energetic enough to affect the electronic structure of molecules by dissociating bonds and ionizing and exciting molecules. These properties mean that UV light can destroy molecules important to abiogenesis \citep{Sagan1973}, but also that UV light can power photochemistry relevant to the synthesis of prebiotically important molecules. UV light has been invoked in prebiotic chemistry as diverse as the origin of chirality \citep{Rosenberg2008}, the synthesis of amino acid precursors \citep{Sarker2013}, and the polymerization of RNA \citep{Mulkidjanian2003}. Most recently, UV light has been shown to play a key role in the first plausible prebiotic synthesis of the activated pyrimidine ribonucleotides \citep{Powner2009}, the synthesis of glycolaldehyde and glyceraldehyde \citep{Ritson2012}, and a reaction network generating precursors for a range of prebiotically important molecules including lipids, amino acids, and ribonucleotides \citep{Patel2015}. 

Simulating UV-sensitive prebiotic chemistry in laboratory contexts requires understanding what the prebiotic UV environment was like, both in overall fluence level and in wavelength dependence. Prebiotic chemistry on Earth is generally assumed to have occurred in aqueous solution at the surface of the planet or at hydrothermal vents deep in the ocean. UV-dependent prebiotic chemistry could not have occurred too deep in the ocean due to attenuation from water, and so must have occurred near the surface. Therefore, in order to understand the fidelity of laboratory simulations of UV-sensitive prebiotic chemistry, it is important to understand what the prebiotic UV environment at the planetary surface was like.

In this work, we use a two-stream multi-layer radiative transfer model to constrain the prebiotic UV environment at the surface. We calculate the surface radiance as a function of solar zenith angle (SZA), surface albedo ($A$), and atmospheric composition. We convolve the calculated surface radiance spectra against action spectra corresponding to two different simple photochemical reactions (one a stressor, the other a eustressor) that may have been important during the era of abiogenesis, and integrate the result to compute the biologically effective dose rate (BED) and estimate the impact of these parameters on prebiotic chemistry. Previous work (e.g., \citealt{Cockell2002}, \citealt{Cnossen2007}, \citealt{Rugheimer2015}) has ignored the effect of SZA and albedo; we demonstrate that taken together, these factors can lead to variations in BED of more than an order of magnitude. Earlier analyses have focused on "case studies" for the atmospheric composition; we step through the plausible\footnote{As well as some deemed implausible, to explore parameter space} range of abundances of CO$_2$, H$_2$O, CH$_4$, SO$_2$, H$_2$S, O$_2$, and O$_3$ to constrain the impact of varying levels of these gases on the surface UV environment. 

In Section~\ref{sec:background}, we discuss previous work on this topic, and available constraints on the prebiotic atmosphere. In Section~\ref{sec:uv2g}, we describe our radiative transfer model and its inputs and assumptions. In Section~\ref{sec:testing}, we describe the tests we performed to validate our model. Section~\ref{sec:resdisc} then presents and discusses the results obtained through use of our model and the implications for the prebiotic UV environment, and Section~\ref{sec:conc} summarizes our findings.

\section{Background\label{sec:background}}
\subsection{Previous Work}
Recognizing the relevance of UV fluence to life (though mostly in the context of a stressor), previous workers have placed constraints on the the surface UV environment of the primitive Earth. In this section, we present a review of some recent work on this topic, and discuss how our work differs from them. 

\citet{Cockell2002} calculate the UV flux received at the surface of the Earth at 3.5 Ga using a monolayer delta-Eddington approach to radiative transfer, assuming a solar zenith angle SZA$=0^\circ$ (i.e. the sun directly overhead)\footnote{\cite{Cockell2002} does not specific the albedo assumed}, for an atmosphere composed of 0.7 bar N$_2$ and 40 mb and 1 bar of CO$_2$, as well as an atmosphere with a sulfur haze. \citet{Cockell2002} found the surface UV flux to be spectrally characterized by a cutoff at $>190$ nm imposed by CO$_2$. They further found the surface UV flux for non-hazy primordial atmospheres to be far higher than for the modern day due to a lack of UV-shielding oxygen and ozone, with hazes potentially able to provide far higher attenuation. 

\citet{Cnossen2007} calculate the UV flux received at the surface of the earth at 4-3.5 Ga at SZA$=0^\circ$. To calculate atmospheric radiative transfer, they partition the atmosphere into layers. They compute absorption using  the Beer-Lambert Law. To account for scattering, they calculate the flux scattered in each layer and assume half of it proceeds up, and half proceeds down. They iterate this process to the surface. They explore the effect of atmospheric composition on surface flux, assuming an N$_2$-CO$_2$ dominated atmosphere with levels of CO$_2$ varying from 0.02-1 bar, levels of CH$_4$ spanning 1 order of magnitude, and levels of O$_3$ spanning 5 orders of magnitude. \citet{Cnossen2007} found that atmospheric attenuation prevented flux at wavelengths shorter than 200 nm from reaching the surface in all the case studies they considered. In all cases, they found the surface flux to be far higher than on modern Earth, again due to lack of UV-shielding oxic molecules. They further found that the surface flux was insensitive to variation in CH$_4$ and O$_3$ concentration at the levels they considered, and that the wavelength cutoff from CO$_2$ rendered the surface flux insensitive to H$_2$O level. \citet{Cnossen2007} also use observations of a flare on an analog to the young Sun, $\kappa$ Ceti, to estimate the impact of solar variability on the surface UV environment; they find the effect to be minor due to strong atmospheric attenuation.

\citet{Rugheimer2015} use a coupled climate-photochemistry model to compute radiative transfer through, among others, an atmosphere corresponding to the Earth at 3.9 Ga. Their model assumes and atmospheric pressure of 1 bar. It assumes atmospheric mixing ratios of 0.9, 0.1, and $1.65 \times 10^{-6}$ for N$_2$, CO$_2$, and CH$_4$, respectively, coupled with modern abiotic outgassing rates of gases such as SO$_2$ and H$_2$S, and iterates to photochemical convergence. They report the resulting actinic fluxes (spherically integrated radiances) at the bottom of the atmosphere. \citet{Rugheimer2015} reiterated the findings of previous workers that overall far more UV flux reached the surface of the primitive Earth compared to the modern day, with a cut-off at 200 nm due to shielding from CO$_2$ and H$_2$O. 

Our work builds on these previous efforts. Like \citet{Rugheimer2015}, we employ a two-stream multilayer approximation to radiative transfer, which consequently accounts for multiple scattering. Proper treatment of scattering is crucial in studies of the anoxic primitive Earth because of the uncovering of an optically thick yet scattering-dominated regime due to the absence of oxic shielding. For example, for a 0.9 bar N$_2$/0.1 bar CO$_2$ atmosphere of the kind considered by \citet{Rugheimer2015}, the N$_2$ column density is $1.88\times10^{25}$ cm$^{2}$ and the CO$_2$ column density is  $2.09\times10^{24}$ cm$^{2}$. At 210 nm, the Rayleigh scattering cross-section due to N$_2$ is $2.9\times10^{-25}$ cm$^{-2}$ and the Rayleigh scattering cross-section due to CO$_2$ is $6.8\times10^{-25}$ cm$^{-2}$, corresponding to a scattering optical depth of $\tau=6.8>1$. This optically thick scattering regime is shielded on Earth by strong O$_2$/O$_3$ absorption, but is revealed under anoxic prebiotic conditions. In this regime, scattering interactions and reflection from the surface become common, and self-consistent calculation of the upward and downward scattered fluence becomes important. We argue that consequently the radiative transfer formalism of \citet{Cnossen2007} is inappropriate, because it implicitly neglects multiple-scattering; it also ignores coupling between the upward and downward streams, and implicitly assumes an albedo of zero. Such an approximation may be reasonable on the modern Earth, where the scattering regime is confined to the optically thin region of the atmosphere by O$_2$ and O$_3$, meaning there are few scattering events and limited backscatter of reflected radiation. However, it is inappropriate for the anoxic prebiotic Earth where much of the prebiotically critical 200-300 nm regime is both scattering and optically thick, especially when treating cases with high albedo (e.g. snowfields). 

Like \citet{Cnossen2007} and \citet{Cockell2002}, we explore multiple atmospheric compositions. However, we treat variations in the abundance of each gas independently, in order to isolate each gas's effect individually, and explore a broader range of gases and abundances. We also explore the effects of albedo and zenith angle, which these earlier works did not. 

\begin{figure}[H]
\centering
\includegraphics[width=16.5 cm, angle=0]{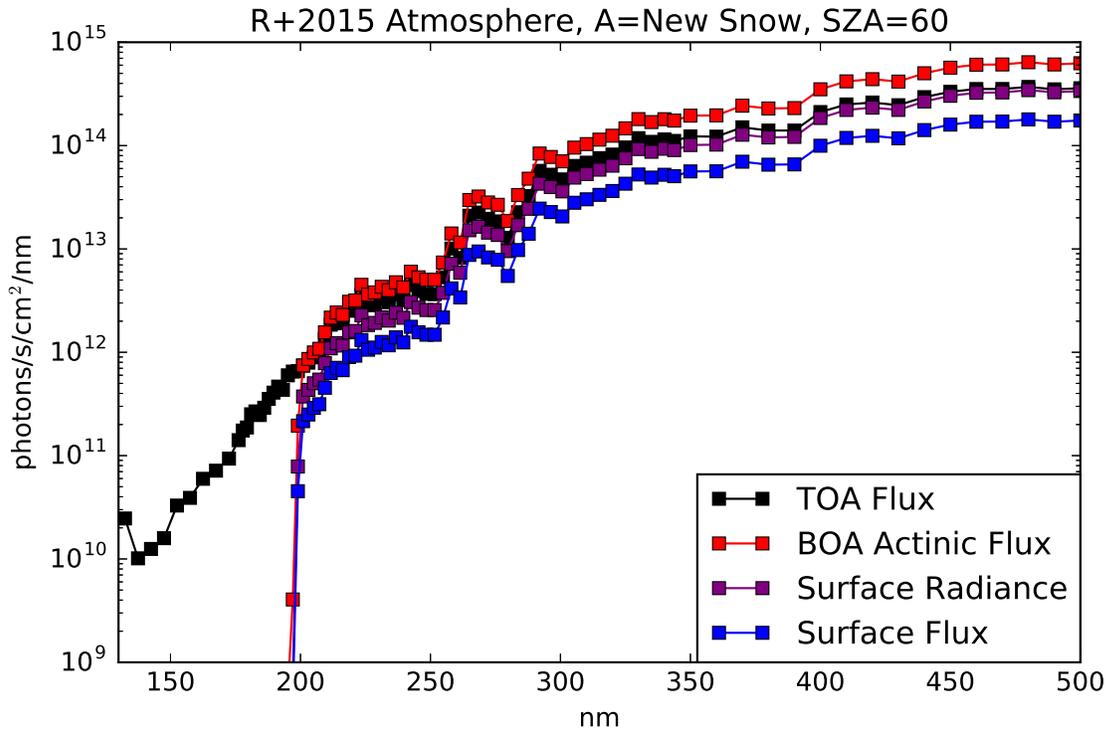}
\caption{TOA (top-of-atmosphere) solar flux, surface flux, BOA actinic flux, reported by \citet{Rugheimer2015}) and surface radiance for a planet with an atmosphere corresponding to that calculated by \citet{Rugheimer2015} for the 3.9 Ga Earth, with an albedo corresponding to fresh snow and a solar zenith angle of 60$^\circ$. In this example, these quantities can vary by up to a factor of 3.6, despite being for identical physical conditions and having the same units. In this paper, we report the surface radiance.  \label{fig:intvsflux}}
\end{figure}

Finally, \citet{Cnossen2007} and \citet{Cockell2002} reported the surface flux. However, as pointed out by \citet{Madronich1987}, the flux "describes the flow of radiant energy though the atmosphere, while the [intensity] concerns the probability of an encounter between a photon and a molecule''. The distinction is often academic from a laboratory perspective, since such in such studies the zenith angle of the source is often 0, meaning that the flux and spherically-integrated radiance\footnote{Also known as intensity; see \citealt{LiouBook}, page 4}  are identical. However, the flux can deviate significantly from the radiance in a planetary context (see, e.g., \citealt{Madronich1987}). \citet{Rugheimer2015} report the actinic flux (i.e. the integral over the unit sphere of the radiance field; see \citealt{Madronich1987}) at the bottom-of-atmosphere (BOA). This quantity, however, includes the upward diffuse reflection from the planet, which a molecule lying on the surface would not be not exposed to. We instead report what we term the \textit{surface radiance}, which is the integral of the radiance field at the planet surface, integrated over the hemisphere defined by positive elevation (i.e. that part of the sky not blocked by the planet surface). Figure~\ref{fig:intvsflux} demonstrates the difference between the surface flux, the BOA actinic flux, and the surface radiance for the model atmosphere of \citet{Rugheimer2015}, with a surface albedo corresponding to fresh snow and SZA=$60^\circ$.

\subsection{Constraints on the Composition of the Atmosphere at 3.9 Ga}
In this section, we briefly summarize available constraints on the terrestrial atmosphere at $\sim3.9$ Ga. A more detailed discussion is available in our earlier paper (\citealt{Ranjan2015}, Appendix B). 

Measurements of oxygen isotopes in zircons suggest the existence of a terrestrial hydrosphere by 4.4 Ga, usually interpreted as evidence that liquid water was stable at Earth's surface \citep{Mojzsis2001, Wilde2001, Catling2007}. Since the Sun was 30\% less luminous in this era (the "Faint Young Sun Paradox"), an enhanced greenhouse effect, e.g. through higher levels of CO$_2$, is usually invoked \citep{Kasting1993, Wordsworth2013, Kasting2014}. The initial nebular atmosphere is thought to have been lost soon after planet formation, and the subsequent atmosphere is thought to have been dominated by volcanic outgassing from high-temperature magmas. Measurements of ancient volcanic rocks suggest that the redox state of the Earth's mantle, and hence the gas speciation from magma melts, has not changed since 4.3 Ga \citep{Trail2011, Delano2001}, suggesting that an outgassed atmosphere would be dominated by CO$_2$, H$_2$O, SO$_2$, with H$_2$S also being delivered. Measurements of N$_2$/Ar fluid inclusions in 3.5 Ga quartz crystals \citep{Marty2013} have been used to demonstrate that N$_2$ was also a major atmospheric constituent, established at levels of 0.5-1.1 bar by 3.5 Ga (comparable to the present day). SO$_2$ and H$_2$S are not expected to have persisted at high levels in the atmosphere due to their tendency to photolyze and/or oxidize; however, it has been suggested that during epochs of high volcanism volcanogenic reductants could exhaust the surface oxidant supply, permitting transient buildup of gases vulnerable to oxidation, e.g. SO$_2$, to the 1-100 ppm level \citep{Kaltenegger2010}. O$_2$ and its by-product O$_3$ are thought to have been rare due to strong sinks from volcanogenic reductants coupled with a lack of the biogenic oxygen source. This low-oxygen hypothesis is reinforced by measurements of mass-independent fractional of sulfur in rocks from $>$2.45 Ga \citep{Farquhar2000}, which suggests atmospheric UV throughput was high and oxygen/ozone content was low in the atmosphere prior to 2.45 Ga \citep{Farquhar2001, Pavlov2002}, and measurements of Fe and U-Th-Pb isotopes from a 3.46 Ga chert, which are consistent with an anoxic ocean \citep{Li2013}.

In summary, available geological constraints are suggestive of an atmosphere at 3.9 Ga with N$_2$ levels roughly comparable to the modern day, with sufficient concentration of greenhouse gases (e.g. CO$_2$) to support surface liquid water. O$_2$ (and hence O$_3$) levels are thought to have been low. If the young Earth were warm, water vapor would have been an important atmospheric constituent. During epochs of high volcanism, reducing volcanogenic gases (e.g. SO$_2$) may also have been  important constituents of the atmosphere. 

\section{Surface UV Radiation Environment Model\label{sec:uv2g}}
\subsection{Model Description\label{sec:uv2gmodeldesc}}
We use the two-stream approximation to compute the radiative transfer of UV radiation through the Earth's atmosphere. We choose this method to follow and facilitate intercomparison with past work on this subject (e.g., \citealt{Cockell2002}, \citealt{Rugheimer2015}). We follow the treatment of \citet{Toon1989}, and we use Gaussian quadrature to connect the diffuse radiance (intensity) to the diffuse flux since \citet{Toon1989} find Gaussian quadrature to be more accurate than the Eddington and hemispheric mean closures at solar (shortwave) wavelengths. We do not include a pseudo-spherical correction because the largest SZA we consider is $66.5^{\circ}$, and radiative transfer studies for the modern Earth suggest the pseudo-spherical correction is only necessary for SZA$>75^{\circ}$ \citep{Kylling1995}. 

While we are most interested in radiative transfer from 100-400 nm due to the prebiotically interesting 200-300 nm range \citep{Ranjan2015}, our code can model radiative transfer out to 900 nm. The 400-900 nm regime where the atmosphere is largely transparent is useful because it enables us to compare our models against other codes (e.g., \citealt{Rugheimer2015}) and observations (e.g. \citealt{Wuttke2006}) which extend to the visible.  We include both solar radiation and blackbody thermal emission in our source function and boundary conditions. Planetary thermal emission is negligible at UV wavelengths for habitable worlds: we include it because the computational cost is modest, and because it may be convenient to those wishing to adapt our code to exotic scenarios where the planetary thermal emission is not negligible compared to instellation at UV wavelengths. We note as a corollary that this makes our model insensitive to the temperature profile and surface temperature. 

Our code includes absorption and scattering due to gaseous N$_2$, CO$_2$, H$_2$O, CH$_4$, SO$_2$, H$_2$S, O$_2$, and O$_3$. We do not include extinction due to atmospheric particulates or clouds, hence our results correspond to clear-sky conditions. Laboratory studies suggest that [CH$_4$]/[CO$_2$]$\geq0.1$ is required to trigger organic haze formation \citep{DeWitt2009}. Such levels of CH$_4$ are unlikely to be obtained in the absence of biogenic CH$_4$ production \citep{Guzman-Marmolejo2013}, hence organic hazes of the type postulated by \citet{Wolf2010} are not expected for prebiotic Earth. Modern terrestrial observations suggest that clouds typically attenuate UV fluence by a factor of $\leq5\times$ under even fully overcast conditions \citep{Cede2002, Calbo2005}. Since our work focuses on the potential of atmospheric and surficial features to drive $\gtrsim10\times$ changes in surface UV, we might expect our conclusions to be only weakly sensitive to the inclusion of clouds. However, clouds on early Earth may have been thicker or had different radiative properties from modern Earth. Further work is required to constrain the potential impact of particulates and clouds on the surface UV environment of prebiotic Earth, and the results presented in this paper should be considered upper bounds.

We take the top-of-atmosphere (TOA) flux to be the solar flux at 3.9 Ga at 1 AU, computed at 0.1 nm resolution from the models of \citet{Claire2012}. \citet{Claire2012} use measurements of solar analogs at different ages to calibrate a model for the emission of the sun through its history. We choose 3.9 Ga as the era of abiogenesis because it coincides with the end of the Late Heavy Bombardment (LHB) and is consistent with available geological and fossil evidence for early life (see, e.g., \citealt{Ohtomo2013, Buick2007, Noffke2013, Hofmann1999, Noffke2006, Javaux2010}). Since two-stream radiative transfer is monochromatic, we integrate spectral parameters (solar flux, extinction and absorption cross-sections, and albedos) over user-specified wavelength bins, and compute the two-stream approximation for each bin independently. We use linear interpolation in conjunction with numerical quadrature to perform these integrals. Our wavelength bin sizes vary depending on the planned application, but in general range from 1-10 nm. 

We set the extinction cross-section of the gases in our model equal to laboratory or observational measurements from the literature when available, and equal to the scattering cross-section when not \footnote{i.e. we assumed no absorption where we lacked constraints}. We assume all scattering is due to Rayleigh scattering, and compute the Rayleigh scattering cross-section for all our molecules. Where total extinction cross-section measurements lie below the Rayleigh scattering prediction (e.g., O$_2$ from 370-400 nm), we set the total extinction cross-sections to the Rayleigh value and the absorption cross-section to zero. This formalism implicitly trusts the Rayleigh scattering calculation over the reported cross-sections; we adopt this step because at such low cross-sections, the measurements are more difficult and the error higher. For example, several datasets reported negative cross-sections in such regimes, which are clearly unphysical. The extinction cross-section measurements and Rayleigh scattering formalism used in our model are described in Appendix~\ref{sec:XCs}. 

Two-stream radiative transfer models require the partitioning of the atmosphere into $N$ homogenous layers, and requires the user to specify the optical depths ($\tau_i$), single-scattering albedo ($\omega_{0_{i}}$), and asymmetry parameter $g_i$ across each layer ($0\leq i \leq N-1$), as well as the solar zenith angle (SZA) $\theta_0$ and the albedo of the planetary surface $A$. Inclusion of blackbody emission further requires the planetary surface temperature $T_{surf}$ and the temperature at the layer boundaries, $T_j$ ($0\leq i \leq N$). Since we consider only Rayleigh scattering, $g_i=0$ for all $i$. We compute $\omega_{0_{i}}$ by computing the molar-concentration-weighted scattering and total extinction cross-sections for the atmosphere in each homogenous layer, $\sigma_{scat_{i}}$ and $\sigma_{tot_{i}}$, and taking their ratio: $\omega_{0_{i}}=\sigma_{scat_{i}}/\sigma_{tot_{i}}$. For reasons of numerical stability, we set the maximum value\footnote{We previously followed \citet{Rugheimer2015} and imposed a maximum on $\omega_{0_{i}}$ of $1-1\times10^{-3}$, but we found that for thick, highly scattering atmospheres (e.g. multibar CO$_2$ atmospheres), this comparatively low upper limit on $\omega_{0_{i}}$ led to spurious absorption at purely scattering wavelengths} of $\omega_{0_{i}}$ to be $1-1\times10^{-12}$. Finally, we compute the the optical depth $\tau_i=d_i\times \sigma_{tot_{i}}\times n_i$, where $d_i$ is the thickness of each layer and $n_i$ is the number density of gas molecules in each layer. $d_i$ and N are chosen by the user. Unless otherwise stated, we followed the example of \citet{Segura2007} and \citet{Rugheimer2015} (their 3.9 Ga Earth case) and partitioned the atmosphere into 64 1-km thick layers. $T_j$, $n_i$, and the molar concentration of the gases are also specified by the user; they may be self-consistently specified through a climate model. $\theta_0$ and $A$ are free parameters. We explored both fixed values of A (see, e.g., \citet{Rugheimer2015}) as well as values of A corresponding to different physical surface media; see Appendix~\ref{sec:Albedos} for details.

We make the following modifications to the \citet{Toon1989} formalism. First, while \citet{Toon1989} adopt a single albedo for the planetary surface for both diffuse and direct streams, we allow for separate values for the diffuse and direct albedos \citep{Coakley2003}.  We use the direct albedo when computing the reflection of the direct solar beam at the surface, and the diffuse albedo for the reflection of the downwelling diffuse flux from the atmosphere off the surface. Appendix~\ref{sec:Albedos} discusses the albedos used in more detail. 

The \citet{Toon1989} two-stream formalism provides the upward and downward diffuse flux in each layer of the model atmosphere as a function of optical depth of the layer, $F^{\uparrow}_{i}(\tau)$ and $F^{\downarrow}_{i}(\tau)$, where $\tau$ is the optical depth within the layer. From these quantities, we can compute the net flux at any point in the atmosphere, $F_{net_{i}}(\tau)=F^{\uparrow}_i(\tau)-F^{\downarrow}_i(\tau)-F_{dir}(\tau_{c_i}+\tau)$,  where $F_{dir}(\tau)=\mu_0\pi F_{s}\exp(-\tau/\mu_0)$\footnote{via Beer-Lambert law}, $\mu_0=\cos(\theta_0)$ and $\theta_0$ is the solar zenith angle,  $\pi F_s$ is the solar flux at Earth's orbit, and $\tau_{c_{i}}$ is the cumulative optical depth from the TOA to the top of layer $i$. We can similarly compute the mean intensity $J$ via $4\pi J=(1/\mu_1)(F^{\uparrow}+F^{\downarrow})+F_{dir}/\mu_0$. We can also calculate the surface radiance $I_{surf}=F^{\downarrow}_{N}/\mu_1+F_{dir}(\tau_{c_{N}})/\mu_0$, where $F^{\downarrow}_{N}$ and $\tau_{c_{N}}$ are the downward diffuse flux and the cumulative optical depth at the bottom edge of the $(N-1)$th layer, respectively. In Gaussian quadrature for the $n=1$ (two-stream) case, $\mu_1=1/\sqrt{3}$ \citep{Toon1989, LiouBook, Liou1974}.

For each run of our model, we verify that the total upwelling flux at TOA $F^{\uparrow}(0)$ was less than or equal to the total incoming flux $F_{dir}(0)$ integrated over all UV/visible wavelengths, which is required for energy conservation since the Earth is a negligible emitter at these wavelengths. The code and auxiliary files associated with this model are available at: \url{https://github.com/sukritranjan/ranjansasselov2016b}.

\section{Model Validation\label{sec:testing}}
In this section, we describe our efforts to test and validate our radiative transfer model. We describe tests of physical consistency in the pure absorption and scattering limiting cases, comparisons of our model to published radiative transfer calculations, and the efficacy of our model at recovering published measurements of surficial UV radiance and irradiance.

\subsection{Tests of Model Physical Consistency: The Absorption and Scattering Limits}
We describe here tests of the physical consistency of our model in the limits of pure absorption and pure scattering. We use the atmospheric model (composition and T/P profile) of \citet{Rugheimer2015} described in Section~\ref{sec:rugheimer2015}, and evaluate radiative transfer through this atmosphere in these two limiting cases. This atmospheric model includes an optically thick regime ($\tau>1$)from 130-332.5 nm and an optically thin regime ($\tau<1$) from 332.5-855 nm. For each of these limiting cases, we evaluate radiative transfer corresponding to a range of albedos and solar zenith angles. We evaluate uniform albedos of 0, 0.20, and 1, corresponding to the extrema of possible albedo values and the albedo assumed by \citet{Rugheimer2015}. We evaluate solar zenith angles of 0$^{\circ}$, 60$^{\circ}$ and 85$^{\circ}$, corresponding to extremal values of the possible solar zenith angle along with the value corresponding to the \citet{Rugheimer2015} findings. We choose 85$^{\circ}$ as our limit instead of 90$^{\circ}$ since the plane-parallel approximation breaks down when the Sun is sufficiently close to the horizon.

\subsubsection{Pure Absorption Limit}
As noted by \citet{Toon1989}, in the limit of a purely absorbing atmosphere, the diffuse flux should vanish and the surface flux should reduce to the direct flux. In exploring the pure absorption limit, we cannot set $\omega_0=0$ as under Gaussian quadrature $\gamma_2$=0 for $\omega_0=0$, leading to a singularity when evaluating $\Gamma$ under the \citet{Toon1989} formalism. We tried values for $\omega_0$ ranging from $10^{-3}$ to $10^{-10}$ for the $A=0.20$, $\theta_0=60^{\circ}$ case. For all values of $\omega_0$, we found the diffuse surface flux to be highly suppressed relative to the direct TOA flux. For $\omega_0=10^{-5}$, the diffuse flux was suppressed relative to the TOA flux by $\gtrsim6$ orders of magnitude in each wavelength bin. The diffuse flux is also strongly suppressed relative to the direct surface flux, except at short wavelengths ($\lambda<198$ nm) where extinction is so strong that that the diffuse layer blackbody flux dominates over the direct solar flux. 

\begin{figure}[H]
\centering
\includegraphics[width=12 cm, angle=0]{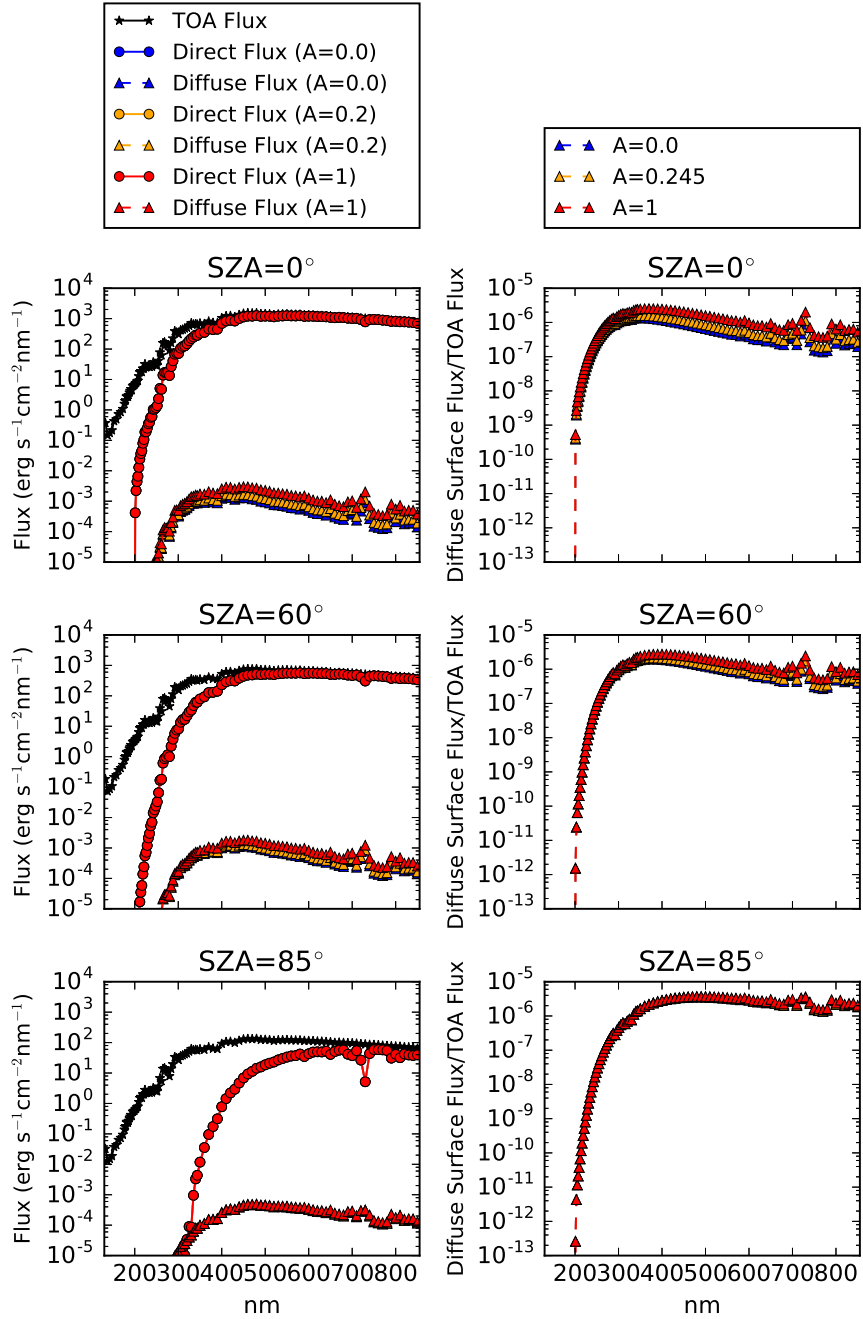}
\caption{Direct, diffuse, and TOA fluxes (left-hand column) and diffuse flux at the surface normalized by TOA flux (right-hand column) for different solar zenith angles and surface albedos, for an atmosphere corresponding to the \citet{Rugheimer2015} 3.9 Ga Earth model with $\omega_0=10^{-5}$. The diffuse flux vanishes in this limit, as expected. \label{fig:pureabs}}
\end{figure}

We evaluate the surface flux for $\omega_0=10^{-5}$ for a range of $A$ and $\theta_0$. Figure~\ref{fig:pureabs} presents the results. In all cases, the diffuse flux is highly suppressed relative to the TOA flux across all optical depths. \citet{Toon1989} report that while the pure absorption limit is satisfied by two-stream approximations with Gaussian closure for $A=0$, for $A>0$ exponential instabilities may lead to anomalous behavior. We do not observe this phenomenon in our model. We conclude that our implementation of the two-stream algorithm passes the absorption limit test.

\subsubsection{Pure Scattering Limit}
In the limit of a purely scattering atmosphere ($\omega_0=1$), $F_{net}$ should be a constant throughout the atmosphere at all wavelengths since radiation is neither absorbed nor emitted by the atmospheric layers \citep{Liou1973, Toon1989}. In exploring this limit, we cannot set $\omega_0=1$: separate solutions are required for the fully conservative case (see e.g. \citealt{Liou1973}). However, in practice we can set $\omega_0$ arbitrarily close to 1 \citep{Toon1989} and ensure that the net flux is constant throughout the atmosphere, or at least that its variations are small compared to the incident flux. We computed $F_{net}$ at layer boundaries in the atmosphere for the $A=0.20$, $\theta_0=60^{\circ}$ case, for values for $\omega_0$ ranging from $1-10^{-3}$ to $1-10^{-12}$. For each wavelength bin, we computed the maximum deviation from the median net flux in the atmospheric column, and normalized this deviation to the incident flux. For $\omega_0=1-10^{-3}$, the variation of $F_{net}$ from the median value ranged from $6\times10^{-2}$ at short wavelengths to $7\times10^{-5}$ at long wavelengths. The increase in deviation towards shorter wavelengths is expected because of the higher opacity at shorter wavelengths. Increasing $\omega_0$ decreased the variation in $F_{net}$.  For $\omega_0=1-10^{-7}$, the fractional deviation of $F_{net}$ from the columnar median varied from $6\times10^{-4}$ to $7\times10^{-9}$, and for $\omega_0=1-10^{-12}$, the deviation of $F_{net}$ varied from $2\times10^{-6}$ to $7\times10^{-14}$. 

\begin{figure}[H]
\centering
\includegraphics[width=12 cm, angle=0]{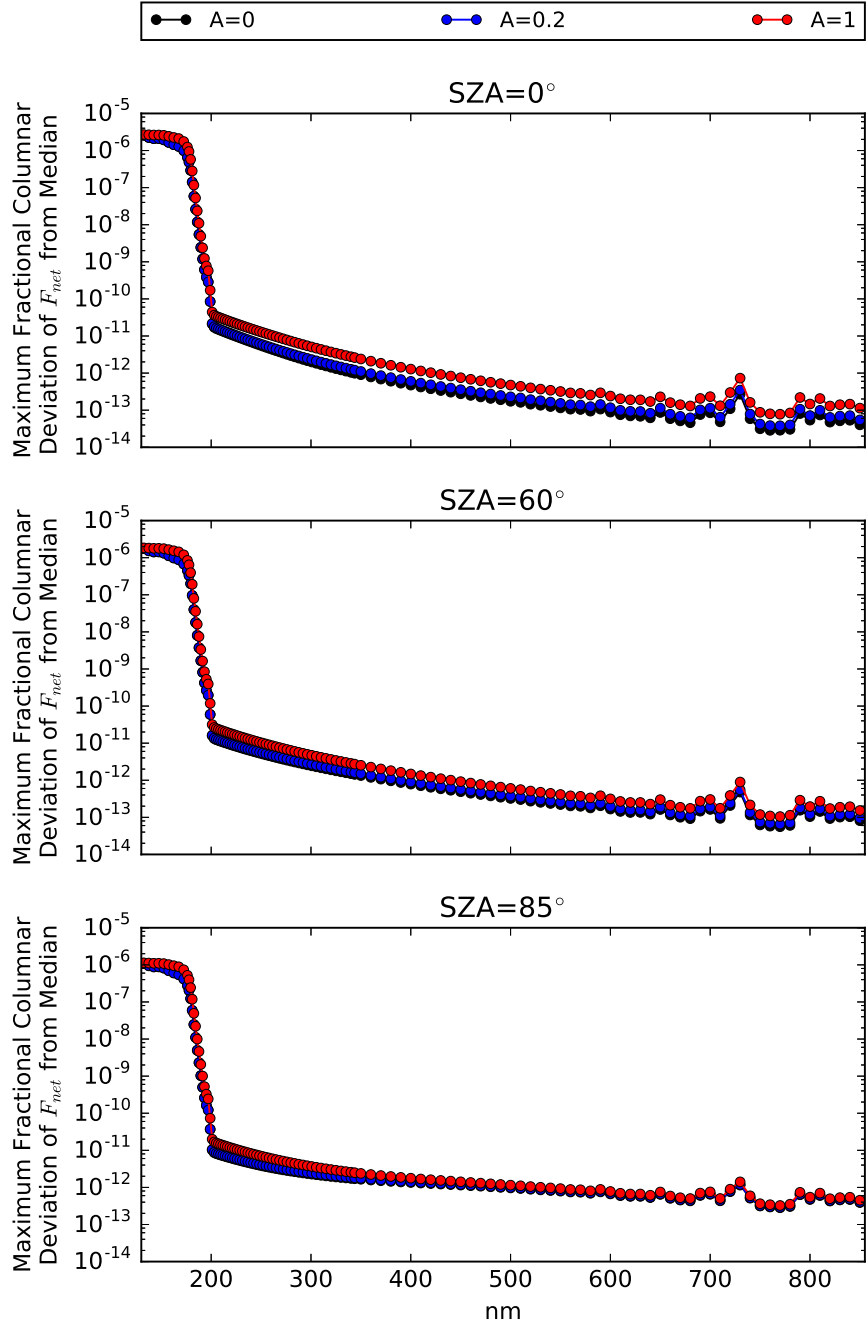}
\caption{The maximum deviation of $F_{net}$ from its median value in a given atmospheric column as a function of wavelength for an atmosphere corresponding to the \citet{Rugheimer2015} 3.9 Ga Earth model, with $\omega_0=1-10^{-12}$ and a variety of surface albedos and solar zenith angles. In the scattering limit, $F_{net}$ approaches a constant value, with the variation in $F_{net}$ decreasing as $\omega_0$ approaches 1. \label{fig:purescat}}
\end{figure}

We compute the maximum deviation of $F_{net}$ at the layer boundaries from the columnar median as a function of wavelength for $\omega_0=1-10^{-12}$ for a range of $A$ and $\theta_0$. Figure ~\ref{fig:purescat} presents the results. At optically thin wavelengths, larger albedos and zenith angles lead to higher deviations; we attribute this to higher levels of flux scattered into the more computationally difficult diffuse stream. At optically thick wavelengths, the magnitude of the deviations is insensitive to the planetary albedo. We attribute this to the extinction of incoming flux higher in the atmosphere, meaning that surface properties have less impact on the flux profile. In the optically thick regime, smaller zenith angles lead to higher deviations. For all values of $\theta_0$ and $A$ considered here, the columnar deviation from uniformity is $<3\times10^{-6}$ of incoming fluence across all wavelengths for $\omega_0=1-10^{-12}$, and the deviation decreases as $\omega_0$ approaches 1 as expected.

\subsection{Reproduction of Results of \citet{Rugheimer2015}\label{sec:rugheimer2015}}
In this section, we describe our efforts to recover the results of \citet{Rugheimer2015} with our code. \citet{Rugheimer2015} present a model for the total BOA actinic flux on the 3.9 Ga Earth orbiting the 3.9 Ga Sun\footnote{Figure 2, "Sun" curve} from 130-855 nm. They couple a 1D climate model \citep{Kasting1986, Pavlov2000, HaqqMisra2008} and a 1D photochemistry model \citep{Pavlov2002, Segura2005, Segura2007} and iterate to convergence. They assume an overall atmospheric pressure of 1 bar and atmospheric mixing ratios of 0.9, 0.1, and $1.65 \times 10^{-6}$ for N$_2$, CO$_2$, and CH$_4$, respectively. For all other gases, their model assumes outgassing rates corresponding to modern terrestrial nonbiogenic fluxes. 

When computing layer-by-layer radiative transfer, \citet{Rugheimer2015} include absorption due to O$_3$, O$_2$, CO$_2$, and H$_2$O and Mie scattering due to sulfate aerosols. Rayleigh scattering is computed via an N$_2$-O$_2$ scattering law \citep{Kasting1982} that is scaled to include the effect of enhanced CO$_2$ scattering. \citet{Rugheimer2015} partition their atmosphere into 64 1-km layers and assume a solar zenith angle of 60$^{\circ}$. As with our model, they compute the solar UV radiative transfer using a \citet{Toon1989} two-stream approximation with Gaussian quadrature closure. The surface albedo $A=0.20$ is tuned to yield a surface temperature of 288K in the modern Earth/Sun system, to approximate the effect of clouds \citep{Rugheimer2015}. 

\begin{figure}[H]
\centering
\includegraphics[width=12 cm, angle=0]{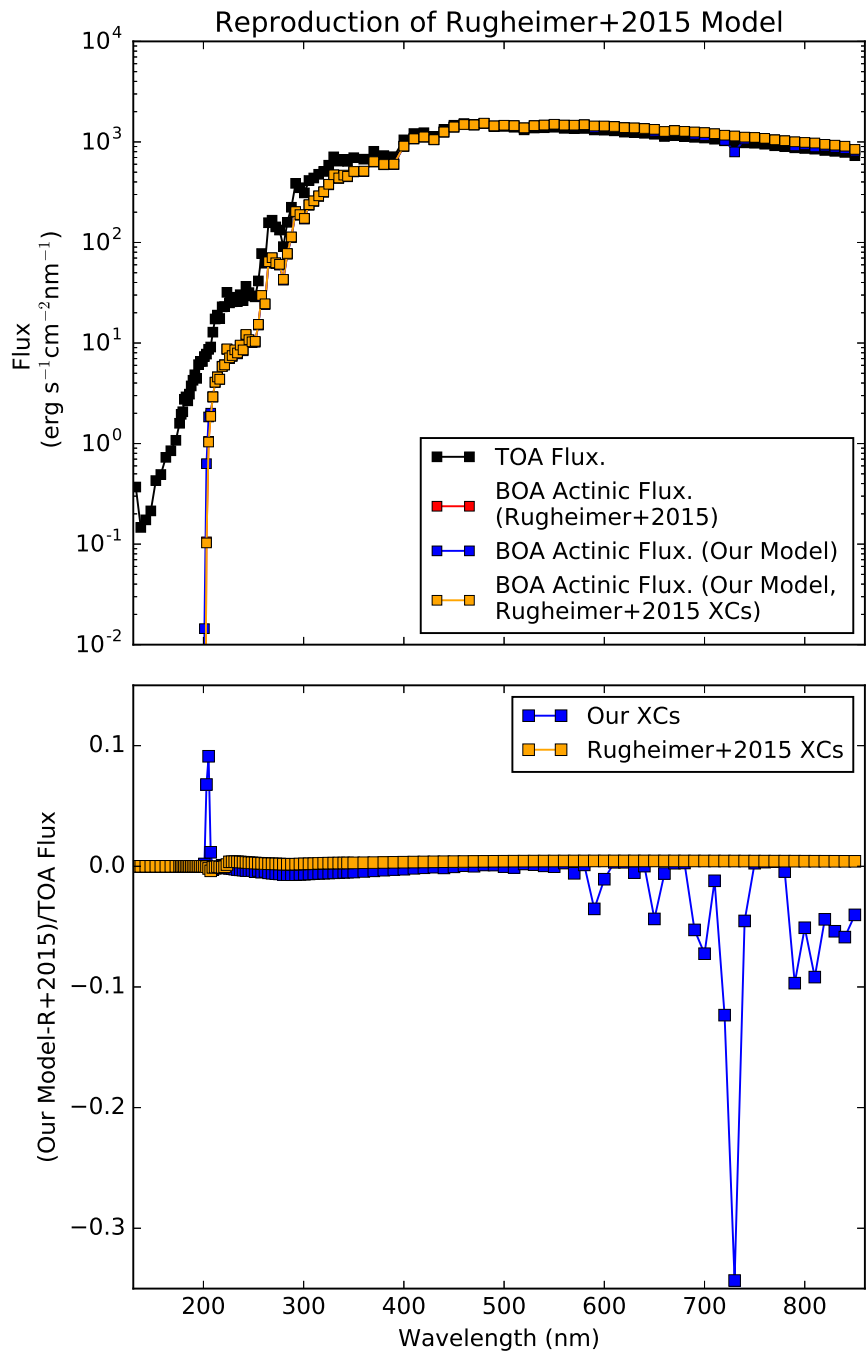}
\caption{Comparison of the BOA actinic fluxes obtained with our radiative transfer model for the updated 3.9 Ga Earth atmosphere of \citet{Rugheimer2015}, versus that computed by the authors themselves. The top shows the computed actinic fluxes and TOA flux, while the bottom shows the fractional difference between our models, normalized by the TOA flux. The difference between our and \citet{Rugheimer2015}'s results is driven by the use of different cross-section compilations; if we use the same cross-sections as \citet{Rugheimer2015} (orange curves), we agree with \citet{Rugheimer2015} to better than 0.45\% of the TOA flux. \label{fig:rugheimerepoch0}}
\end{figure}

We obtain the metadata\footnote{i.e., wavelength bins, mixing ratios as a function of altitude, atmospheric profile, and emergent spectra normalized to the TOA flux, i.e. $4\pi J_{N-1}/F_\sun$, where $J_{N-1}$ is the mean intensity in the middle of the lowest layer of their atmospheric model and $F_\sun$ is the flux of solar radiation incident on the TOA} for an updated version of the 3.9 Ga Earth model of \citet{Rugheimer2015}, courtesy of the authors. We use this metadata to run our radiative transfer model on the \citet{Rugheimer2015} atmospheric model. Figure~\ref{fig:rugheimerepoch0} summarizes the results. The top row compares the incident flux at TOA (black) with the \citet{Rugheimer2015} results (red) and our model computations (blue, orange). The \citet{Rugheimer2015} results are not visible due to the close correspondence between our models. The bottom row gives the difference between our model and the \citet{Rugheimer2015} results, normalized to the TOA flux.

Our model reproduces the \citet{Rugheimer2015} results to within 34\% of the of the TOA flux. Much of the difference between our model and \citet{Rugheimer2015}'s can be accounted for by differences in the absorption cross-sections we use. Our cross-section lists are more complete than \citet{Rugheimer2015}; for example, our H$_2$O cross-section tabulation includes absorption at wavelengths longer than 208.3 nm, whereas \citet{Rugheimer2015} do not. Further, we include absorption due to SO$_2$ and CH$_4$, compute explicitly Rayleigh scattering on a gas-by-gas basis and include blackbody emission from atmospheric layers and the planetary surface, whereas \citet{Rugheimer2015} do not (though this last is not a significant factor given the paucity of planetary radiation at UV wavelengths). 

If we run our model using the \citet{Rugheimer2015} cross-sections and scattering formalism and include only absorption due to O$_2$, O$_3$, CO$_2$, and H$_2$O, we arrive at the orange curve. This curve matches the \citet{Rugheimer2015} results to within 0.45\% of the TOA flux. Our model, both with and without the \citet{Rugheimer2015} absorption, scattering, and emission formalism, can reproduce the scientific conclusions of \citet{Rugheimer2015} such as the 204 nm irradiance cutoff due to atmospheric CO$_2$. We conclude that our model is capable of reproducing the results of \citet{Rugheimer2015}.

\subsection{Comparison To Modern Earth Surficial Measurements}
We describe here comparisons of our radiative transfer model calculations to surface measurements of UV reported in the literature. 

\subsubsection{Reproduction of Antarctic Diffuse Spectral Radiance Measurements\label{sec:wuttke}}
\citet{Wuttke2006} report measurements of the diffuse spectral radiance (observed in the zenith direction) in Antartica.  We compare our model to their measurements of the diffuse radiance collected under low-cloud conditions (since our model does not include scattering processes due to clouds). The measurement site was flat and uniformly covered by snow,  and the solar zenith angle during the measurements was 51.2$^{\circ}$. When running our model, we assume the same solar zenith angle, and take the albedo of the site to match fresh-fallen snow. We run our model at a spectral resolution matching the \citet{Wuttke2006} measurements, i.e. from 280-500 nm at 0.25 nm resolution and from 501-1050 nm at 1 nm resolution. We run our model from 0-60 km of altitude, at 600 meter resolution (i.e. 100 layers evenly spaced in altitude), and assume a surface pressure of 1 bar. We assume composition and T/P profiles matching that of \citet{Rugheimer2013} for the modern Earth. In order to reproduce the \citet{Wuttke2006} measurements, we are obliged to reduce the water abundance by a factor of 10 relative to the \citet{Rugheimer2013} models. This makes sense, since Antartica is a dry desert environment. Similarly, the \citet{Rugheimer2013} model has an ozone total column depth of  $5.3\times10^{18}$ cm$^{-2}$ or 200 Dobson units (DU). For comparison, the Earth's typical ozone total column density is around 300 DU \citep{Patel2002} and a column depth of 220 DU is considered to be the start point for an ozone hole\footnote{See, e.g., \url{http://ozonewatch.gsfc.nasa.gov/}}. It is therefore unsurprising that matching the observed diffuse radiance in Antarctica requires scaling up the ozone abundance of \citet{Rugheimer2013}, by a factor of 1.25. While our simple model, which excludes trace absorbers, clouds and aerosols and is based on a globally averaged composition profile, cannot be expected to precisely replicate this measurement, we can reasonably expect it to identify major features of the modern surface UV environment.

\begin{figure}[H]
\centering
\includegraphics[width=12 cm, angle=0]{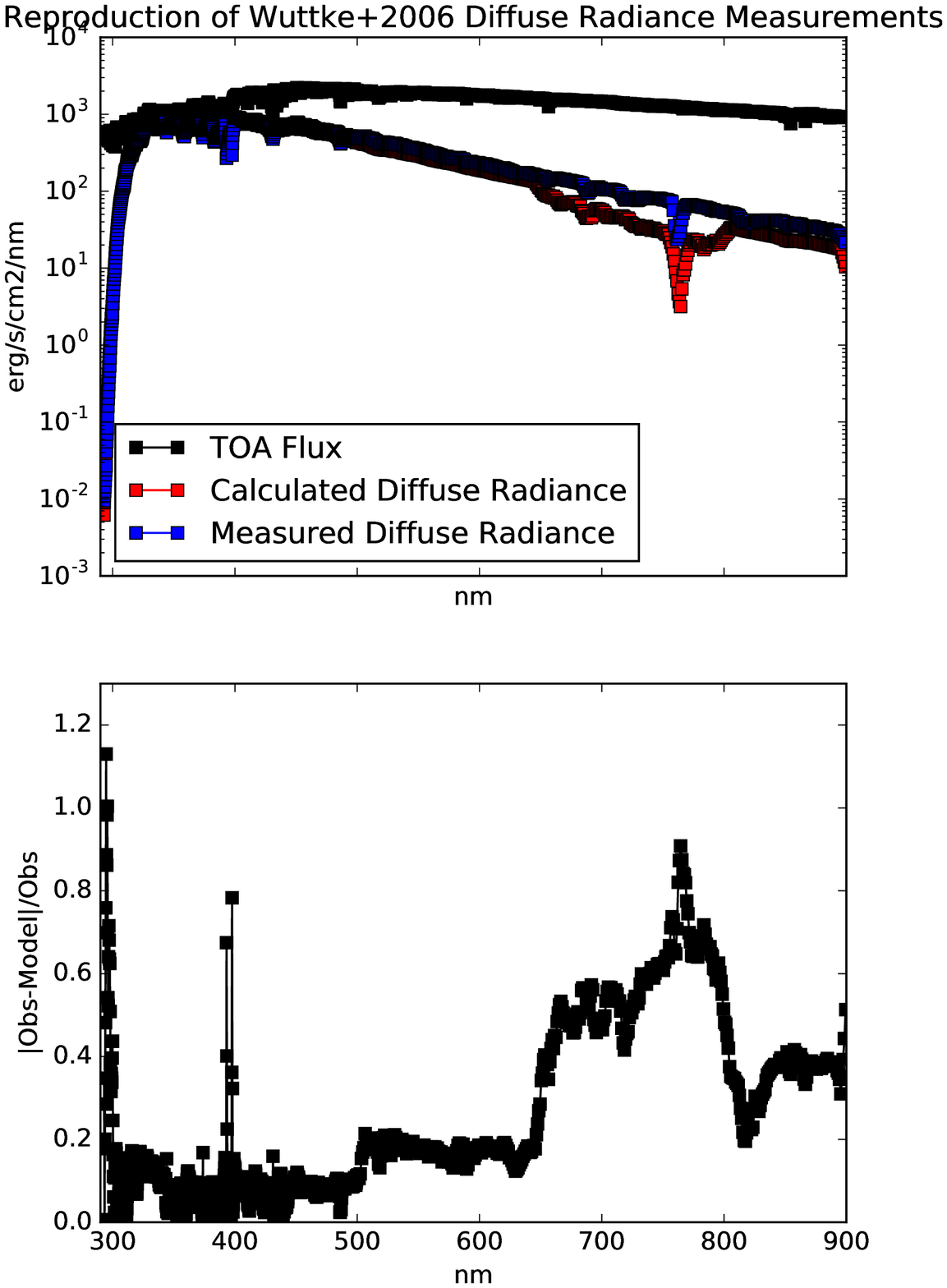}
\caption{Top: Zenith diffuse radiance measurement by \citet{Wuttke2006} in Antarctica, compared to the diffuse radiance calculated by our model for an atmosphere corresponding to the \citet{Rugheimer2013} modern Earth model, with the H$_2$O levels scaled down by a factor of 10 and the ozone column density scaled up by a factor of 1.25. Bottom: fractional difference between the measurements of \citet{Wuttke2006} and our model calculation. Our code recovers key features of the terrestrial UV environment, such as the shortwave cutoff due to ozone.\label{fig:wuttke}}
\end{figure}

Figure~\ref{fig:wuttke} presents the measured diffuse radiance observed by \citet{Wuttke2006} and our model calculation smoothed by a 10-point moving average (boxcar) filter. Our code correctly replicated the major features of the modern terrestrial UV environment, such as the existence and location of the shortwave cutoff due to ozone. Figure~\ref{fig:wuttke}also presents the fractional difference between our model prediction and the measurement. The difference is within a factor of 2.2, and is highest in regions of strong atmospheric attenuation of UV. This accuracy is sufficient to distinguish between spectral regions of low and high ($>100\times$) atmospheric attenuation, i.e. to identify the UV fluence that is suppressed by atmospheric absorbers (e.g., Section~\ref{sec:co2lim}). It is similarly sufficient to identify order-of-magnitude-or-greater changes in dose rates due to varying columns of a given absorber (e.g.,  Section~\ref{sec:so2}), particularly because the highest error occurs at the lowest throughput and hence has the least weight in the dose rate calculation. We conclude that our code is sufficiently accurate for the applications considered in this work.

\subsubsection{Reproduction of Toronto Surface Flux Measurements}

The World Ozone and UV Data Centre (WOUDC; \url{woudc.org}) compiles measurements of UV surface flux. We compare our radiative transfer model calculations to a measurement of the UV surface flux at Toronto from 292-360 nm on 6/21/2003 at 11:54:06 (solar time). We chose this measurement to compare to as it corresponded approximately to the data shown in \citet{Kerr2008} (i.e. a measurement in Toronto at noon in summer). The solar zenith angle at time of measurement was 20.376$^\circ$. We take the UV surface albedo to be 0.04, following the typical value suggested by \citet{Kerr2008}. We run our model from 292-360 nm at 0.5 nm resolution, matching the resolution of the measurements, from 0-60 km of altitude, at 600 meter vertical resolution. We assume a T/P profile and atmospheric composition profile matching that of \citet{Rugheimer2013} for the modern Earth. As in Section~\ref{sec:wuttke}, we note that the \citet{Rugheimer2013} model has a total ozone column density of 200 Dobson units (DU), while the ozone column measured for this observation was 354 DU. We consequently scale our ozone mixing ratios by a factor of 1.77 to match the true column depth. As with Section~\ref{sec:wuttke}, while we cannot expect our simple model to precisely replication the surface flux measurement,  we can reasonably expect it to identify major features of the modern surface UV environment.

\begin{figure}[H]
\centering
\includegraphics[width=12 cm, angle=0]{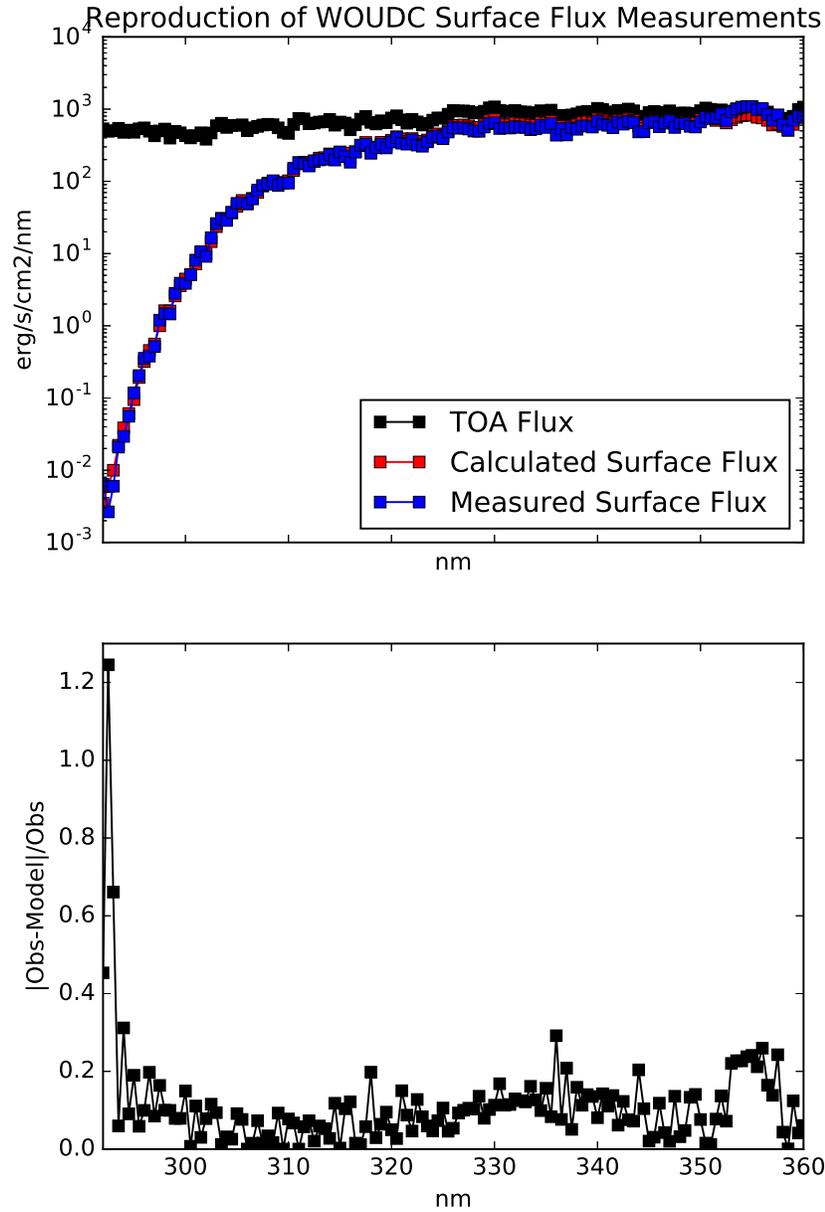}
\caption{Top: Surface flux reported by a WOUDC station near summer noon in Toronto, compared to the surface flux calculated by our model for an atmosphere corresponding to the \citet{Rugheimer2013} modern Earth model with the ozone column density scaled up by a factor of 1.77. Bottom: fractional difference between the measurements and our model. Our code clearly identifies and locates the shortwave cutoff due to ozone. \label{fig:woudc}} 
\end{figure}

Figure~\ref{fig:woudc} presents the measured UV flux compared to our model prediction, and the fractional difference between the two. Our model correctly replicates the shortwave UV cutoff due to ozone, which is characteristic of the modern surface UV environment. The relative difference between our model and the measurement is within a factor of 2.3, with the difference highest where the fluence is strongly suppressed   This performance is similar to that of our reproduction of \citet{Wuttke2006}, and is sufficient for the purposes of this paper.

\section{Results and Discussion\label{sec:resdisc}}
In this section, we apply our two-stream radiative transfer model to the \citet{Rugheimer2015} 3.9 Ga Earth atmospheric model and variants. We explore the impact of albedo, zenith angle, and atmospheric composition on the surface radiance. We again note that unlike \citet{Rugheimer2015}, we do not self-consistently calculate the photochemistry. Rather, we adopt ad-hoc values for these parameters to place bounds on the surface radiance environment. Our objective is to enable prebiotic chemists to correlate hypothesized prebiotic atmospheric composition (e.g. high levels of water vapor on a warm, wet young Earth) to the range of surficial UV environments that such gases would permit in a planetary context. 

In calculating our models, we step from 100 to 500 nm of wavelength, at a resolution of 1 nm. This wavelength range includes the prebiotically crucial 200-300 nm range \citep{Ranjan2015} and the onsets of CO$_2$, H$_2$O, and CH$_4$ absorption. Unless stated otherwise, we assume the atmospheric composition and T/P profile calculated for the 3.9 Ga Earth by \citet{Rugheimer2015}. We calculate radiative transfer in 1 km layers starting at the planet surface and ending at a high of 64 km, which corresponds to 7.7 scale heights for this atmosphere. 

\subsection{Action Spectra and UV Dose Rates\label{sec:dosimeters}}
To quantify the impact of the surface radiation environments on prebiotic chemistry, we follow the example of \citet{Cockell1999} in computing biologically weighted UV dose rates. Specifically, we compute the biologically effective relative dose rate $$D=(\int_{\lambda_{0}}^{\lambda_{1}} d\lambda A(\lambda)I_{surf}(\lambda))/(\int_{\lambda_{0}}^{\lambda_{1}} d\lambda A(\lambda)I_{space}(\lambda)),$$ where $A(\lambda)$ is an action spectrum, $\lambda_0$ and $\lambda_1$ are the limits over which $A(\lambda)$ is defined, $I_{surf}(\lambda)$ is the hemispherically-integrated total UV surface radiance, and $I_{space}(\lambda)$ is the the solar flux at the Earth's orbit. An action spectrum parametrizes the relative impact of radiation on a given photoprocess as a function of wavelength, with a higher value of $A$ meaning that a higher fraction of the incident photons are being used in said photoprocess. Hence, $D$ measures the relative rate of a given photoprocess for a single molecule at the surface of a planet, relative to in space at the location of the planet. If we compute the dose rate $D_i$ corresponding to two UV surface radiance spectra $I_{surf, 1}$ and $I_{surf, 2}$ on a molecule that undergoes a photoprocess characterized by an action spectra $A$ and find $D_1>D_2$, we can say that the photoprocess encoded by $A$ proceeds at a higher rate under $I_{surf, 1}$  than $I_{surf, 2}$.

Previous workers used the modern DNA damage action spectrum \citep{Cockell2002, Cnossen2007, Rugheimer2015} as a gauge of the level of stress imposed by UV fluence on the prebiotic environment. However, this action spectrum is based on studies of highly evolved modern organisms. Modern organisms have evolved sophisticated methods to deal with environmental stress, including UV exposure, that would not have been available to the first life. Further, this approach presupposes that UV light is solely a stressor, and ignores its potential role as a eustressor for abiogenesis. 

In this work, we use the action spectra corresponding to the production of aquated electrons from photoionization of tricyanocuprate and to the cleavage of the N-glycosidic bond in uridine monophospate (UMP, an RNA monomer) to compute our biologically effective doses. These processes are simple enough to have plausibly been in operation at the dawn of the first life, particularly in the RNA world hypothesis  \citep{Gilbert1986, Copley2007, McCollom2013}\footnote{The RNA world is the hypothesis that RNA was the original autocatalytic information-bearing molecule. Under this hypothesis, the problem of abiogenesis reduces to an abiotic synthesis of autocatalytic RNA polymers}. In the following sections, we discuss in more detail our rationale for choosing these pathways, and how we construct the action spectra associated with them. 

\subsubsection{Eustressor Pathway: Production of Aquated Electrons From Photoionization of CuCN$_3^{2-}$} 
\citet{Ritson2012} outline a synthesis of glycolaldehyde and glyceraldehyde from HCN and formaldehyde. This pathway depends on UV light for the photoreduction of HCN mediated by the metallocatalyst tricyanocuprate (CuCN$_3^{2-}$, and \citet{Ritson2012} hypothesize this photoreduction is driven by photoionization of the tricyanocuprate, generating aquated electrons ($e^{-}_{aq}$). Such aquated electrons are useful in a variety of prebiotic chemistry, participating generally in the reduction of nitriles to amines, aldehydes to hydroxyls, and hydroxyls to alkyls\footnote{J. Szostak, private communication, 2/5/16}; see \citet{Patel2015} for an example of a potential prebiotic reaction network that leverages aquated electrons in numerous reactions. 

We define an action spectrum for the generation of aquated electrons from the irradiation of tricyanocuprate by multiplying the absorption spectrum of tricyanocuprate by the quantum yield (QY, number of $e^{-}_{aq}$ produced per photon absorbed) of $e^{-}_{aq}$ from the system. We take our absorption spectrum from the work of \citet{Magnani2015}, via \citet{Ranjan2015}. The QY of $e^{-}_{aq}$ production, $\Phi_{e^{-}_{aq}}$, is not known. Following \citet{Ritson2012}'s hypothesis that $e^{-}_{aq}$ production is driven by tricyanocuprate photoionization, we assume the QY to be characterized by a step function with $\Phi_{e^{-}_{aq}}(\lambda\leq\lambda_{0})=\Phi_{0}$  and $\Phi_{e^{-}_{aq}}(\lambda>\lambda_{0})=0$ otherwise. We choose $\Phi_{0}=0.06$, consistent with the QY for tricyanocuprate measured by \citet{Horvath1984} at 254 nm. Empirically, we know $\lambda_{0}>254$ nm; to explore a range of possible $\lambda_{0}$, we consider $\lambda_0=254$ nm and $\lambda_0=300$ nm. The action spectrum is defined over the range $190-351$ nm, corresponding to the range of the absorption spectrum measured by \citet{Magnani2015}. As shorthand, we refer to this photoprocess under the assumption that $\lambda_0=$X nm by CuCN3-X. 

\subsubsection{Stressor Pathway: Cleavage of N-Glycosidic Bond of UMP} 
UMP is a monomer of RNA, a key product of the \citet{Powner2009} pathway, and critical molecule for abiogenesis in the RNA-world hypothesis for the origin of life. Shortwave UV irradiation of UMP cleaves the glycosidic bond joining the nucleobase to the sugar \citep{Gurzadyan1994}, destroying the biological effectiveness of this molecule. The reaction is difficult to reverse; indeed, the key breakthrough of the \citet{Powner2009} pathway was determining how to synthesize and incorporate this bond into the RNA monomers abiotically. Hence, this pathway represents a stressor for abiogenesis in the RNA-world hypothesis.

Glycosidic bond cleavage is not the only process operating in UMP at UV wavelengths. The (wavelength, QY) measured by \citet{Gurzadyan1994} for glycosidic bond cleavage in UMP in anoxic aqueous solution are (193 nm, $4.3\times10^{-3}$) and (254 nm, $(2-3)\times10^{-5}$). By comparison, the (wavelength, QY) they measure for chromophore loss (a measure of unaltered UMP abundance, based on absorbance at 260 nm) are (193 nm, $4\times10^{-2}$) and (254 nm,  $1.2\times10^{-3}$) -- 1-2 orders of magnitude higher. The chromophore loss at 254 nm is well studied; for UMP, it is mostly due to photohydration, with a minor contribution from photodimer formation. The photohydration can be reversed with 90-100\% efficiency via heating or lowering the pH \citep{Sinsheimer1954}, whereas further UV light (especially shortwave ~230 nm) can cleave the photodimers. Since these processes are reversible via dark reactions, the UMP in some sense is not fully "lost", unlike the glycosidic bond cleavage. We therefore argue the bond cleavage is more important than photohydration/photodimerization in measuring UV stress on UMP.

We define action spectra for the cleavage of the glycosidic bond in UMP by multiplying the absorption spectrum of UMP by the quantum yield for the process. We take the absorption spectra from the work of \citet{Voet1963}, which gives the absorption spectra of UMP at pH=7.6. The QY of glycosidic bond cleavage as a function of wavelength has not been measured. To gain traction on this problem, we use the work of \citet{Gurzadyan1994}, which found the QY of N-glycosidic bond cleavage in UMP in neutral aqueous solution saturated with Ar (i.e. anoxic) to be $4.3\times10^{-3}$ at 193 nm and $(2-3)\times10^{-5}$ for 254 nm. We therefore represent the QY curve as a step function with value $4.3\times10^{-3}$ for $\lambda\leq\lambda_0$ and $2.5\times10^{-5}$ for $\lambda>\lambda_0$. We consider $\lambda_0$ values of 193 and 254 nm, corresponding to the empirical limits from \citet{Gurzadyan1994}, as well as 230 nm, which corresponds to the end of the broad absorption feature centered near 260 nm corresponding to the $\pi-\pi^{*}$ transition and also to the transition to irreversible decomposition suggested by \citet{Sinsheimer1949}. As shorthand, we refer to this photoprocess under the assumption that $\lambda_0=$Y nm by CuCN3-Y.

\begin{figure}[H]
\centering
\includegraphics[width=16.5 cm, angle=0]{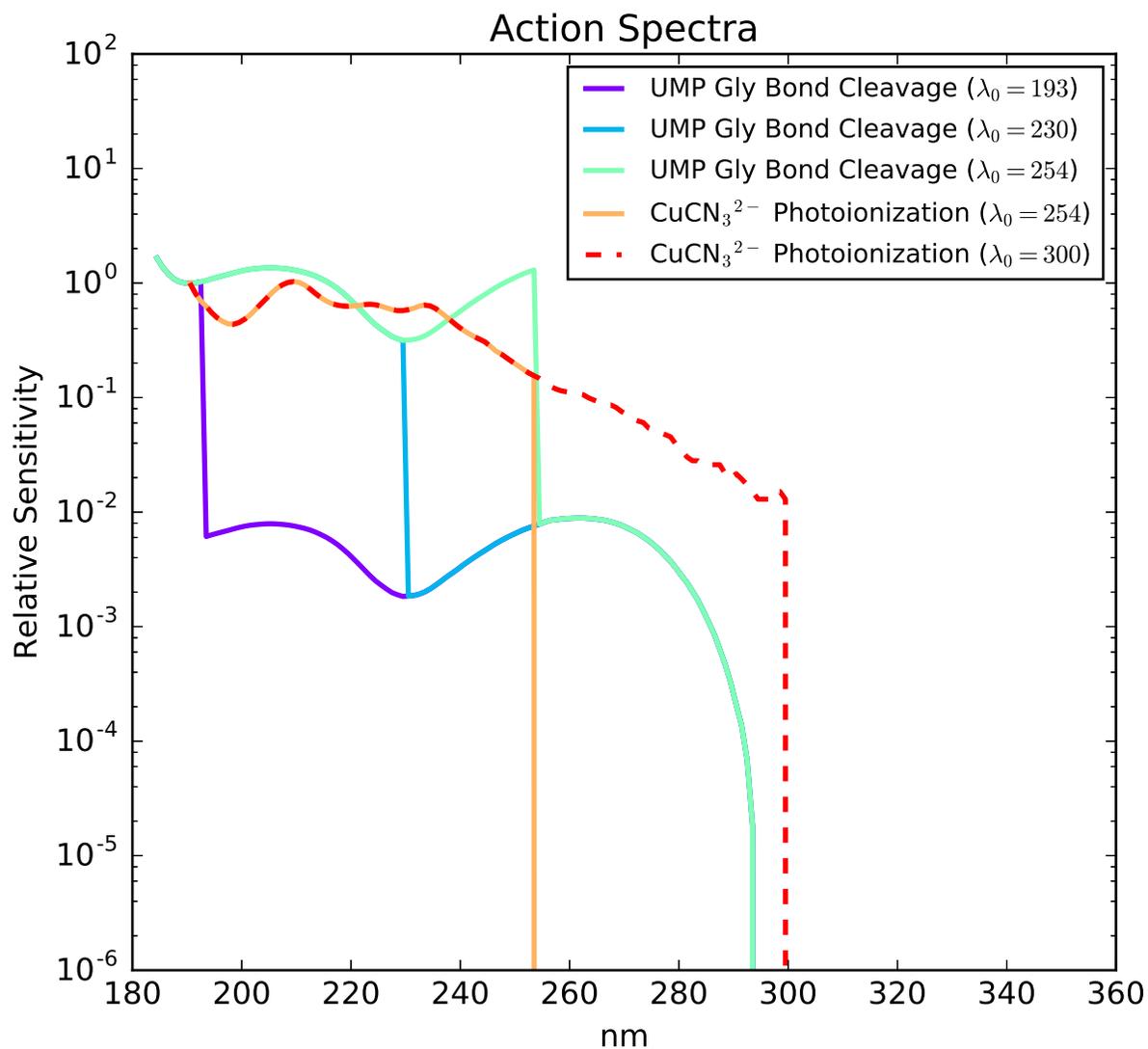}
\caption{Action spectra for photolysis of UMP and photoionization of CuCN$_3^{2-}$, assuming a step-function form to the QE for both processes with step at $\lambda=\lambda_0$ The spectra are arbitrarily normalized to 1 at 190 nm. \label{fig:actspec}}
\end{figure}

Figure~\ref{fig:actspec} shows the action spectra considered in our study. Action spectra are normalized arbitrarily (see, e.g., \citealt{Cockell1999} and \citealt{Rugheimer2015}), hence they encode information about relative, not absolute, UV dose rate. We arbitrarily normalize these spectra to 1 at 190 nm. 

\subsection{Impact of Albedo and Zenith Angle On Surface Radiance \& Prebiotic Chemistry\label{sec:albzen}}
In this section, we quantify the impact of varying albedo and zenith angle on surface radiance, and on prebiotic chemistry as measured by our action spectra.

The atmospheric radiative transfer computed for the prebiotic (3.9 Ga) Earth by \citet{Rugheimer2015} assumed a spectrally uniform albedo of 0.20 and a solar zenith angle of 60$^\circ$. However, much broader ranges of albedos and zenith angles are available in a planetary context. In this section, we explore the impact of different albedos ($A$)and solar zenith angles (SZA) on the surface radiance (azimuthally integrated) on the 3.9 Ga Earth. 

We calculate the surface radiance at SZA$=0^\circ$, 48.2$^\circ$, and $66.5^\circ$, for a range of different surface albedos. SZA$=0^\circ$ is the smallest possible value for SZA and corresponds to the shorted possible path through the Earth's atmosphere. It is achieved at tropical latitudes. SZA$=48.2^\circ$ corresponds to the insolation-weighed mean zenith angle on the Earth \citep{Cronin2014}.   SZA$=66.5^\circ$ corresponds to the maximum zenith angle experienced at the poles (noon at the summer solstice)\footnote{For simplicity, we assume here the modern terrestrial obliquity of 23.5$^\circ$. We are not aware of any evidence suggesting terrestrial obliquity was much different at 3.9 Ga; in fact, dynamical modelling suggests that the Earth's obliquity is stabilized by the Moon \citep{Laskar1993} so that it varies with an amplitude of only $\sim1.3^{\circ}$. Our results are insensitive to this magnitude of variation in obliquity}. Our choices of zenith angle thus encapsulate the minimum possible zenith angles, and hence the shortest possible atmospheric path lengths, over the Earth's surface. As such, they may be understood as corresponding to the range of maximum possible UV surface radiances accessible at different latitudes on the Earth's surface. It is of course possible to achieve arbitrarily low zenith angles (and hence arbitrarily low surface radiances) anywhere on Earth through the diurnal cycle and through seasonal variations at polar latitudes.

When considering albedos, we consider fixed uniform albedos of 0, 0.2, and 1. Albedos of 0 and 1 correspond to the lowest and highest possible values of $A$, and hence the lowest and highest\footnote{By virtue of backscattering of the upward diffuse radiance} surface radiance, respectively. The $A=0.2$ case corresponds to the \citet{Rugheimer2015} base case. We also consider albedos corresponding to different physical surface environments, including ocean, tundra, desert, and old and new snow, including the dependence on $z$ (see Appendix~\ref{sec:Albedos} for details). We consider this wide range of possible surface albedos because the climate state of the young Earth is minimally constrained by the available evidence, and climate states different than modern Earth are plausible. For example, \citet{Sleep2001} argue for a cold, ice-covered Hadean/early Archaean climate, which would imply high-albedo conditions even at low latitudes.

\begin{figure}[H]
\centering
\includegraphics[width=12 cm, angle=0]{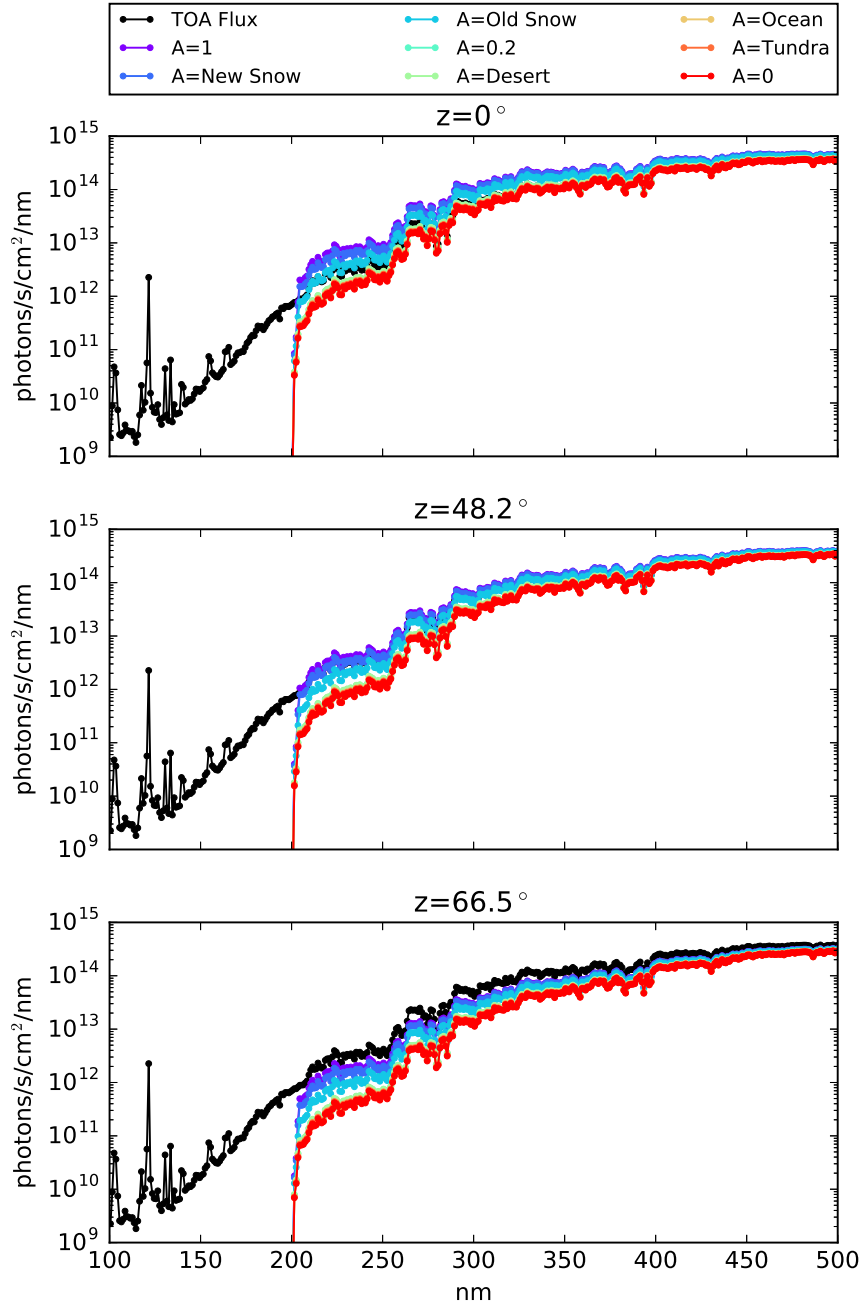}
\caption{Surface radiance for the Earth at 3.9 Ga, assuming an atmosphere corresponding to the model of \citet{Rugheimer2015} and a range of solar zenith angles and surface albedos. Taken together, albedo and zenith angle can drive variations in spectral surface radiance as high as a factor of 20.6 for $\lambda>204$ nm at 1 nm resolution.\label{fig:albzen}}
\end{figure}

Figure~\ref{fig:albzen} presents the surface radiance computed for the \citet{Rugheimer2015} atmospheric model for these different zenith angles and surface albedos. The results match our qualitative expectations. Low albedo surfaces correspond to lower surface radiances, with spectral contrast ratios as high as a factor of 7.4 between the $A=1$ and $A=0$ cases for $\lambda>204$ nm (i.e. the onset of the CO$_2$ cutoff, see \citealt{Ranjan2015}). Similarly, small zenith angles correspond to higher surface radiances, with spectral contrast ratios as high as high as 4.1 between SZA$=0^\circ$ and $z=66.5^\circ$ for $\lambda>204$ nm. Taken together, the effect is even stronger: for $\lambda>204$ nm, the $A=1$, SZA$=0^\circ$ case (the highest-radiance case in the parameter space we considered) has spectral contrast ratios as high as a factor of 30 with respect to the $A=0$, SZA$=66.5^\circ$ case (the lowest-radiance case considered in our parameter space). Using more physically motivated albedos, the spectral contrast ratio between a model with SZA$=0^\circ$ and albedo corresponding to fresh snow (i.e. the brightest natural surface included in our model) and a model with SZA$=66.5^\circ$ and albedo corresponding to tundra (i.e. the darkest natural surface included in our model)  were as high as a factor of 21. This means that at some wavelengths, 21 times more fluence would have been available on the equator of a high-albedo snow-and-ice covered "snowball Earth" compared to the polar regions of a warmer world with tundra or open ocean at the poles, for identical planetary atmospheres.

We note that, somewhat non-intuitively, for some values of albedo and zenith angle, surface radiances exceeding the incident TOA radiance are possible (see, e.g., the $A=1$, SZA$=0^\circ$ case in Figure~\ref{fig:albzen}). This is due to two factors. First, the downwelling diffuse radiance is enhanced by the backscatter of the upwelling diffuse radiance. Second, our flux conservation requirement coupled with our assumption of an isotropically scattering surface means that the upwelling radiance field is enhanced over the downwelling radiance field for low values of SZA and high values of $A$, meaning even more radiance is available to be backscattered. For a more thorough discussion of this phenomenon, see Appendix~\ref{sec:enhancedupwellingradiance}. For a discussion of a similarly non-intuitive result for surface flux, see \citet{Shettle1970}.

We quantify the impact of albedo and zenith angle from a biological perspective by computing the biologically effective relative dose rates $D_i$ for the photoprocesses described in Section~\ref{sec:dosimeters}, for the hemispherically-integrated surface radiances corresponding to the different surface types and zenith angles considered in this study. These values are reported in Table~\ref{tbl:albzendose}. Variations in albedo can affect the biologically effective dose rates of UV by factors of 2.7-4.4, depending on the zenith angle and the action spectrum used to compute the dose rate. Variations in zenith angle can affect the biologically effective dose rate of UV by factors of 3.6-4.1, depending on the surface albedo and action spectrum. Taken together, variations in albedo and zenith angle can change the biologically effective dose of UV by factors of 10.5-17.5, depending on the action spectrum used to compute the dose rate. We conclude that local conditions like albedo and latitude could impact the availability of UV photons for prebiotic chemistry by an order of magnitude or more.

\begin{table}[H]
\begin{center}
\caption{Abbreviations: T=Tundra, OS=Old Snow, NS=New Snow. \label{tbl:albzendose}}
\begin{tabular}{p{1.2 cm}p{1.5 cm}p{1.8 cm}p{1.8 cm}p{1.8 cm}p{1.8 cm}p{1.8 cm}}
\tableline\tableline
Zenith Angle & Albedo &UMP-193 &UMP-230 &UMP-254&  CuCN3-254 & CuCN3-300\\
66.5	&Tundra	&0.09	&0.08	&0.12	&0.11	&0.15\\
66.5	&Ocean	&0.10	&0.09	&0.12	&0.11	&0.15\\
66.5	&Desert	&0.11	&0.10	&0.13	&0.13	&0.17\\
66.5	&OS	&0.19	&0.21	&0.27	&0.27	&0.32\\
66.5	&NS	&0.25	&0.36	&0.43	&0.44	&0.47\\

\tableline
0	&Tundra	&0.34	&0.34	&0.46	&0.45	&0.56\\
0	&Ocean	&0.35	&0.34	&0.47	&0.45	&0.56\\
0	&Desert	&0.39	&0.39	&0.53	&0.51	&0.63\\
0	&OS	&0.71	&0.85	&1.09	&1.09	&1.23\\
0	&NS	&0.99	&1.47	&1.73	&1.79	&1.85\\

\tableline
66.5	&NS/T	&2.69	&4.31	&3.63	&3.96	&3.15\\
0	&NS/T	&2.88	&4.38	&3.72	&4.02	&3.34\\
\tableline
0/66.5	&Tundra	&3.64	&3.99	&3.96	&4.01	&3.72\\
0/66.5	&New Snow	&3.89	&4.05	&4.06	&4.07	&3.95\\
\tableline
0/66.5	&NS/T	&10.48	&17.47	&14.75	&16.13	&12.42\\
\tableline

\tableline
\end{tabular}
\end{center}
\end{table}

\subsection{Impact of Varying Levels of CO$_2$ On Surface Radiance \& Prebiotic Chemistry\label{sec:co2lim}}
In \citet{Ranjan2015}, we argued that shortwave UV light would have been inaccessible on the young Earth due to shielding from atmospheric CO$_2$. Specifically, we noted that atmospheric attenuation decreased fluence by a factor of 10 by 204 nm, with the damping increasing rapidly with the atmospheric cross-section (driven by CO$_2$) at shorter wavelengths. This shielding is important because it screens out photons shortward of 200 nm that are expected to be harmful, while still permitting the potentially biologically useful flux in the $\gtrsim200$ nm range (see, e.g., \citealt{Guzman2008}, \citealt{Barks2010}, \citealt{Patel2015}) to reach the planetary surface. 

However, this argument is based on the models elucidated in \citet{Rugheimer2015}. \citet{Rugheimer2015} assumed a CO$_2$ partial pressure at 3.9 Ga of 0.1 bar. This value is ad hoc: no direct geological constraints on CO$_2$ levels are available, and a variety of models with a wide range of CO$_2$ levels have been proposed that are consistent with the available climate constraints. Proposed CO$_2$ levels for the $\sim3.9$ Ga Earth range from $8\times10^{-4}$ bar \citep{Wordsworth2013} to $7$ bar \citet{Kasting1987}. 

In this section, we examine the sensitivity of the shielding of UV shortwave fluence due to varying levels of CO$_2$ in the atmosphere. We compute radiative transfer through an atmosphere with varying levels of CO$_2$ and N$_2$ under irradiation by the 3.9 Ga Sun. We evaluate radiative transfer at (zenith angle, albedo) combinations of (SZA$=0^\circ$, $A=$fresh snow) and (SZA$=66.5^\circ$, $A=$tundra), corresponding to the extremal values of the range of plausible maximal surface radiances accessible on the Earth assuming present-day obliquities. We omit attenuation due to other gases in our model (H$_2$O, CH$_4$, etc) in order to isolate the influence of CO$_2$. We emphasize, therefore, that the UV throughput we calculate should not be taken to correspond to the surface radiance plausibly expected on the 3.9 Ga Earth for a given CO$_2$ column, since they do not include attenuation from other gases and/or particulates/hazes that may have been present. Rather, these calculations represent the upper limits on surface radiance that are imposed by a given CO$_2$ column. 

We assume a fixed background of N$_2$ gas, with column density $N_{N_{2}}=1.88\times10^{25}$ cm$^{-2}$, corresponding to the 0.9 bar N$_2$ column assumed by the \citet{Rugheimer2015} atmospheric model. We choose this value for consistency with the model of \citet{Rugheimer2015}, noting that it is also consistent with the deepest constraint on N$_2$ abundance in the Archaean, i.e. the finding of \citet{Marty2013} that pN$_2$ at 3-3.5 Ga was 0.5-1.1 bar based on analysis of fluid inclusions trapped in Archaean hydrothermal quartz samples. Since N$_2$ does not absorb at UV wavelengths longer than 108 nm \citep{Huffman1969, Chan1993n2} and N$_2$ scattering is weak compared to e.g. CO$_2$ scattering, our results should be insensitive to the precise N$_2$ level. 

\begin{figure}[H]
\centering
\includegraphics[width=12 cm, angle=0]{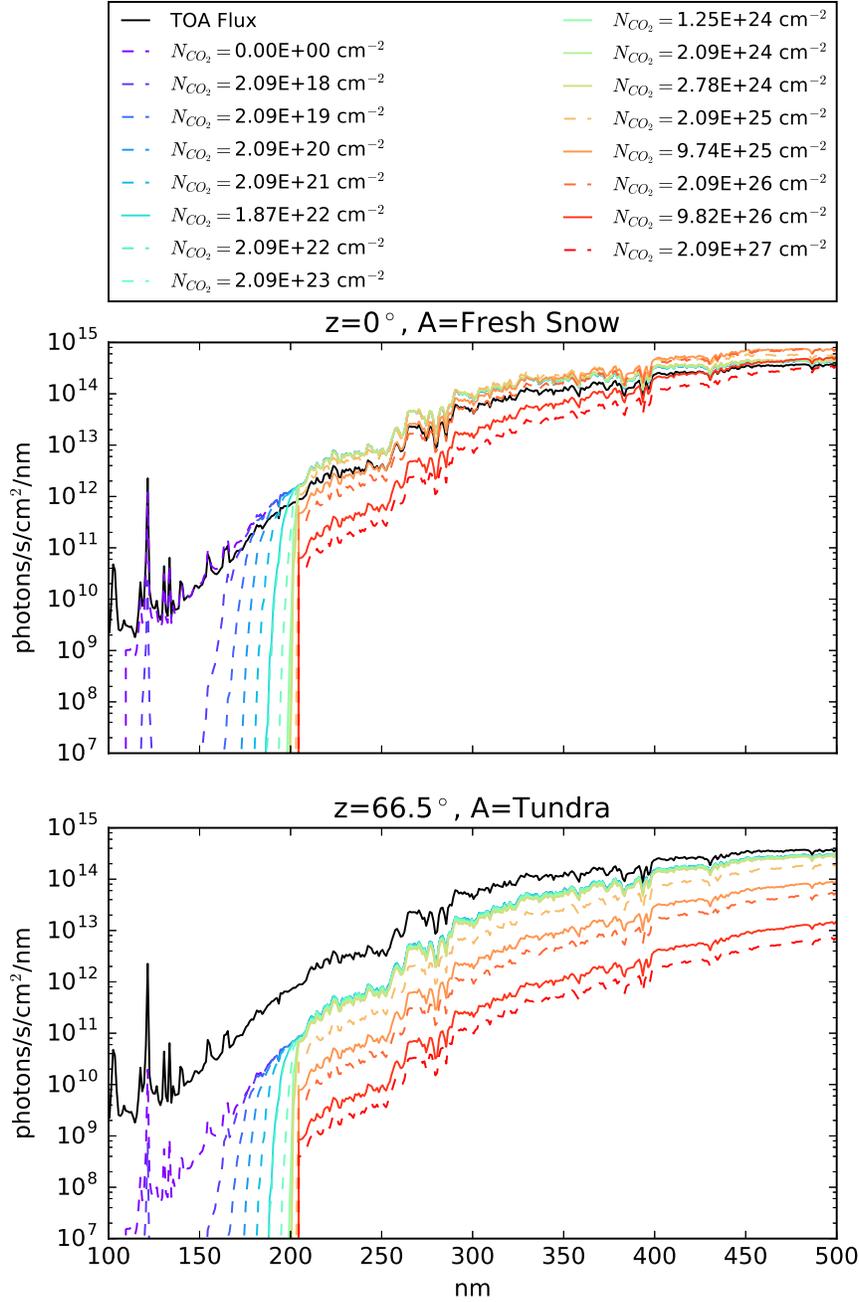}
\caption{Surface radiances for an atmosphere with $N_{N_{2}}=1.88\times10^{25}$ cm$^{-2}$ and varying levels of CO$_2$, for surface albedo and solar zenith angle combinations of (tundra, $66.5^\circ$) and (new snow, $0^\circ$), corresponding to the range of maximum surface radiances available across the planetary surface. Solid lines correspond to CO$_2$ levels with motivation from climate models in the literature. The line corresponding to $N_{CO_{2}}=2.09\times10^{24}$ cm$^{-2}$ (i.e. the fiducial 0.1-bar CO$_2$ level in the \citet{Rugheimer2015} model) is highlighted with a thicker line. \label{fig:co2lim}}
\end{figure}

We parametrize CO$_2$ abundance by scaling the CO$_2$ column calculated by \citet{Rugheimer2015} corresponding to 0.1 bar of CO$_2$, with total column density $N_{CO_{2}}=2.09\times10^{24}$ cm$^{-2}$. We calculate the UV radiance through CO$_2$ columns equal to the \citet{Rugheimer2015} column scaled by factors in the range from $10^{-6}-10^{3}$. To link these columns to CO$_2$ partial pressures, we employ the relation that the partial pressure of a well-mixed gas in an atmosphere is $p_i=g\bar{m} N_i$, where $g$ is the acceleration due to gravity ($g=981$ cm s$^{-2}$ for Earth), $\bar{m}$ is the atmospheric mean molecular mass, and $N_i$ is the column density of the gas. Figure~\ref{fig:co2lim} presents the UV surface radiances for atmospheres with $N_{N_{2}}=1.88\times10^{25}$ cm$^{-2}$, and varying levels of CO$_2$. Table~\ref{tbl:co2lim} presents the optical and atmospheric parameters associated with each model. 

\begin{table}[H]
\begin{center}
\caption{Atmospheric parameters defining the models shown in Figure~\ref{fig:co2lim}. Also given are the mean molecular weight and surface partial pressures of CO$_2$ and N$_2$ associated with each model atmosphere. A background column density of $N_{N_{2}}=1.88\times10^{25}$ cm$^{-2}$  is assumed throughout. Each model atmosphere had radiative transfer computed for both ($A$=tundra, SZE=$66.5^\circ$) and ($A$=fresh-fallen snow, SZE=$0^\circ$). \label{tbl:co2lim}}
\begin{tabular}{p{2 cm}p{2 cm}p{1.7 cm}p{1.7 cm}p{8 cm}}
\tableline\tableline
N$_{CO{_2}}$ (cm$^{-2}$) & $\bar{m}$ (g) & pN$_2$ (bar) & pCO$_2$ (bar) &Note \\
\tableline
0.00 & 4.65$\times10^{-23}$ & 0.860 & 0.00 & \\	
2.09$\times10^{18}$ & 4.65$\times10^{-23}$ & 0.860 & 9.55$\times10^{-8}$ & \\
2.09$\times10^{19}$ & 4.65$\times10^{-23}$ & 0.860 & 9.55$\times10^{-7}$ & \\
2.09$\times10^{20}$ & 4.65$\times10^{-23}$ & 0.860 & 9.55$\times10^{-6}$ & \\
2.09$\times10^{21}$ & 4.65$\times10^{-23}$ & 0.860 & 9.55$\times10^{-5}$ & \\
1.87$\times10^{22}$ & 4.65$\times10^{-23}$ & 0.860 & 8.53$\times10^{-4}$ & Corresponds to \citealt{Wordsworth2013} $2\times$PAL CO$_2$ model\\
2.09$\times10^{22}$ & 4.66$\times10^{-23}$ & 0.860 & 9.56$\times10^{-4}$ & \\
2.09$\times10^{23}$ & 4.68$\times10^{-23}$ & 0.865 & 9.61$\times10^{-3}$& \\
1.26$\times10^{24}$ & 4.82$\times10^{-23}$ & 0.891 & 5.97$\times10^{-2}$ & Corresponds to \citealt{vonParis2008} pCO$_2=0.06$ bar lower limit)\\
2.09$\times10^{24}$ & 4.92$\times10^{-23}$ & 0.909 & 0.101 & \\
2.78$\times10^{24}$ & 5.00$\times10^{-23}$ & 0.923 & 0.136 & Corresponds to \citealt{Kasting1987} pCO$_2=0.2$ bar lower limit\\
2.09$\times10^{25}$ & 6.05$\times10^{-23}$ & 1.12 & 1.24 & \\
9.76$\times10^{25}$ & 6.88$\times10^{-23}$ & 1.27 & 6.58	& Corresponds to \citealt{Kasting1987} pCO$_2=7$ bar upper limit\\
2.09$\times10^{26}$ & 7.09$\times10^{-23}$ & 1.31 & 14.6 & \\	
9.84$\times10^{26}$ & 7.26$\times10^{-23}$ & 1.34 & 70.1	&  Corresponds to volatilization of crustal carbon inventory of C as CO$_2$\\
2.09$\times10^{27}$ & 7.28$\times10^{-23}$ & 1.35 & 150 & \\

\tableline
\end{tabular}
\end{center}
\end{table}

To place these column densities in context, we compute the CO$_2$ columns associated with different climate models in the literature. 
\begin{itemize}
\item \citet{Kasting1987} calculate the range of CO$_2$ partial pressures required to sustain a plausible (i.e. consistent with an ice-free planet with liquid water oceans) climate on Earth throughout its history assuming a CO$_2$-H$_2$O greenhouse with 0.77 bar of N$_2$ as a background gas. Interpolating between model calculations, for 3.9 Ga the authors suggest a plausible CO$_2$ pressure range of $.2-7$ bar (calculated using a pure-CO$_2$ atmosphere equivalence). The lower limit corresponds to a surface temperature of 273K, whereas the upper point is interpolated between the CO$_2$ level required to sustain a temperature of 293K at 2.5 Ga, and the 10-bar limit for pCO$_2$ proposed by \citet{Walker1986} at 4.5 Ga. This corresponds to CO$_2$ column density range of $2.79\times10^{24}-9.76\times10^{25}$  cm$^{-2}$. 
\item If the requirement on global mean temperature is relaxed to 273 K from the $>278$ K of \cite{Kasting1987} \citep{HaqqMisra2008}, \citet{vonParis2008} find only $0.06$ bar of CO$_2$ is required at an insolation corresponding to 3.8 Ga \citep{Gough1981} in a CO$_2$-H$_2$O greenhouse with 0.77 bar of N$_2$ as a background gas. This corresponds to a CO$_2$ column density of $1.26\times10^{24}$ cm $^{-2}$. 
\item More dramatically, \citet{Wordsworth2013} model a N$_2$-H$_2$-CO$_2$ atmosphere, including the effects of collision-induced absorption (CIA) of N$_2$ and H$_2$ under the assumption of high levels of N$_2$ and H$_2$ relative to present atmospheric levels (PAL). By including N$_2$-H$_2$ CIA, for a solar constant of 75\% the modern value (corresponding to $3.8$ Ga using the methodology of \citealt{Gough1981}) they are able to maintain global mean surface temperatures suitable for liquid water with dramatically lower CO$_2$ levels than H$_2$O-CO$_2$ greenhouses. For an atmosphere with 3$\times$PAL N$_2$ and an H$_2$ mixing ratio of $0.1$, only 2$\times$ PAL of CO$_2$ ($7.8\times10^{-5}$ bar)is required, corresponding to $1.87\times10^{22}$ cm$^{-2}$. \\
\item Finally, as an extreme upper bound, we consider the observation of \citet{Kasting1993} based on \citet{Ronov1969} and \citet{Holland1978} that Earth has $\sim10^{23}$ g of carbon stored in crustal carbonate rocks. If this entire carbon inventory were volatilized as CO$_2$, it would correspond to a CO$_2$ column of $9.84\times10^{26}$ cm $^{-2}$ . 
\end{itemize}

We compute the surface radiance for N$_2$-CO$_2$ model atmospheres with $N_{CO_{2}}$ corresponding to the CO$_2$ columns computed for the above literature models, with a fixed N$_2$ background of $N_{N_{2}}=1.88\times10^{25}$ cm$^{-2}$ as before. These models and the parameters associated with them are also shown in Figure~\ref{fig:co2lim} and Table~\ref{tbl:co2lim}.

\subsubsection{Surface Radiance}
An N$_2$-CO$_2$ atmosphere with even small amounts of CO$_2$ is enough to form a strong shield to extreme-UV (EUV) radiation. A column density as low as $N_{CO{2}}=2.09\times10^{19}$ cm$^{-2}$, $10^{-5}$ of the \citet{Rugheimer2015} level, is enough to reduce the maximum possible surface radiance below 1\% of the TOA flux for wavelengths shorter than 167 nm. In \citet{Ranjan2015}, we argued that variations in solar UV output due to variability and flaring would have minimal impact on prebiotic chemistry because most of the variability was confined to wavelengths shorter than 165 nm, and incoming photons at these wavelengths were strongly attenuated by both the atmosphere and water. Here, we have shown that this atmospheric shielding exists for atmospheres with $N_{CO{2}}\geq2.09\times10^{19}$ cm$^{-2}$. In practice, this condition is satisfied by all climatologically plausible models for the 3.9 Ga Earth we are aware of. Therefore, even molecules that are removed from aqueous environments, e.g. through drying, are not expected to be vulnerable to solar UV variability for any plausible primitive atmosphere.

In \citet{Ranjan2015}, we argued that atmospheric CO$_2$ would have cut off fluence at wavelengths shorter than 204 nm, and hence that UV laboratory sources like ArF excimer lasers with primary emission at 193 nm were inappropriate for simulations of prebiotic chemistry. This was based on the assumption that $N_{CO{2}}=2.09\times10^{24}$ cm$^{-2}$. However, lower CO$_2$ levels are plausible for the young Earth. The lowest CO$_2$ atmosphere model that we are aware of that is consistent with an ice-free Earth at 3.9 Ga is that of \citet{Wordsworth2013}, with $N_{CO{2}}=1.87\times10^{22}$ cm$^{-2}$; at these levels, CO$_2$ extinction is enough to reduce the surface radiance shortward of 189 nm anywhere on Earth to less than 1\% of the TOA flux, but longer-wavelength photons might have been accessible in the absence of absorption from other species. Therefore, we revise our earlier statement: while photons shortward of 189 nm would have been inaccessible to prebiotic chemistry, photons longward of 189 nm might have been available if $N_{CO{2}}\leq1.87\times10^{22}$ cm$^{-2}$ and if there are no other major UV absorbers in the atmosphere (e.g.,  H$_2$O vapor, see Section~\ref{sec:h2o}). In such a regime, sources with primary emission in the 190-200 nm regime, like ArF eximer lasers, may be appropriate sources for prebiotic chemistry studies.

\citet{Kasting1987} suggest a climatologically plausible upper limit of 7 bars of pure CO$_2$ at 3.9 Ga, interpolating between the upper bound of \citet{Walker1986} at 4.5 Ga and the CO$_2$ level required to sustain $T\approx293$ K at 2.5 Ga. This corresponds to $N_{CO_{2}}=9.76\times10^{25}$ cm$^{-2}$. At this CO$_2$ level, surface fluences remain above 1\% of that incident at the TOA at wavelengths $>210$ nm even in the minimum fluence (low albedo, high zenith angle) case. Consequently, for CO$_2$ levels corresponding to mean temperatures similar to that of modern Earth, photons of wavelength $\lambda>245$ nm would have been accessible on the 3.9 Ga Earth, provided no other major UV absorbers besides CO$_2$  (e.g. SO$_2$, H$_2$S) were present in the atmosphere. Therefore, the use of sources with primary emission at wavelengths longer than 210 nm, e.g., 254 nm Hg lamps, is appropriate for simulations of prebiotic chemistry assuming a climate similar to the present day (with the obvious caveat that monochromatic sources risk missing crucial wavelength-dependent processes, see, e.g., \citealt{Ranjan2015}). 

Overall, the UV surface fluence is relatively insensitive to the level of atmospheric CO$_2$. The  conventional H$_2$O-CO$_2$-N$_2$ minimal greenhouses  with pCO$_2$=0.06-0.2 bar ($N_{CO_{2}}=1.26-2.78\times10^{24}$ cm$^{-2}$) \citep{vonParis2008, Kasting1987} feature virtually identical UV surface fluence environments. The surface fluence is suppressed only modestly in the scattering regime ($\lambda>204$ nm), even for optically thick atmospheres; for example, for $N_{CO{2}}=2.09\times10^{27}$ cm$^{-2}$, the surface fluence is suppressed by $\lesssim$ 3 orders of magnitude despite an optical depth of $\sim$ 1000. This behavior is a consequence of the random walk photons undergo in highly scattering atmospheres, and illustrates the importance of accurately including the effects of multiple scattering when calculating radiative transfer in such atmospheres.

\subsubsection{Biologically Effective Doses}
\begin{figure}[H]
\centering
\includegraphics[width=16.5 cm, angle=0]{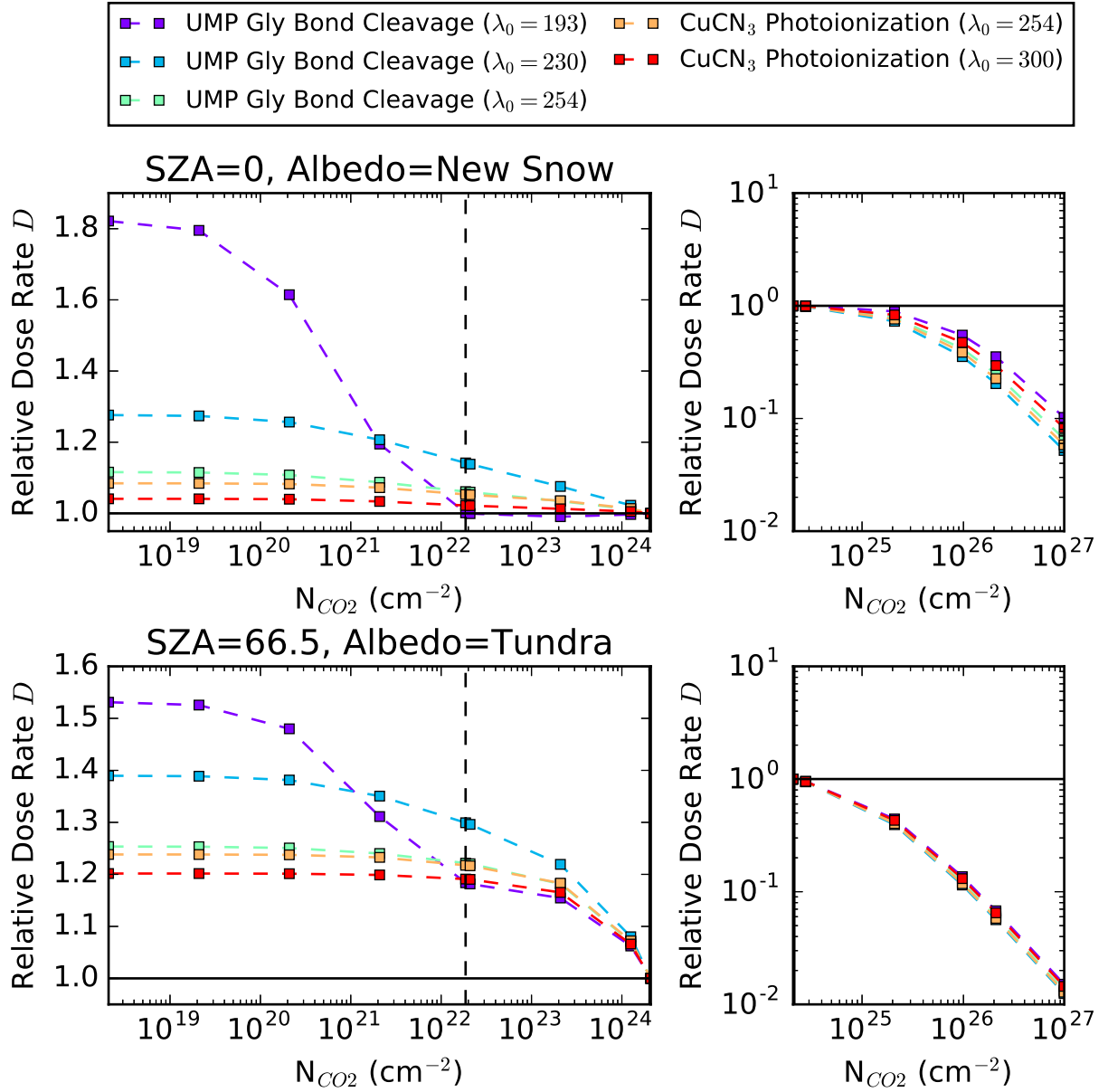}
\caption{Biologically effective dose rates for UMP-X and CuCN3-Y as a function of $N_{CO_{2}}$, normalized to their values at $N_{CO{2}}=2.09\times10^{24}$ cm$^{-2}$. The dashed line demarcates the level of CO$_2$ required by the climate model of \citet{Wordsworth2013}, the lowest proposed so far. The plot truncates at the CO$_2$ level corresponding to volatilization of all crustal carbonates as CO$_2$, an upper bound on plausible CO$_2$ levels. \label{fig:co2bed}}
\end{figure}

Figure~\ref{fig:co2bed} presents the biologically effective dose rates for the photoprocesses considered in our study under irradiation by surface fluences corresponding to attenuation by different levels of CO$_2$, normalized by the dose rates corresponding to $N_{CO{2}}=2.09\times10^{24}$ cm$^{-2}$ (0.1 bar CO$_2$). With this normalization, a dose rate $D>1$ means a higher dose rate than the 0.1 bar CO$_2$ case and hence a higher photoreaction rate, and $D<1$ the opposite. Both the maximum radiance ($A$=new snow, $SZA=0^\circ$) and the minimum radiance ($A$=tundra, SZA$=66.5^\circ$) cases are presented. 

For the minimum radiance case, we observe that the biologically effective dose rates uniformly decrease with increasing $N_{CO{2}}$, corresponding to uniforming decreasing fluence levels, as expected. $D_{i}>1$ for  $N_{CO{2}}<2.09\times10^{24}$ cm$^{-2}$ and $D_{i}<1$ for  $N_{CO{2}}>2.09\times10^{24}$ cm$^{-2}$, meaning that both stressor and eustressor photoprocesses are slowed by increasing $N_{CO{2}}$.

In the maximum radiance case, the biologically effective dose rates generally also follow the same trend with $N_{CO{2}}$, i.e. decreasing as $N_{CO{2}}$ increases. However, from $N_{CO{2}}=2.09\times10^{23}-10^{24}$ cm$^{-2}$, the dose rate of UMP-193 increases slightly, by $0.6\%$. This is because for $N_{CO{2}}\geq2.09\times10^{23}$ cm$^{-2}$, photons shortward of $\lambda_0=193$ nm are effectively completely blocked from the surface, leaving only the longwave photons at the lower QY available to power the reaction. For high albedo surfaces, this longwave fluence increases slightly with $N_{CO{2}}$ due to enhanced backscattering of reflected light from the surface by atmospheric CO$_2$\footnote{For low optical depth; at high optical depth (i.e. in the shortwave), so little fluence reaches the ground that this effect is lost. When ground reflection is turned off by setting the albedo to that of tundra (i.e. $\approx0$), while maintaining SZA=0, the radiance in all wavebands declines monotonically with $N_{CO{2}}$}, increasing slightly the effective dose. For $N_{CO{2}}>2.09\times10^{24}$ cm$^{-2}$, this effect is overwhelmed by the overall decrease in fluence levels, and the UMP-193 dose rate returns to the general trend.

High levels of atmospheric CO$_2$ can suppress prebiotically relevant photoprocesses as measured by our action spectra. For $N_{CO{2}}=9.84\times10^{26}$ cm$^{-2}$, corresponding to the outgassing of all CO$_2$ from crustal carbonates, the dose rates are suppressed by a factor of $10-100$ relative to the dose rates at $N_{CO{2}}=2.09\times10^{24}$ cm$^{-2}$, depending on the surface conditions. UV-sensitive pathways relevant to prebiotic chemistry may be photon-limited in high-CO$_2$ cases. If assuming a high-CO$_2$ atmosphere for UV-dependent prebiotic pathways, it may be important to characterize the sensitivity of the prebiotic pathways to fluence levels, to make sure that thermal backreactions will not retard the photoprocesses at low fluence levels.

Low levels of atmospheric CO$_2$ can modestly enhance prebiotically relevant photoprocesses as measured by our action spectra. For  $N_{CO{2}}=1.87\times10^{22}$ cm$^{-2}$ (corresponding to the low-CO$_2$ model of \citet{Wordsworth2013}), in the maximum radiance case the dose rates are $1.006-1.16$ of the dose rates at $N_{CO{2}}=2.09\times10^{24}$ cm$^{-2}$. For the corresponding minimum radiance case, the dose rates are $1.19-1.31$ of the dose rates at $N_{CO{2}}=2.09\times10^{24}$ cm$^{-2}$. If one allows CO$_2$ levels to decrease without regard for climatological or geophysical plausibility, higher dose rates are possible, but only to a point: assuming no CO$_2$ at all (only scattering from N$_2$), dose rates of 1.05-1.76 relative to the $N_{CO{2}}=2.09\times10^{24}$ cm$^{-2}$ base case are plausible (maximum radiance conditions). Hence, the variation in dose rate due to reducing CO$_2$ is less than a factor of 2. Pathways derived assuming attenuation from more conventional levels of CO$_2$ should function even with lower levels of CO$_2$ shielding. 

Overall, across the range of possible CO$_2$ levels ($N_{CO{2}}\leq9.84\times10^{26}$ cm$^{-2}$), the variation in biologically effective dose rates is $<$ 2 orders of magnitude, for both stressor and eustressor pathways and all values of $\lambda_0$, assuming no other UV absorbers. Hence, as measured by these action spectra, UV-sensitive prebiotic photochemistry is relatively insensitive to the level of CO$_2$ in the atmosphere, assuming no other absorbers to be present.

\subsection{Alternate shielding gases\label{sec:altgasres}}

In the previous section, we considered the constraints placed by varying levels of atmospheric CO$_2$ on the surficial UV environment. However, CO$_2$ is not the only plausible UV absorber in the atmosphere of the young Earth. Other photoactive gases that may have been present in the primitive atmosphere include SO$_2$, H$_2$S, CH$_4$ and H$_2$O. These gases have absorption cross-sections in the 100-500 nm range, and as such if present at significant levels could have influenced the surficial UV environment. O$_2$ and O$_3$, while expected to be scarce in the prebiotic era, are strongly absorbing in the UV, and might have an impact even at low abundances.

In this section, we explore the potential impact of varying levels of gases other than CO$_2$ on surficial UV fluence on the 3.9 Ga Earth. We consider each gas species $G$ individually, computing radiative transfer through two-component atmospheres under insolation by the 3.9 Ga Sun, with varying levels of the photoactive gas $G$ and a fixed column of N$_2$ as the background gas. We consider a range of column densities of $G$ corresponding the levels computed in \citet{Rugheimer2015} scaled by factors of 10. Table~\ref{tbl:rughabundances} gives the abundance of each gas computed by \citet{Rugheimer2015}. 

As in the case of CO$_2$ (Section~\ref{sec:co2lim}), we assume that $G$ is well-mixed, and we omit attenuation due to other gases in order to isolate the effect of the specific molecule $G$. We evaluate radiative transfer for an ($A$, SZA) combination corresponding to (fresh snow, $0^\circ$) only; hence, the surface radiances we compute may be interpreted as planetwide upper bounds. We assume a fixed N$_2$ background with column density of $N_{N_{2}}=1.88\times10^{25}$ cm$^{-2}$, corresponding to the 0.9 bar N$_2$ column in the \citet{Rugheimer2015} atmospheric model. We again emphasize that the UV throughput we calculate should not be taken to correspond to the surface radiance plausibly expected on the 3.9 Ga Earth for a given column of $G$, since they do not include attenuation from other gases that may have been present. Rather, these calculations represent the upper limits on surface radiance that are imposed by various levels of these gases. 

The model metadata we secured from \citet{Rugheimer2015} do not include H$_2$S abundances. We estimate an upper bound on the H$_2$S abundances by assuming the relative abundance of H$_2$S compared to SO$_2$ traces their emission ratio from outgassing. \citet{Halmer2002} find the outgassing emission rate ratios of [H$_2$S]/[SO$_2$]=0.1-2 for subduction zone-related volcanoes and 0.1-1 for rift-zone related volcanoes for the modern Earth. Since the redox state of the mantle has not changed from 3.6 Ga and probably from 3.9 Ga \citep{Delano2001}, we can expect the outgassing ratio to have been similar at 3.9 Ga. Therefore, we assign an upper bound to the H$_2$S column of $2\times$ the SO$_2$ column. 

\begin{table}[H]
\begin{center}
\caption{Gas abundances in the \citet{Rugheimer2015} 3.9 Ga Earth model. The H$_2$S abundances listed are upper bounds estimated from SO$_2$ levels.\label{tbl:rughabundances}} %
\begin{tabular}{p{2 cm}p{2 cm}p{2 cm}}
\tableline\tableline
$G$ & N$_{G}$ (cm$^{-2}$) & Molar Concentration \\
\tableline
N$_2$ & $1.88\times10^{25}$ & 0.9\\
CO$_2$ & $2.09\times10^{24}$ & 0.1\\
H$_2$O & $9.96\times10^{22}$ & $4.76\times10^{-3}$\\
CH$_4$ & $3.45\times10^{19}$ & $1.65\times10^{-6}$\\
SO$_2$ & $7.05\times10^{14}$ & $3.37\times10^{-11}$\\
O$_2$ & $5.66\times10^{19}$ & $2.71\times10^{-6}$\\
O$_3$ & $1.92\times10^{15}$ & $9.16\times10^{-11}$\\
H$_2$S $^\star$ & $\leq1.41\times10^{15}$ & $\leq6.7\times10^{-11}$\\

\tableline
\end{tabular}
\end{center}
\end{table}

\subsection{CH$_4$}
\begin{figure}[H]
\centering
\includegraphics[width=16.5 cm, angle=0]{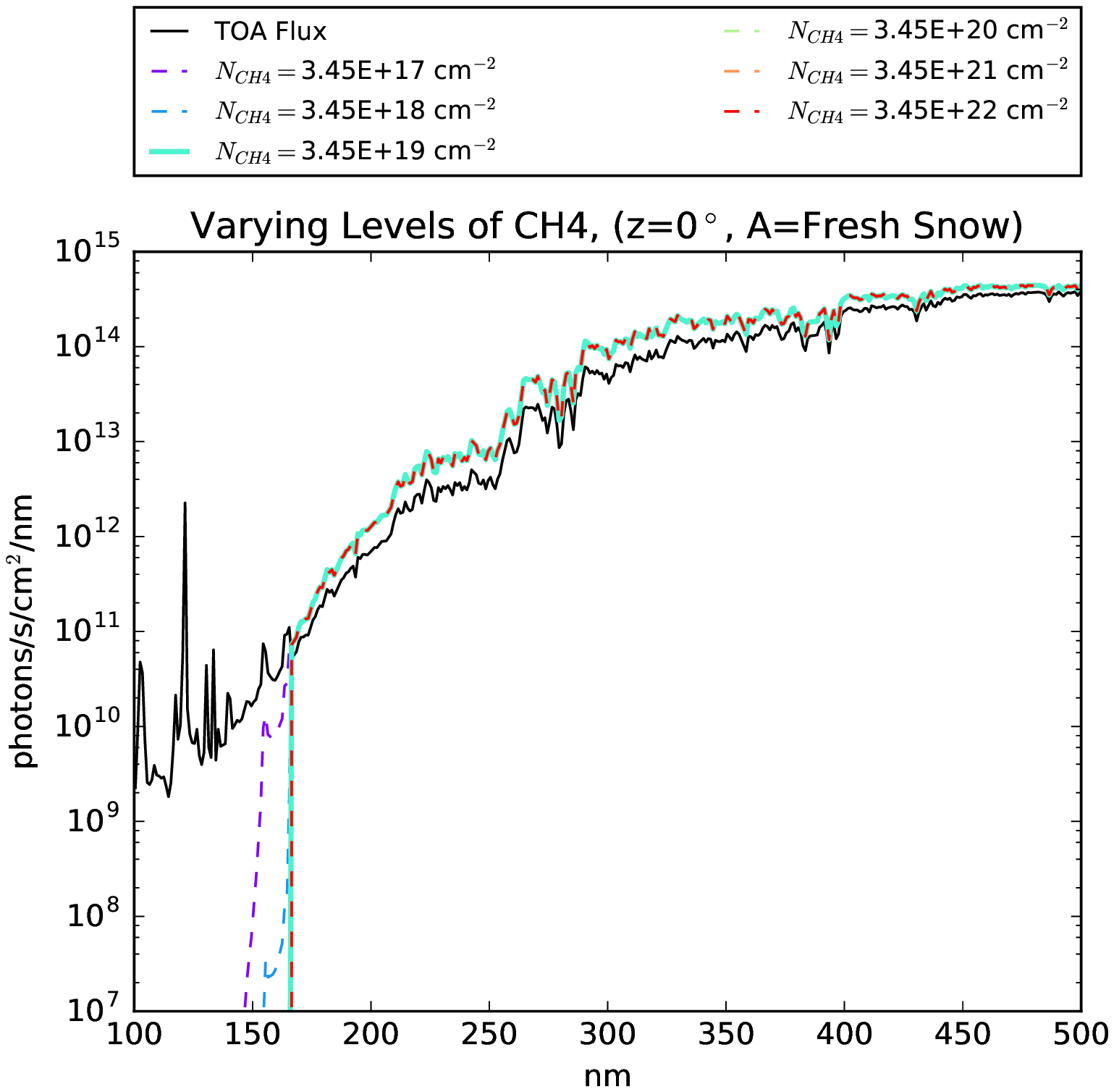}
\caption{Surface radiances for an atmosphere with $N_{N_{2}}=1.88\times10^{25}$ cm$^{-2}$ and varying levels of CH$_4$, for a surface albedo corresponding to new snow and SZA=$0^\circ$. The line corresponding to the CH$_4$ level assumed in the \citet{Rugheimer2015} model is presented as a solid, thicker line.\label{fig:ch4lim}}
\end{figure}

\citet{Rugheimer2015} fix by assumption a uniform CH$_4$ mixing ratio of $1.65\times10^{-6}$ throughout their 1-bar atmosphere, corresponding to $N_{CH_{4}}=3.45\times10^{19}$ cm$^{-2}$. However, other authors have postulated a broad range of abundances for CH$_4$. \citet{KastingBrown1998} estimate a methane abundance of 0.5 ppm (mixing ratio of $5\times10^{-7}$) for a 1-bar N$_2$-CO$_2$ atmosphere assuming conversion of 1\% carbon flux from mid-ocean ridges to CH$_4$ prior to the rise of life. \citet{Kasting2014} echo this estimated mixing ratio for a CH$_4$ source of serpentinization of ultramafic rock with seawater at present-day levels, while \citet{Guzman-Marmolejo2013} estimate serpentinization can drive CH$_4$ levels up to 2.1 ppmv. However, \citet{Kasting2014} also note that methane production from impacts could have outpaced the supply from serpentinization multiple orders of magnitude, and could have sustained abiotic CH$_4$ levels up to 1000 ppmv. \citet{Shaw2008} also postulates that such high CH$_4$ levels might be plausible, hypothesizing that conversion of dissolved carbon compounds to methane might have produced methane fluxes an order of magnitude higher than the present day, enough to sustain an atmosphere with sufficient methane to maintain a habitable climate. Such estimates correspond to a broad range of CH$_4$, from $N_{CH_{4}}=9.83\times10^{18}$ cm$^{-2}$ to $N_{CH_{4}}=1.97\times10^{22}$ cm$^{-2}$. We consequently explore a range of CH$_4$ values corresponding to $10^{-2}-10^{3}\times$ the \citet{Rugheimer2015} value, encompassing this range. Figure~\ref{fig:ch4lim} shows the resultant spectra.

Absorption of UV light by CH$_4$ is negligible in an atmosphere with even trace amounts of CO$_2$, due to the extremely shortwave onset of absorption by CH$_4$ (165 nm). At CH$_4$ levels of $N_{CH_{4}}=3.45\times10^{22}$ cm$^{-2}$, 3 orders of magnitude higher than those assumed in \citet{Rugheimer2015} and at the upper end of what has been proposed in the literature, CH$_4$ extincts fluence shortward of 165 nm. A CO$_2$ level of $N_{CO_{2}}=2.09\times10^{19}$ cm$^{-2}$ is enough to extinct the fluence shortward of 167. This CO$_2$ level is 5 orders of magnitude less than the \citet{Rugheimer2015} value and 2 orders of magnitude less than the lowest level CO$_2$ suggested in the literature based on climatic constraints. We conclude that in an atmosphere with even trace amounts of CO$_2$, plausible levels of CH$_4$ do not further constrain the surface UV environment.

\subsection{H$_2$O\label{sec:h2o}}
There exist few constraints on primordial water vapor levels. The terrestrial water vapor profile is set by the temperature profile; evaporation rates and H$_2$O saturation pressures increase with temperature, meaning that a hot planet is likely to be steamier than a cold planet. \citet{Knauth2005} use oxygen isotope data from cherts to argue that the ocean temperature was 328-358K (55-85$^\circ$) at 3.5 Ga, though this is not universally accepted \citep{Kasting2010}, in part due to the extraordinary inventory of greenhouse gases that would be required to sustain such high temperatures. By contrast, \citet{Rugheimer2015} computed a surface temperature of 293K for the Earth at 3.9 Ga.

\begin{figure}[H]
\centering
\includegraphics[width=16.5 cm, angle=0]{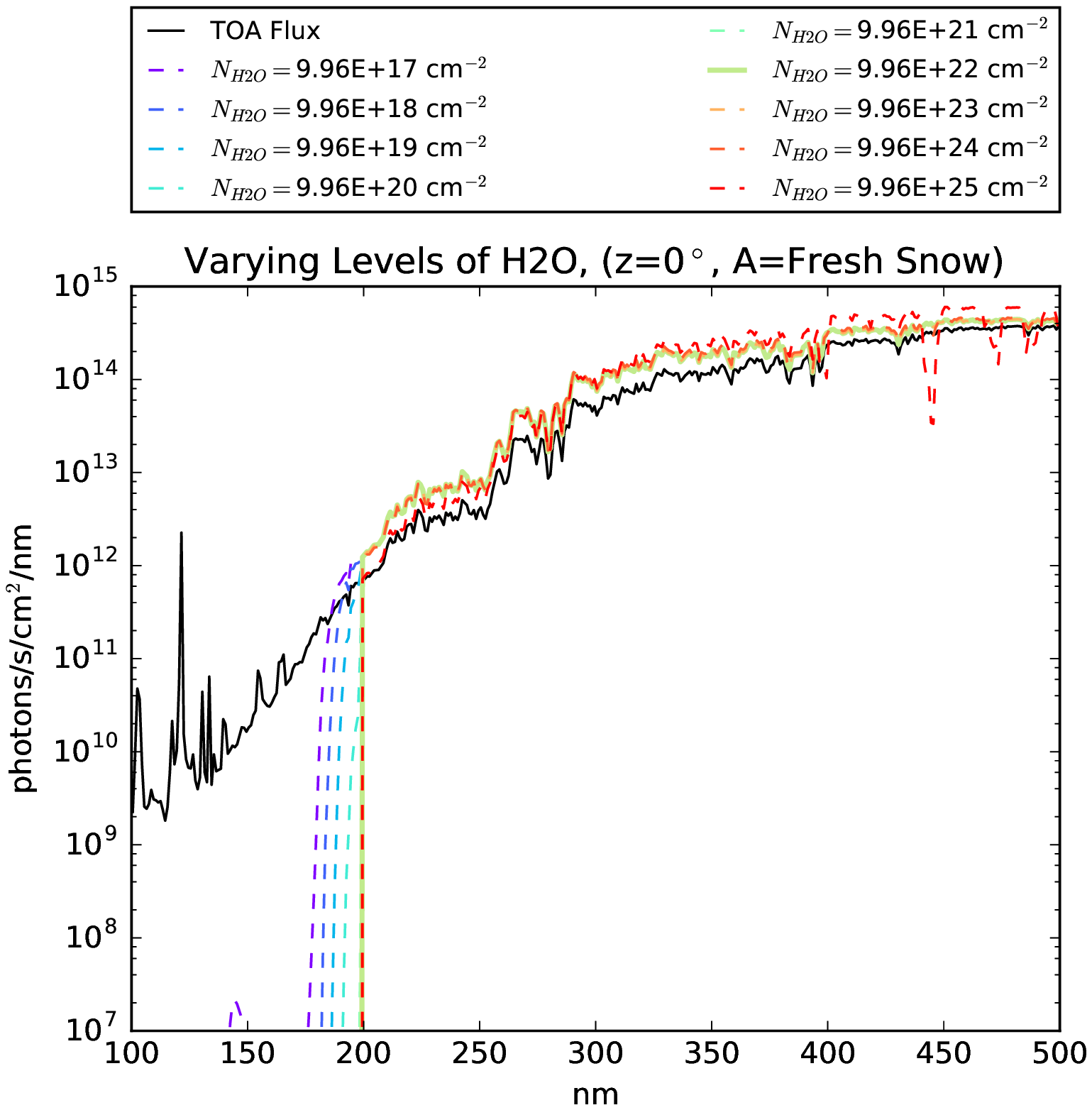}
\caption{Surface radiances for an atmosphere with $N_{N_{2}}=1.88\times10^{25}$ cm$^{-2}$ and varying levels of H$_2$O, for a surface albedo corresponding to new snow and SZA=$0^\circ$. The line corresponding to the H$_2$O level calculated in the \citet{Rugheimer2015} model is presented as a solid, thicker line. \label{fig:h2olim}}
\end{figure}

The model of \citet{Rugheimer2015} corresponds to an atmosphere with an H$_2$O column density of $N_{H_{2}O}=9.96\times10^{22}$ cm$^{-2}$. However, arbitrarily low H$_2$O abundances are possible depending on how cold the atmosphere and surface are. What of the upper limit? \citet{Kasting1984b} computed among other parameters the H$_2$O mixing ratio for a planet with atmospheric composition matching the modern Earth under varying levels of insolation. In the case of 1.45$\times$ modern insolation, \citet{Kasting1984b} compute a surface temperature of 384.2 K (greater than that suggested by \citet{Knauth2005}), a surface pressure of 2.481 bars, and H$_2$O volume mixing ratio of $\sim0.5$. This corresponds to an atmosphere with a mean molecular mass of $3.89\times10^{-23}$ g and an H$_2$O column of $3.25\times10^{25}$ cm$^{-2}$.We interpret this value as an extreme upper limit for H$_2$O column density. We evaluate radiative transfer through N$_2$-H$_2$O atmospheres with total H$_2$O columns corresponding to $10^{-5}-10^{3}$ of the \citet{Rugheimer2015} value, encompassing this upper bound. Figure~\ref{fig:h2olim} shows the resultant spectra.

H$_2$O is a robust UV absorber, with strong absorption for $\lambda_0<198$ nm.  $N_{H_{2}O}=9.96\times10^{21}$ cm$^{-2}$ ($0.1\times$ the \citet{Rugheimer2015} level) is enough to block fluence for $\lambda_0<198$ nm; for comparison, $N_{CO_{2}}=2.09\times10^{24}$ cm$^{-2}$, the fiducial \citet{Rugheimer2015} level, blocks fluence for $\lambda_0<201$ nm. Even in the absence of attenuation from CO$_2$, attenuation from plausible levels of H$_2$O blocks deep UV flux with $\lambda<198$ nm. Hence, no matter what the level of CO$_2$, fluence shortward of 198 nm would have been inaccessible to prebiotic chemistry, assuming warm enough surface temperatures to sustain $N_{H_{2}O}=9.96\times10^{21}$ cm$^{-2}$ of atmospheric water vapor.

\begin{figure}[H]
\centering
\includegraphics[width=16.5 cm, angle=0]{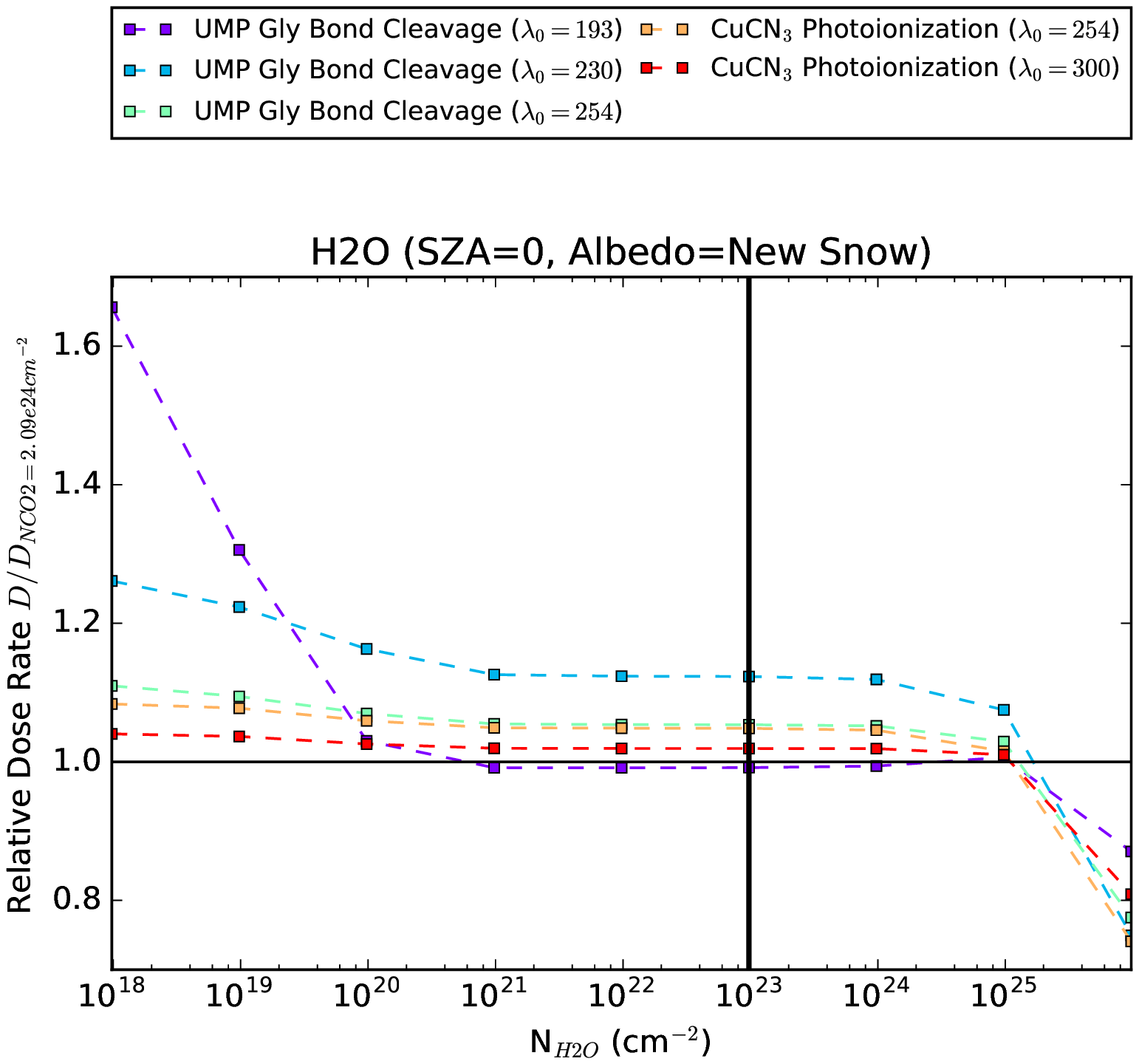}
\caption{Biologically effective dose rates for UMP-X and CuCN3-Y as a function of $N_{H_{2}O}$, normalized to the dose rates at $N_{CO{2}}=2.09\times10^{24}$ cm$^{-2}$. The solid line corresponds to the H$_2$O level of \citet{Rugheimer2015}.\label{fig:h2obed}}
\end{figure}

Figure~\ref{fig:h2obed} presents the dose rates $D_i(N_{H_{2}O})/D_i(N_{CO_{2}}=2.09\times10^{24} \texttt{cm}^{-2})$, i.e. the biologically effective dose rates for UMP-193, -230, and -254, and CuCN3-254 and -300 as a function of $N_{H_{2}O}$\footnote{assuming an H$_2$O-N$_2$ atmosphere} normalized by the corresponding dose rates calculated for an CO$_2$-N$_2$ atmosphere with 0.1 bar of CO$_2$. With this normalization, a value of $D_i/D_i(N_{CO_{2}}=2.09\times10^{24} \texttt{cm}^{-2})>1$ means that the given photoreaction is proceeding at a higher rate than for an atmosphere with $N_{CO_{2}}=2.09\times10^{24} \texttt{cm}^{-2}$. Across the range of $N_{H_{2}O}$ considered here, the dose rates are similar (within a factor of 1.7) of the dose rates in the $N_{CO_{2}}=2.09\times10^{24} \texttt{cm}^{-2}$ case. We therefore argue that the constraints on UV imposed by H$_2$O are similar to those imposed by $N_{CO_{2}}=2.09\times10^{24} \texttt{cm}^{-2}$. Prebiotic chemistry studies derived assuming a surface radiance environment primarily shaped by $N_{CO_{2}}=2.09\times10^{24} \texttt{cm}^{-2}$ should be robust to the level of water vapor in the atmosphere.

\subsection{SO$_2$\label{sec:so2}}
\begin{figure}[H]
\centering
\includegraphics[width=16.5 cm, angle=0]{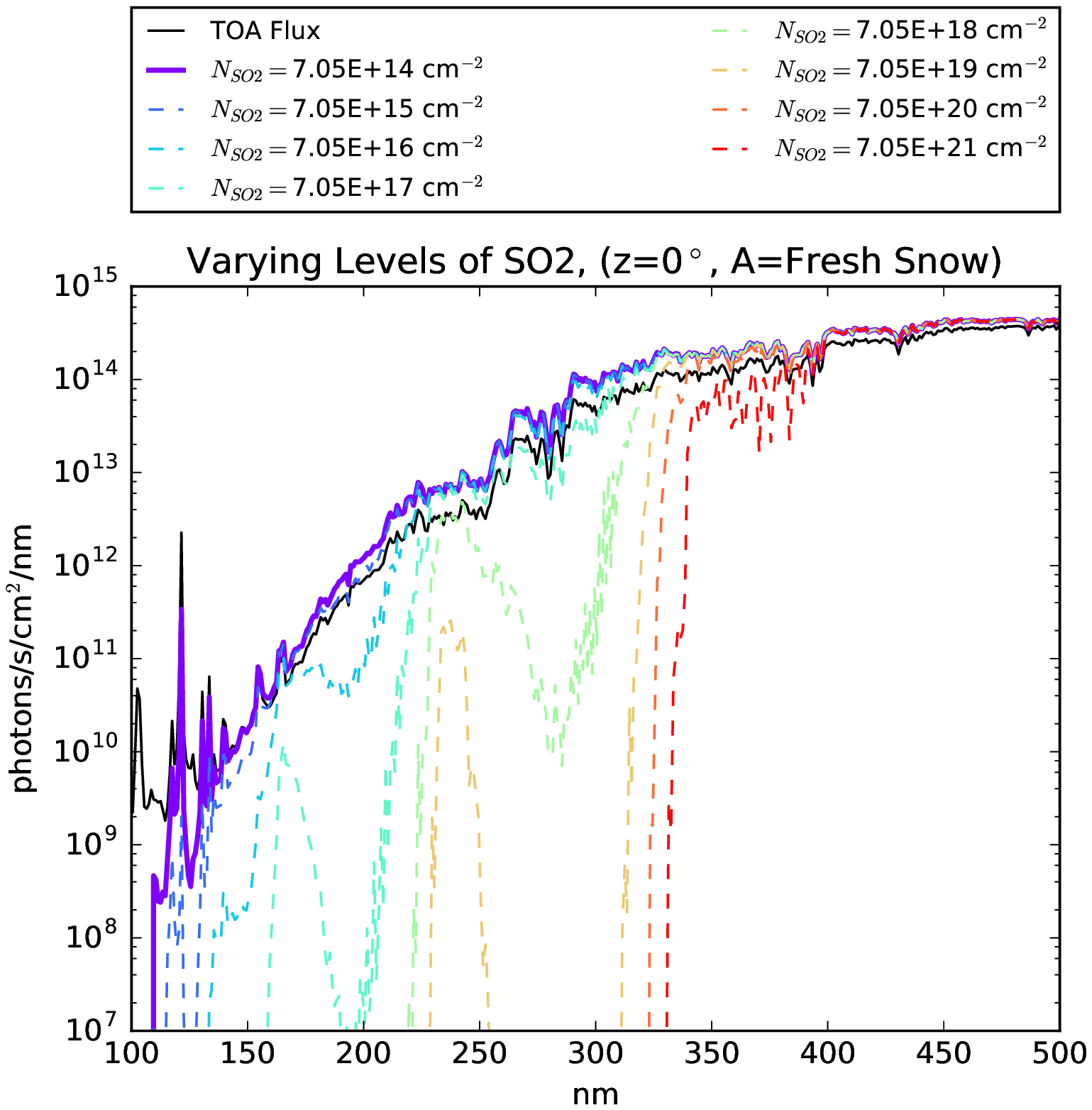}
\caption{Surface radiances for an atmosphere with $N_{N_{2}}=1.88\times10^{25}$ cm$^{-2}$ and varying levels of SO$_2$, for a surface albedo corresponding to new snow and SZA=$0^\circ$. The line corresponding to the SO$_2$ level calculated in the \citet{Rugheimer2015} model is presented as a solid, thicker line. \label{fig:so2lim}}
\end{figure}

SO$_2$ absorbs more strongly and over a much wider range than CO$_2$ (Figure~\ref{fig:so2xc}). However, SO$_2$ is vulnerable to loss processes such as photolysis and reaction with oxidants \citep{Kaltenegger2010}; consequently, SO$_2$ levels are usually calculated to be very low on the primitive Earth. Assuming levels of volcanic outgassing at 3.9 Ga comparable to the present day, as did \citet{Rugheimer2015}, SO$_2$ levels are low, with $N_{SO_{2}}=7.05\times10^{14}$ cm$^{-2}$ in the calculation of \citet{Rugheimer2015}. At these levels, SO$_2$ does not significantly modify the surface UV environment (Figure~\ref{fig:so2lim}). 

However, relatively little is known about primordial volcanism. During epochs of high enough volcanism on the younger, more geologically active Earth, volcanic reductants might conceivably deplete the oxidant supply. While an extreme scenario, if it occurred, SO$_2$ might plausibly build up to the 1-100 ppm level \citep{Kaltenegger2010}, at which point it might begin to supplant CO$_2$ as the controlling agent for the global thermostat \footnote{The outgassing-weathering-subduction feedback loop theorized to regulate planetary temperature.} (c.f. the model of \citealt{Halevy2007} for primitive Mars). Assuming a background atmosphere of 0.9 bar $N_2$ and 0.1 bar CO$_2$, this corresponds to column densities of $N_{SO_{2}}=2.09\times10^{19}-10^{21}$ cm$^{-2}$. We therefore explore a range of SO$_2$ values corresponding to $1-10^{7}\times$ the \citet{Rugheimer2015} value, encompassing this range. Figure~\ref{fig:so2lim} shows the resultant spectra. We considered lower SO$_2$ levels as well, but the resultant spectra were indistinguishable from the $1\times$ case and so are not shown.

Such high SO$_2$ levels would be transient, and would subside with volcanism. But while they persisted, SO$_2$ could dramatically modify the surface UV environment. At $N_{SO_{2}}=7.05\times10^{17}$ cm$^{-2}$, SO$_2$ starts to sharply reduce fluence at wavelengths $\lambda\lesssim200$ nm. For $N_{SO_{2}}\geq7.05\times10^{20}$ cm$^{-2}$ ($\sim36$ ppm in the \citealt{Rugheimer2015} model), all fluence shortward of $\sim327$ nm is shielded out. Such levels of SO$_2$ would correspond to very low-UV epochs in Earth's history.

\begin{figure}[H]
\centering
\includegraphics[width=16.5 cm, angle=0]{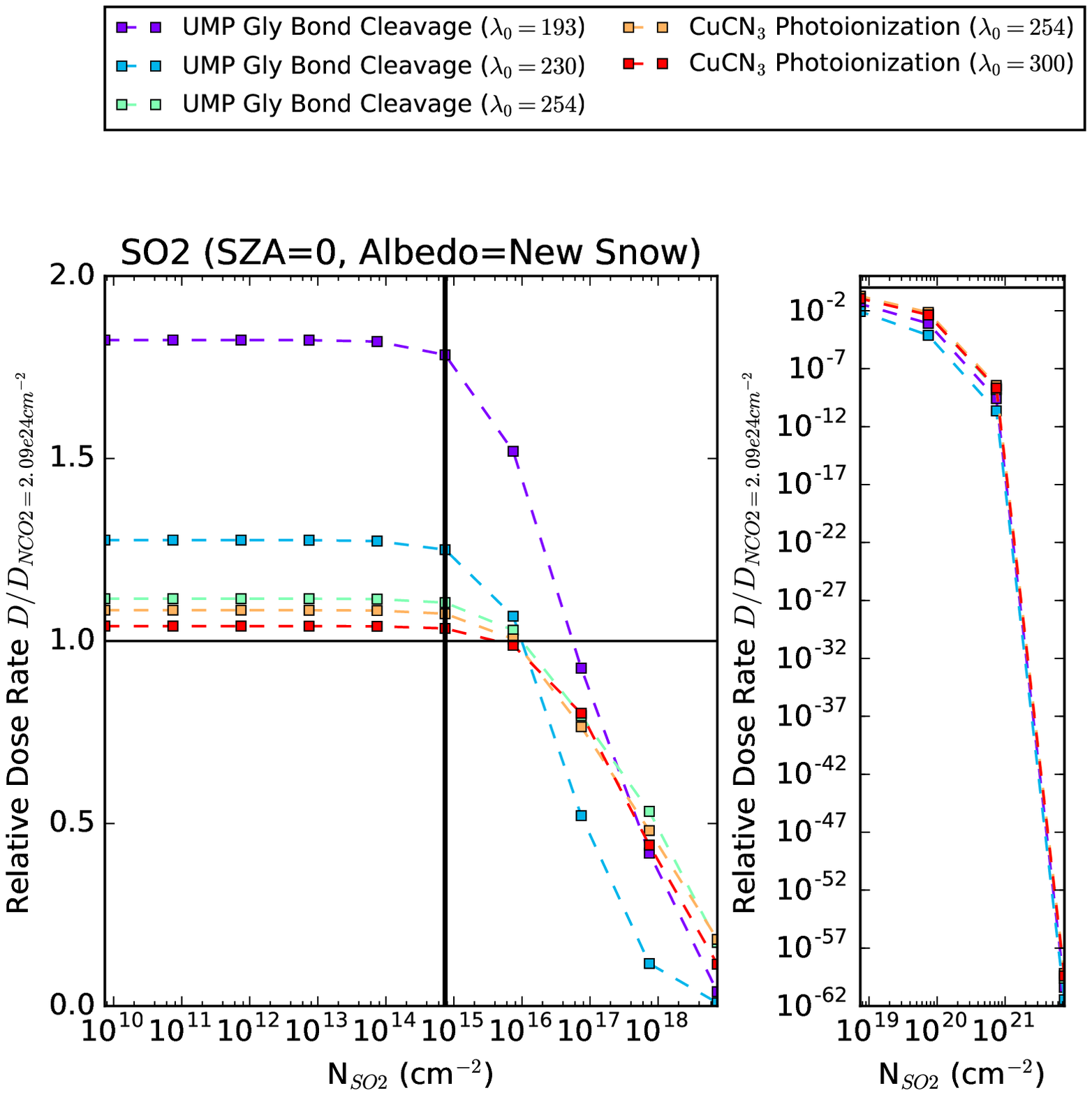}
\caption{Biologically effective dose rates for UMP-X and CuCN3-Y as a function of $N_{SO_{2}}$, normalized to the dose rates at $N_{CO_{2}}=2.09\times10^{24}$ cm$^{-2}$. The solid line corresponds to the SO$_2$ level of \citet{Rugheimer2015}\label{fig:so2bed}}
\end{figure}

We quantify the impact of high SO$_2$ levels on UV-sensitive prebiotic chemistry by again convolving our surface radiance spectra for an SO$_2$-N$_2$ atmosphere against our action spectra and computing the BEDs as functions of $N_{SO_{2}}$. Figure~\ref{fig:so2bed} presents these values, normalized by the corresponding dose rates for the $N_{CO_{2}}=2.09\times10^{24} \texttt{cm}^{-2}$ case. As expected, higher levels of SO$_2$ reduce the BEDs. For $N_{SO_{2}}=7.05\times10^{21}$ cm$^{-2}$, the BEDs are suppressed by over 60 orders of magnitude, implying the corresponding photoreactions would have been quenched. UV-dependent prebiotic chemistry could not function in such high-SO$_2$ epochs. 

\begin{figure}[H]
\centering
\includegraphics[width=16.5 cm, angle=0]{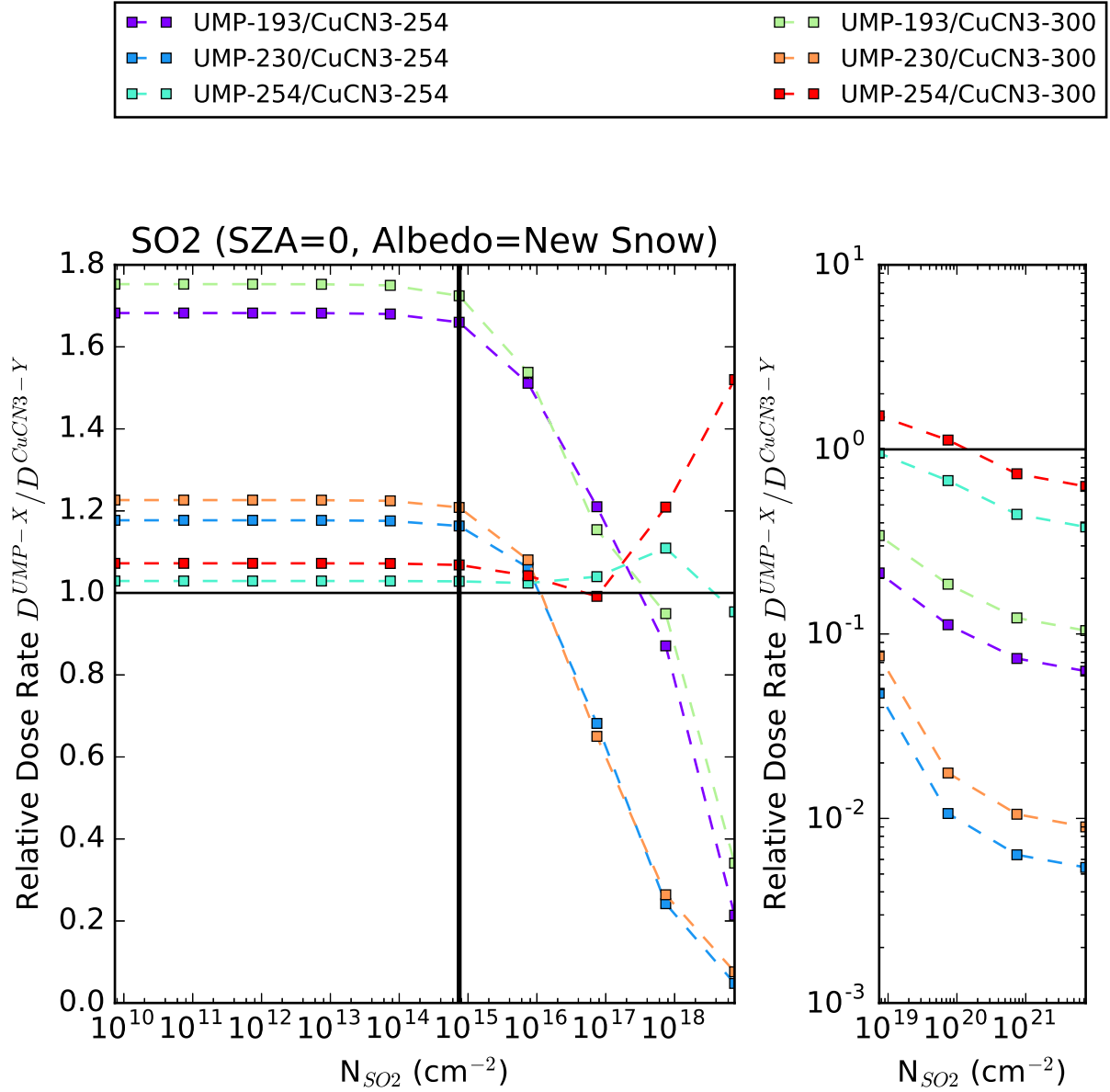}
\caption{Ratio of biologically effective dose rates UMP-X/CuCN3-Y, for X=193, 230 and 254 nm and Y=254 and 300 nm, as a function of $N_{SO_{2}}$. \label{fig:so2bedreldiff}}
\end{figure}

These BEDs do not fall off at the same rates. Figure~\ref{fig:so2bedreldiff} plots the ratio between UMP-X and CuCN3-Y, for all X and Y, as a function of $N_{SO_{2}}$. For $N_{SO_{2}}\geq7.05\times10^{18}$ cm$^{-2}$, for all values of $\lambda_0$ the CuCN3 photoionization BEDs decrease at lower rates than the UMP glycosidic bond cleavage BEDs. For $N_{SO_{2}}\geq7.05\times10^{19}$ cm$^{-2}$, the UMP-230 BED is suppressed two orders of magnitude more than the CuCN3-X BEDs, relative to $N_{SO_{2}}\leq7.05\times10^{15}$ cm$^{-2}$. Since the UMP-X BEDs measure a stressor for abiogenesis while CuCN3-X BEDs measure an eustressor for abiogenesis, one might argue that high-SO$_2$ environments present more clement environments for abiogenesis compared to low-SO$_2$ ones. However, these results are sensitive to the details of the action spectrum and photoprocess chosen. This is especially challenging given the paucity of data on the QY curves for these processes and the assumptions therefore required to construct them. Further, we have considered here only one stressor and eustressor photoprocess, whereas in reality there should have been many more photoprocesses. We therefore conclude that as measured solely by UMP-photolysis and CuCN-photoionization, high SO$_2$ epochs might have been more clement environments for abiogenesis, but further laboratory constraints on the action spectra of these processes are required to reduce the sensitivity of this statement to assumptions. 

\subsection{H$_2$S}

H$_2$S shares similar properties to SO$_2$. Like SO$_2$, H$_2$S is a stronger and broader absorber in the UV than CO$_2$. Like SO$_2$, H$_2$S is primarily generated through outgassing from volcanic sources, and is lost from the atmosphere due to vulnerability to photolysis and reactions with oxidants. Consequently, H$_2$S is is not expected to have been a major constituent of the prebiotic atmosphere.

\begin{figure}[H]
\centering
\includegraphics[width=16.5 cm, angle=0]{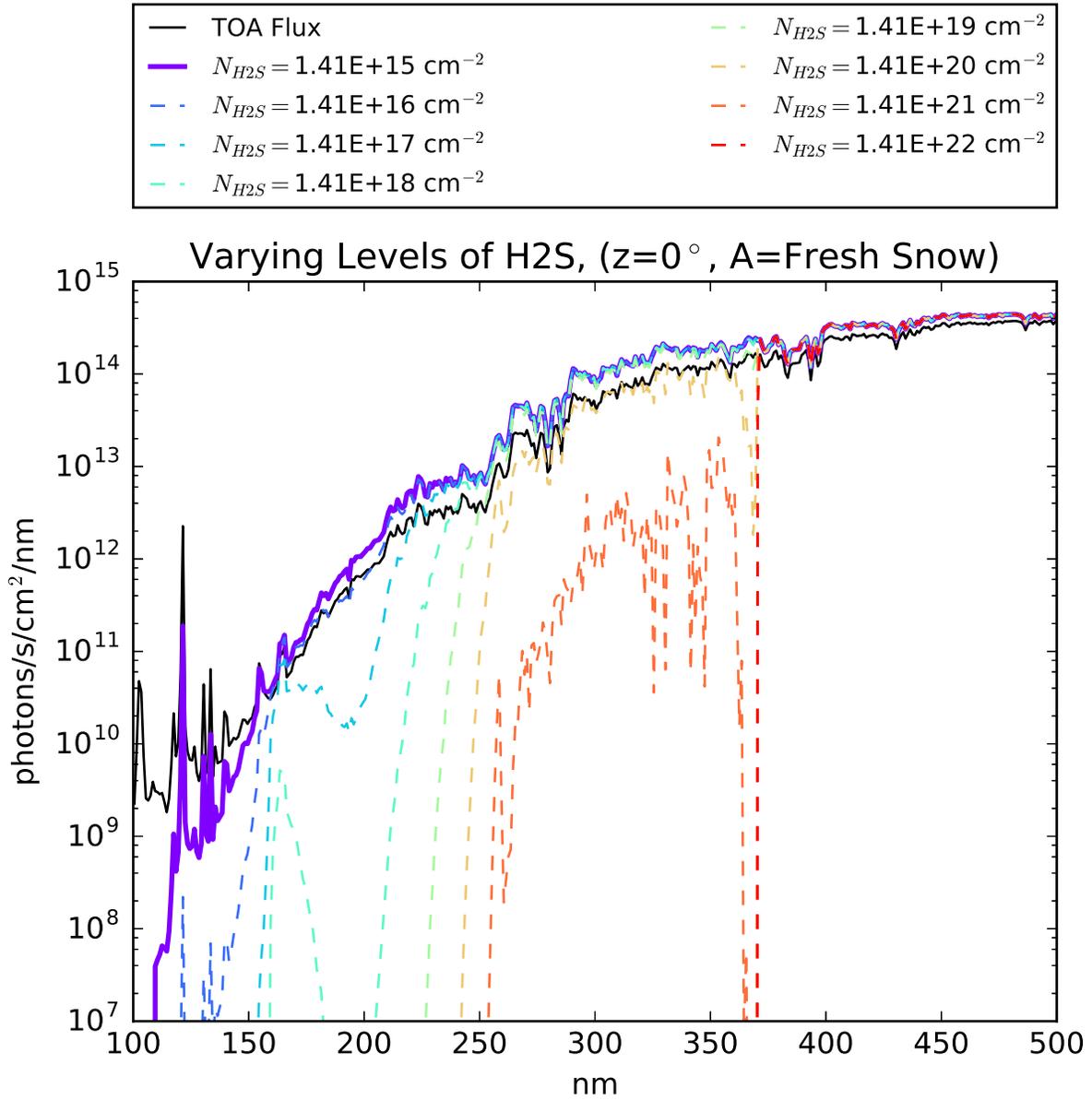}
\caption{Surface radiances for an atmosphere with $N_{N_{2}}=1.88\times10^{25}$ cm$^{-2}$ and varying levels of H$_2$S, for a surface albedo corresponding to new snow and SZA=$0^\circ$. The line corresponding to the $2\times$SO$_2$ level calculated in the \citet{Rugheimer2015} model is presented as a solid, thicker line.\label{fig:h2slim}}
\end{figure}

\citet{Rugheimer2015} do not calculate an abundance for H$_2$S in their model. We estimate an upper bound on $N_{H_{2}S}\leq2\times N_{SO_{2}}$ by assuming that [H$_2$S]/[SO$_2$] traces their relative outgassing ratios, which vary from 0.1-2 in the modern day \citep{Halmer2002}. The redox state of the terrestrial mantle has likely not changed since 3.9 Ga \citep{Delano2001, Trail2011}, suggesting the proportion of outgassed gases from volcanogenic sources should not have changed either. Based on this reasoning, we assign an upper bound to $N_{H_{2}S}$ in an atmosphere corresponding to the \citet{Rugheimer2015} model of $1.41\times10^{15}$ cm$^{-2}$, corresponding to $2\times N_{SO_{2}}$. We explore a range of H$_2$S values corresponding to $1-10^{7}\times$ this value, corresponding to the range of SO$_2$ values we explore. Figure~\ref{fig:h2slim} shows the resultant spectra. We considered lower SO$_2$ levels as well, but the resultant spectra were indistinguishable from the $1\times$ case and so are not shown.

\begin{figure}[H]
\centering
\includegraphics[width=16.5 cm, angle=0]{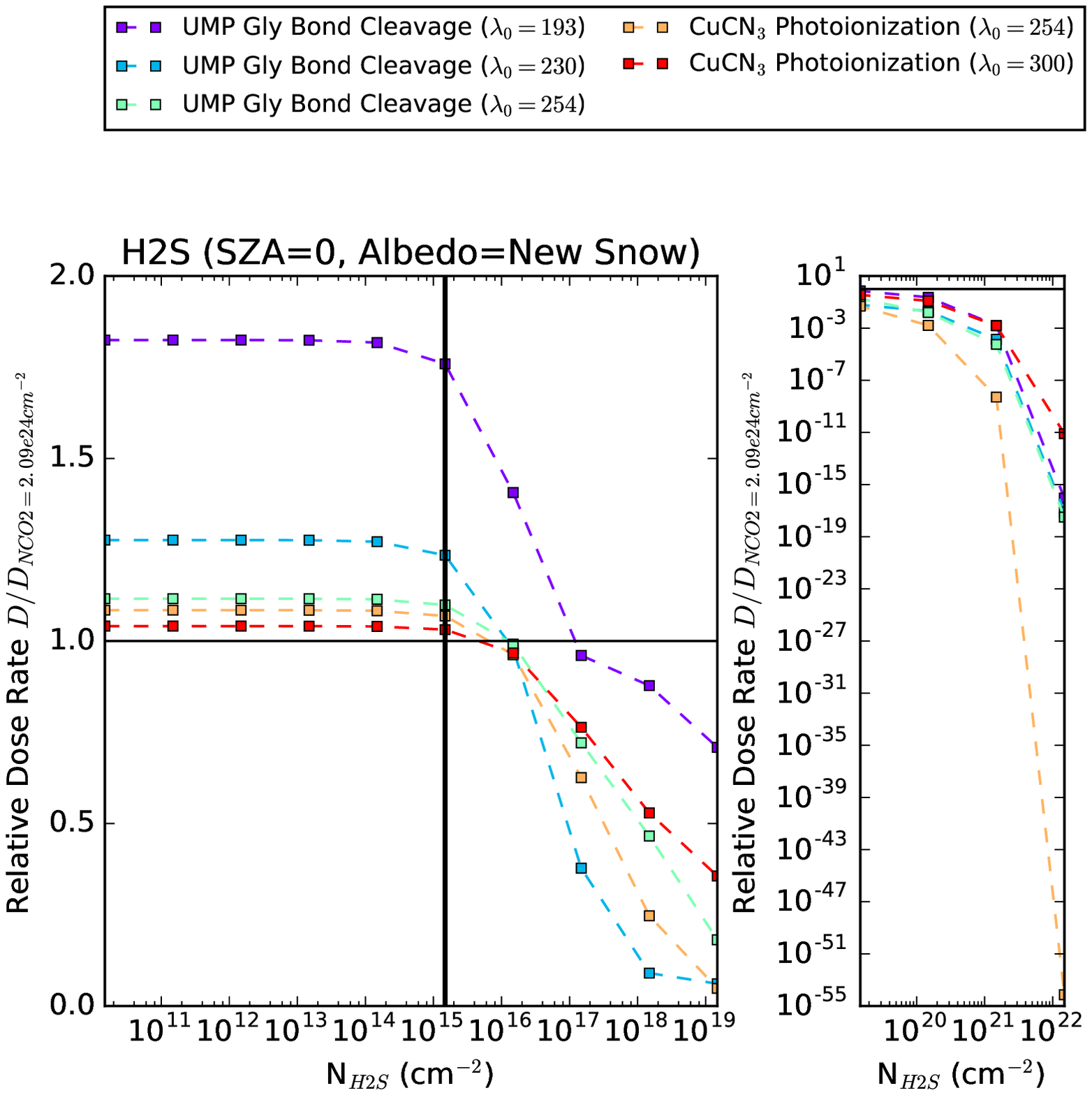}
\caption{Biologically effective dose rates for UMP-X and CuCN3-Y as a function of $N_{H_{2}S}$, normalized to the dose rates at $N_{CO_{2}}=2.09\times10^{24}$ cm$^{-2}$. The solid line corresponds to $2\times$ the SO$_2$ level of \citet{Rugheimer2015}. \label{fig:h2sbed}}
\end{figure}

For $N_{H_{2}S}=1.41\times10^{15}$ cm$^{-2}$, H$_2$S has minimal impact on the surface UV environment. However, as with SO$_2$, epochs of high volcanism, possible on a younger Earth, might have depleted the oxidant supply, permitting the buildup of H$_2$ to higher levels. In this case, H$_2$S could have affected the surface UV environment. For $N_{H_{2}S}\geq1.41\times10^{19}$ cm$^{-2}$, H$_2$ suppresses the fluence at wavelengths shorter than $235$ nm. For $N_{H_{2}S}\geq1.41\times10^{22}$ cm$^{-2}$, fluence shortward of 369 nm is shielded out. This reduction in fluence results in the expected decrease in BED with $N_{H_{2}S}$, as shown in Figure~\ref{fig:h2sbed}. For $N_{H_{2}S}=1.41\times10^{22}$ cm$^{-2}$, all dose rates are suppressed by $\gtrsim$11 orders of magnitude relative to the dose rates at  $N_{H_{2}S}=1.41\times10^{15}$ cm$^{-2}$. As with SO$_2$, high H$_2$S epochs in Earth's history would be low-UV epochs, and UV-sensitive prebiotic pathways might find themselves photon-starved.

\begin{figure}[H]
\centering
\includegraphics[width=16.5 cm, angle=0]{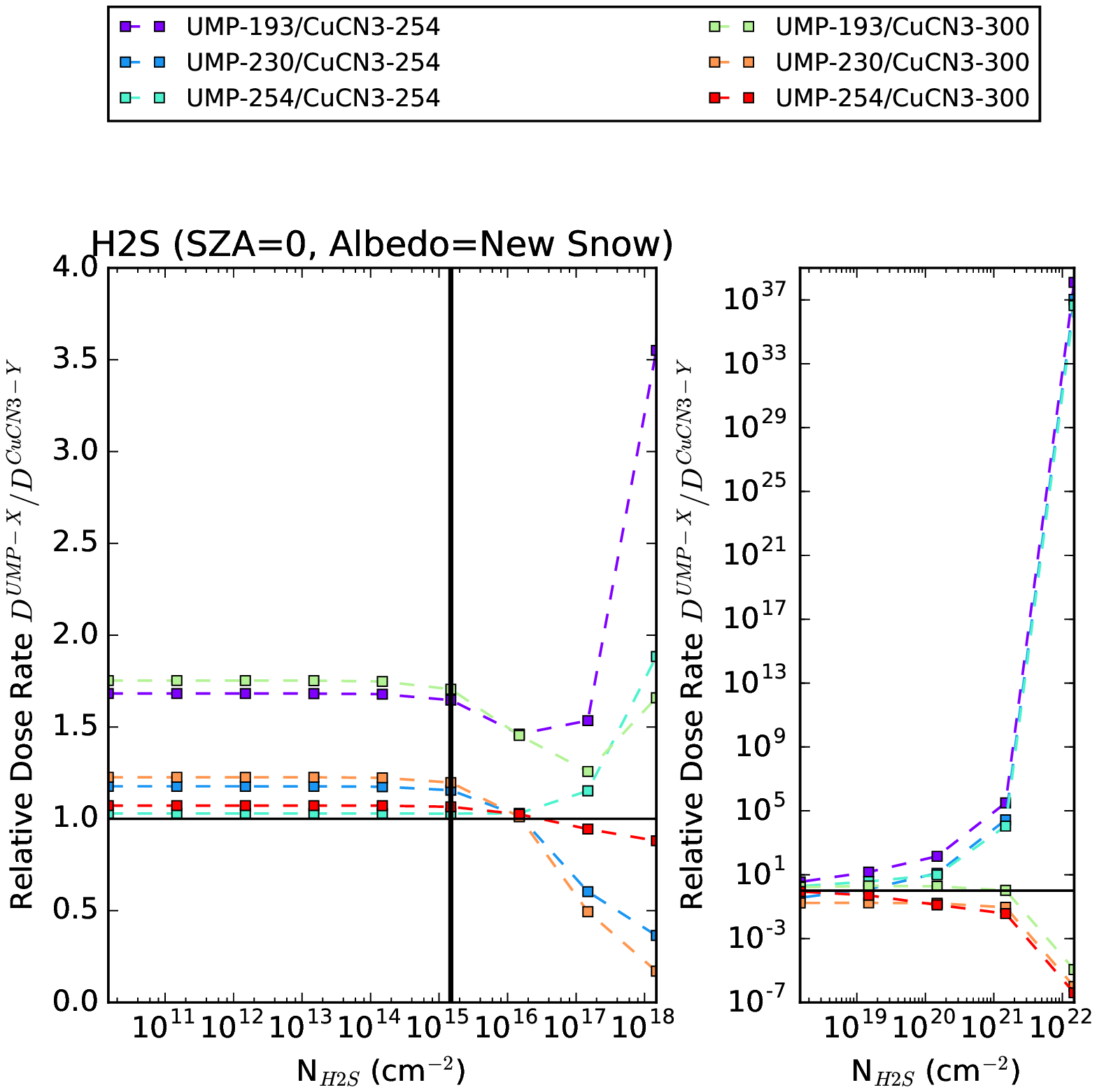}
\caption{Ratio of biologically effective dose rates UMP-X/CuCN3-Y, for X=193, 230 and 254 nm and Y=254 and 300 nm, as a function of $N_{H_{2}S}$ \label{fig:h2sbedreldiff}}
\end{figure}

Similarly as with SO$_2$, the various BED fall off at different rates with increasing $N_{H_{2}S}$. Figure~\ref{fig:h2sbedreldiff} plots the ratio between UMP-X and CuCN3-Y, for all X and Y. The CuCN3-254 dose rate falls off faster than the other dose rates and the CuCN3-300 dose rate falls off slower. This is because H$_2$S absorbs much more strongly at $\lambda<254$ nm than for $\lambda>254$ nm. As presently defined, the CuCN3-254 dose rate can only utilize $\lambda<254$ nm radiation, the UMP-X dose rates can make use of $\lambda>254$ nm fluence but at a much lower efficiency, and the CuCN3-300 dose rate can fully utilize the $\lambda>254$ nm radiation. Hence, $D_{UMP-X}/D_{CuCN3-254}$ increases with $N_{H_{2}S}$ for all $X$, while $D_{UMP-X}/D_{CuCN3-300}$ decreases. Depending on the value of the CuCN3 photoionization step function, high H$_2$S environments may have been much more or much less clement environments for abiogenesis, as measured by the balance between glyocosidic bond cleave in UMP and aquated electron production from CuCN3 photoionization. We conclude that it is critical to further constrain empirically the action spectra of these photoprocesses in order to robustly use them to estimate the favorability of high-versus-low H$_2$S environments for abiogenesis.

\subsection{O$_2$}
\begin{figure}[H]
\centering
\includegraphics[width=16.5 cm, angle=0]{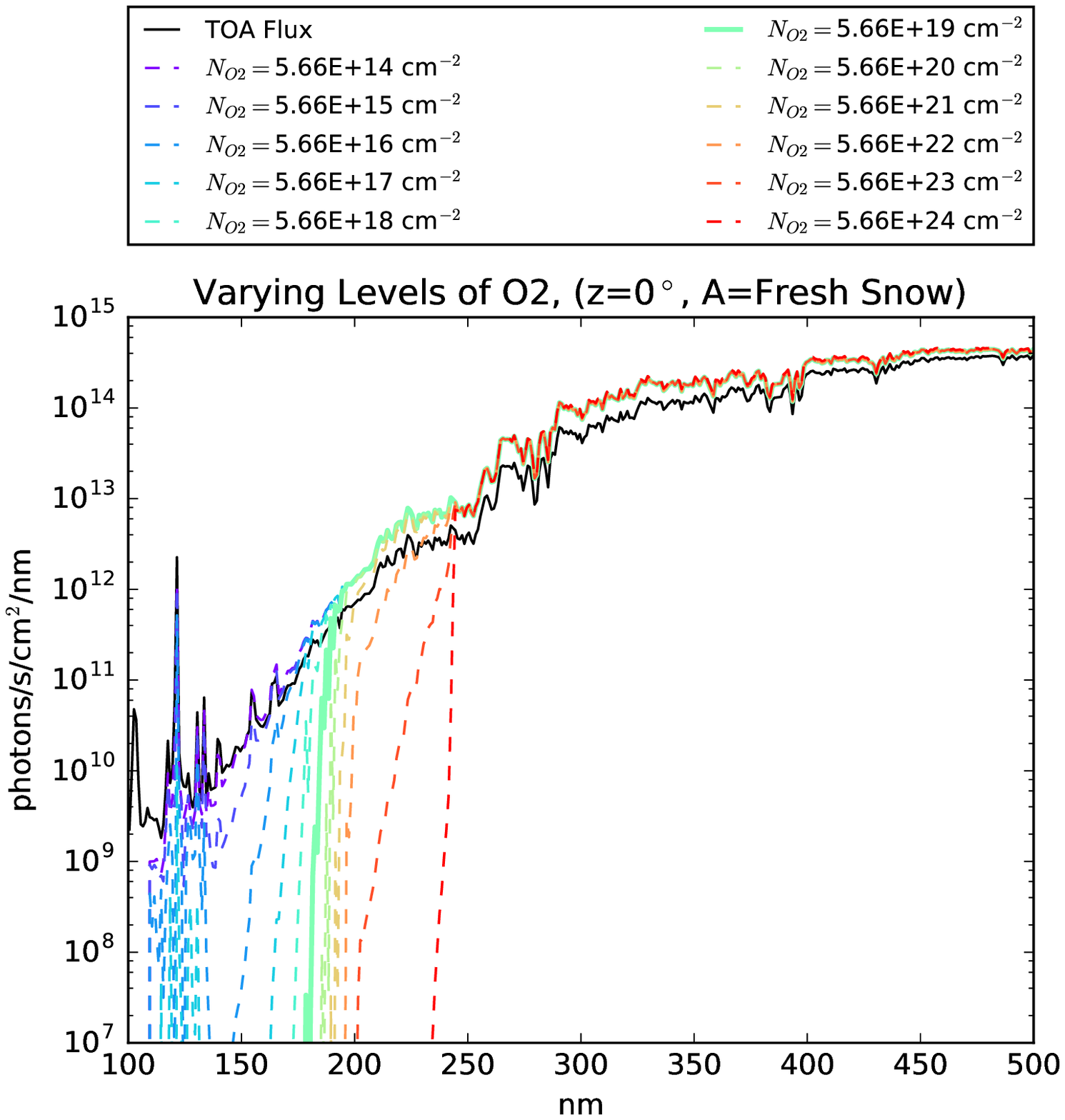}
\caption{Surface radiances for an atmosphere with $N_{N_{2}}=1.88\times10^{25}$ cm$^{-2}$ and varying levels of O$_2$, for a surface albedo corresponding to new snow and SZA=$0^\circ$. The line corresponding to the O$_2$ level calculated in the \citet{Rugheimer2015} model is presented as a solid, thicker line. \label{fig:o2lim}}
\end{figure}

O$_2$ is a robust UV shield. In the aphotosynthetic prebiotic era, O$_2$ is thought to have been generated from photolysis of CO$_2$ and H$_2$O, with sinks from reactions with reductants. Measurements of sulfur mass-independent isotope fractionation (SMIF) imply O$_2$ concentrations $<1\times10^{-5}$ PAL prior to 2.3 Ga \citep{Pavlov2002}, and Fe and U-Th-Pb isotopic measurements from a 3.46 Ga chert is consistent with an anoxic ocean during that era \citep{Li2013}. \citet{Rugheimer2015} calculate an O$_2$ abundance corresponding to $N_{O_{2}}=5.66\times10^{19}$ cm$^{-2}$, consistent with the \citet{Pavlov2002} limit. We explore a range of O$_2$ columns corresponding to $10^{-5}-10^{5}\times$ the \citet{Rugheimer2015} column. We note the O$_2$ columns exceeding the \citet{Rugheimer2015} column are disfavored by the SMIF modeling of \citet{Pavlov2002}, and should therefore be taken as strictly illustrative. Figure~\ref{fig:o2lim} shows the resultant spectra.

At the \citet{Rugheimer2015} level of $N_{O_{2}}=5.66\times10^{19}$ cm$^{-2}$, O$_2$ blocks fluence at $\lambda<183$ nm. $N_{CO_{2}}=2.09\times10^{21}$ cm$^{-2}$ is required to achieve a similar cutoff; $N_{CO_{2}}\geq1.87\times10^{22}$ is expected based on the climate models we have found so far. Therefore, at the O$_2$ levels expected from photochemical models, O$_2$ does not further constrain the surface UV environment assuming a climatically plausible CO$_2$ inventory. But, O$_2$ is a substantially stronger UV absorber than CO$_2$, and it is possible that in comparatively low-CO$_2$/high-O$_2$ scenarios, it could affect the surface UV environment. $N_{O_{2}}=5.66\times10^{21}$ cm$^{-2}$ of O$_2$ cuts off more fluence than $N_{CO_{2}}=1.87\times10^{22}$ (the minimum proposed from climate models).  Hence, at O$_2$ levels of $N_{O_{2}}\geq5.66\times10^{21}$ cm$^{-2}$, O$_2$ may begin to have noticeable impact on the surface UV environment. However, it is unlikely that such high O$_2$ inventories were achieved at 3.9 Ga, given the strong reductant sink from vulcanogenic gases, the lack of a strong O$_2$ source, and the empirical constraint from SMIF. It is most likely that attenuation from O$_2$ would have been negligible compared to that from CO$_2$.

\subsection{O$_3$}
\begin{figure}[H]
\centering
\includegraphics[width=16.5 cm, angle=0]{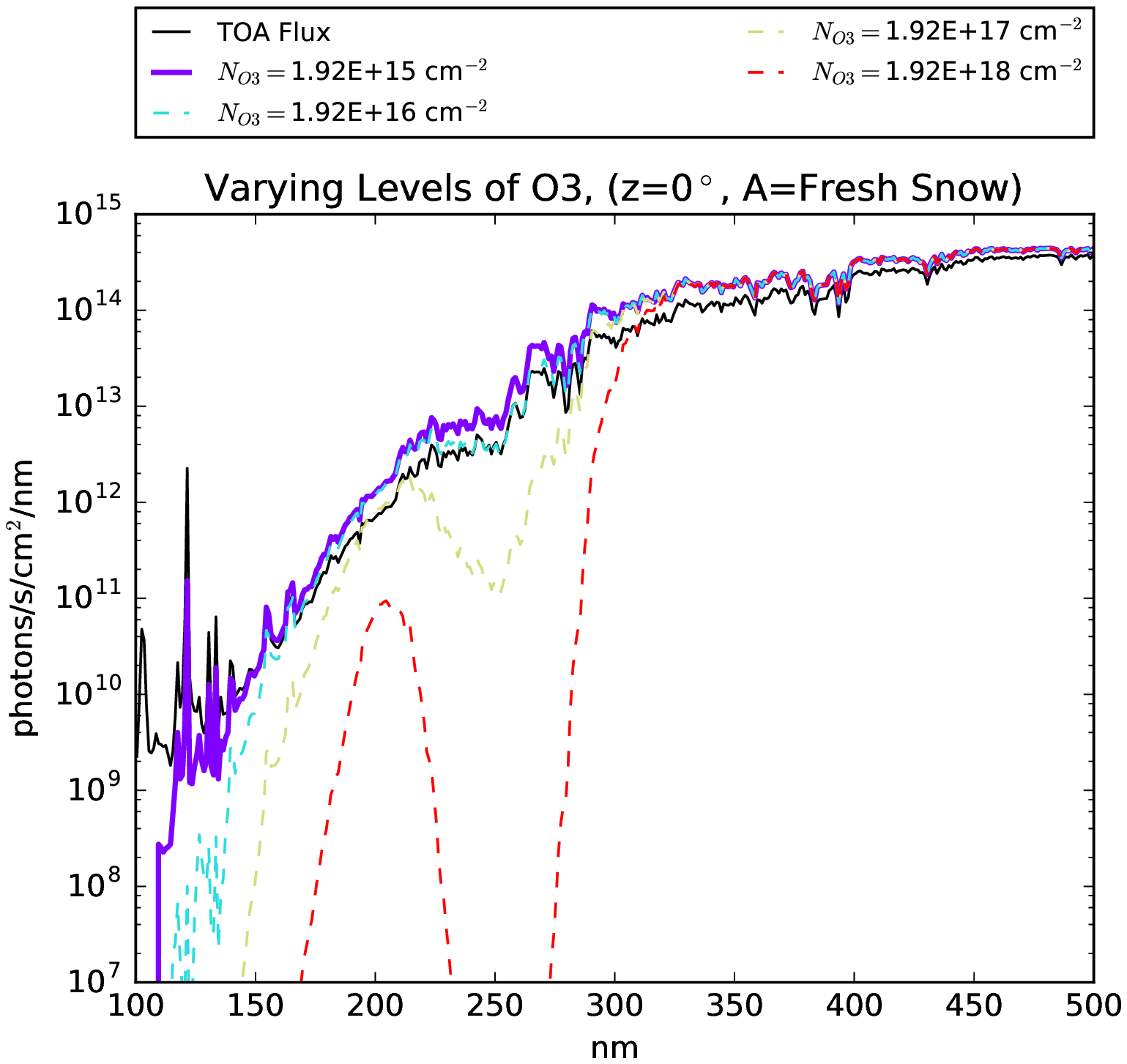}
\caption{Surface radiances for an atmosphere with $N_{N_{2}}=1.88\times10^{25}$ cm$^{-2}$ and varying levels of O$_3$, for a surface albedo corresponding to new snow and SZA=$0^\circ$. The line corresponding to the O$_3$ level calculated in the \citet{Rugheimer2015} model is presented as a solid, thicker line. \label{fig:o3lim}}
\end{figure}

Similarly to O$_2$, O$_3$ is a well-known strong UV shield. It is generated from 3-body reactions involving photolysis of O$_2$ and hence its abundance is sensitive to O$_2$ levels, with sinks from photolysis and reactions with reducing gases. \citet{Rugheimer2015} estimate an O$_3$ column density of $N_{O_{3}}=1.92\times10^{15}$ cm$^{-2}$ for the 3.9 Ga Earth.  At this level, O$_3$ does not significantly modify the surface UV environment. Figure~\ref{fig:o3lim} shows the surface radiance spectra for $N_{O_{3}}=1.92\times10^{15}-10^{18}$ cm$^{-2}$\footnote{We considered lower O$_3$ levels as well, but the resultant spectra were indistinguishable from the $N_{O_{3}}=1.92\times10^{15}$ cm$^{-2}$ case}. As the atmospheric O$_3$ inventory increases, so does its attenuation of UV fluence; at levels roughly comparable to the modern day ($N_{O_{3}}=1.92\times10^{18}$ cm$^{-2}$), it suppresses the BEDs of the photoprocesses we consider by 2 orders of magnitude relative to $N_{CO_{2}}=2.09\times10^{24}$ cm$^{-2}$. However, such O$_3$ levels would require a very high O$_2$ inventory. It is most likely that attenuation from O$_3$ would have been negligible compared to that from CO$_2$ and/or H$_2$O.

\section{Conclusions\label{sec:conc}}
We have used a two-stream radiative transfer model to calculate the hemisphere-integrated surface radiance at UV wavelengths for different surface albedos, solar zenith angles, and atmospheric compositions for the 3.9 Ga Earth. To estimate the effect of these different variables on UV-sensitive prebiotic chemistry, we have convolved the surface radiance spectra with action spectra for glycosidic bond cleavage in UMP (a stressor for abiogenesis) and production of solvated electrons from photoionization of CuCN$_3^{2-}$ (an eustressor for abiogenesis) formed from absorption spectra and assumed QY curves, and integrated the result over wavelength to compute the biologically effective dose rate (BED). 

Our findings demonstrate the importance of considering albedo and zenith angle in calculations of surface UV fluence. For the model atmosphere calculated by \citet{Rugheimer2015} for the 3.9 Ga Earth, variations in albedo (tundra vs new snow) can affect BEDs by a factor of 2.7-4.3, and variations in zenith angle ($0-66.5^\circ$, corresponding to the range of minimum SZA available on Earth) can affect BEDs by a factor of 3.7-4.3, depending on the photoprocess and the assumptions we make about its quantum yield curve. Taken together, albedo and zenith angle can affect BEDs by a factor of 10.4-17.1, meaning that local conditions like latitude and surface type can drive variations in prebiotic photoreaction rate by an order of magnitude or more, independent of atmospheric composition.

While CO$_2$ levels on the 3.9 Ga Earth are debated, even minute amounts of CO$_2$ ($N_{CO_{2}}\geq2.09\times10^{19}$ cm$^{-2}$) are enough to extinguish fluence shortward of 167 nm. Since $N_{CO_{2}}\geq1.87\times10^{22}$ cm$^{-2}$ based on proposed climate models in the literature, CO$_2$ extinction should have shielded prebiotic molecules from solar activity even if removed from the shortwave UV shield of liquid water (e.g. through drying cycles). 

The BEDs vary by less than an order of magnitude as a function of CO$_2$ level for $N_{CO_{2}}\leq9.76\times10^{25}$ cm$^{-2}$ (the limit on CO$_2$ for conventional climate calculations), and less than two orders of magnitude for $N_{CO_{2}}\leq9.82\times10^{26}$ cm$^{-2}$ (the limit on CO$_2$ from the crustal carbon inventory). This implies prebiotic photoreaction rates are insensitive to CO$_2$ levels, assuming no other absorbers to be present.

For climatically reasonable levels of CO$_2$ ($N_{CO_{2}}\geq1.87\times10^{22}$), plausible levels of CH$_4$, O$_2$, and O$_3$ do not significantly modify the UV surface fluence. CH$_4$ is a weak absorber. O$_2$ levels high enough to modify the surface UV fluence are ruled out by SMIF measurements, which means that O$_3$ is expected to be similarly low. 

SO$_2$ and H$_2$S also do not impact the UV surface fluence at concentrations derived assuming modern levels of volcanism. However, it has been hypothesized that SO$_2$ and H$_2$S could have built up to higher (1-100 ppm in a 1-bar atmosphere) levels during epochs of high sustained volcanism. If this scenario occurred, such epochs were low-UV eras in Earth's history. At SO$_2$ levels of $N_{SO_{2}}\geq7.05\times10^{19}$ cm$^{-2}$  and H$_2$S levels of $N_{H_{2}S}=1.41\times10^{21}$ cm$^{-2}$, corresponding to mixing ratios of 3 and 70 ppm respectively in a 1-bar N$_2$/CO$_2$ atmosphere, the photoreactions considered in our study are suppressed by multiple orders of magnitude relative to $N_{CO_{2}}=2.09\times10^{24}$ cm$^{-2}$. If considering high SO$_2$/H$_2$S atmospheres, it is important to characterize the dependence of hypothesized prebiotic pathways on fluence levels to ensure they will not be quenched. It is possible that high-SO$_2$/H$_2$S atmospheres might be more clement environments for abiogenesis as measured by the change in tricyanocuprate photoionization rates compared to UMP glycosidic bond cleavage as a function of SO$_2$/H$_2$S level, but further measurements of the action spectra of these photoprocesses are required to assess this possibility.

At the lowest CO$_2$ level proposed in the literature from climate constraints ($N_{CO_{2}}=1.87\times10^{22}$ cm$^{-2}$) atmospheric CO$_2$ admits light to the surface at wavelengths as short as 190 nm. This raises the question whether sources like ArF eximer lasers (primary emission 193 nm) might be viable for the study of prebiotic chemistry. However, it transpires that H$_2$O vapor is a strong UV shield; at levels as low as $N_{H_{2}O}=9.96\times10^{21}$ cm$^{-2}$ (10\% of that computed by \citet{Rugheimer2015}), it blocks fluence shortward of 198 nm. Therefore, such shortwave sources remain unsuitable for prebiotic chemistry studies, unless assuming exceptionally cold, dry environments (e.g. polar deserts, snowball Earths) with no other major UV absorbers in the atmosphere. 

Conversely, water vapor does not absorb at 254 nm, and for $N_{CO_{2}}\lesssim9.76\times10^{25}$ cm$^{-2}$, fluence at 254 nm remains available. Therefore, photochemical mechanisms isolated by laboratory studies using mercury lamps should have been extant on the prebiotic Earth across a wide range of CO$_2$ and H$_2$O abundances, though characterizing the wavelength dependence of these processes remains key to understanding whether they could function on the early Earth as well as they do in the lab.

In summary, variations in surface albedo and solar zenith angle can, taken together, affect prebiotically relevant photochemical reaction rates by an order of magnitude or more. Surficial prebiotic photochemistry is insensitive to the precise levels of CO$_2$, H$_2$O, O$_2$, O$_3$, and CH$_4$, across the levels of these gases permitted by available constraints. However, it is sensitive to the inventories of SO$_2$ and H$_2$S, if these gases are able to build up to the ppm levels (e.g. during epochs of enhanced volcanism). Surface fluence shortward of 198 nm is available only for a very narrow range of parameter space. However, fluence at 254 nm is available across most of parameter space, meaning that mercury lamps are good candidates for initial studies of prebiotic chemistry (though with the caveat that their use might miss wavelength-dependent effects).

\section{Acknowledgements}
We thank Sarah Rugheimer for providing model metadata for testing, for insightful discussion, and for comments on this article. We thank C. Magnani and Z. Todd for helpful comments and discussion. We thank R. Ramirez, E. Schwieterman, R. Wordsworth, T. Laakso, A. Segura, F. Violotev, J. Szostak, A. Gonzalo, R. Kelley, R. Spurr, I. Cnossen, and L. Zhu for sharing their insights and knowledge with us in discussions. We thank two anonymous referees for comments which greatly improved this article.

This research has made use of NASA's Astrophysics Data System Bibliographic Services, and the MPI-Mainz UV-VIS Spectral Atlas of Gaseous Molecules.

S. R. and D. D. S. gratefully acknowledge support from the Simons Foundation, grant no. 290360.

\section{Author Disclosure Statement}
The authors declare no competing financial interests.

\clearpage

\bibliography{atm_v7.bbl}{}

\clearpage

\appendix
\section{Extinction And Rayleigh Scattering Cross-Sections\label{sec:XCs}}
This Appendix specifies the sources of the total extinction and Rayleigh scattering cross-sections for the gases used in the surface UV environment model. Total extinction cross-sections were taken from literature measurements, while Rayleigh scattering cross-sections were computed from theoretical formalisms. Unless otherwise stated, all measurements were collected near room temperature ($\sim295$ K) and 1 bar of atmospheric pressure, and the digitized data files of the empirical measurements were taken from the MPI-Mainz UV/Vis Spectral Atlas. As discussed in Section~\ref{sec:uv2gmodeldesc}, when the Rayleigh scattering cross-sections exceeded the total measured cross-section from a literature source, the total cross-section was set equal to the Rayleigh scattering value, and the absorption cross-section was set equal to zero. When integrating the cross-sections over a wavelength bin, we linearly interpolated within a given dataset when computing the integral.

\subsection{N$_2$} 
We compute the Rayleigh scattering cross-section of N$_2$ using the formalism of \citet{Vardavas1984}: $\sigma=4.577\times10^{-21} \times KCF\times [A(1+B/\lambda^2)]/\lambda^4$, where $\lambda$ is in $
\mu$ m and $KCF$ is the King correction factor, where $KCF=(6+3\delta)/(6-7\delta)$, where $\delta$ is the depolarization factor. This approach accounts for the wavelength dependence of the index of refraction but assumes a constant depolarization factor. We take the values of the coefficients $A$ and $B$ from \citet{Cox2000}, and the depolarization factor of $\delta=0.0305$ from \citet{Penndorf1957}. 

\begin{figure}[H]
\centering
\includegraphics[width=16.5 cm, angle=0]{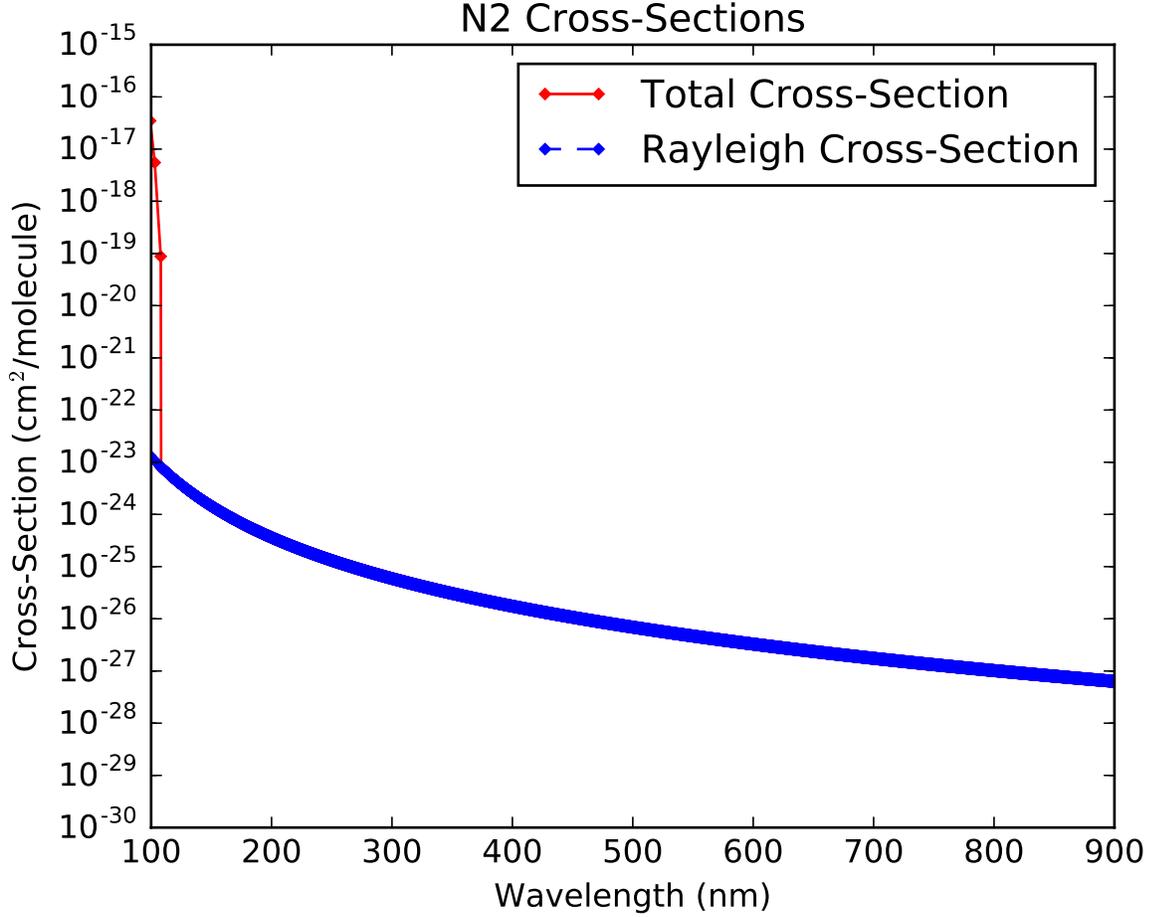}
\caption{Total extinction and Rayleigh scattering cross-sections for N$_2$. \label{fig:n2xc}}
\end{figure}

We take empirically measured N$_2$ extinction cross-sections shortward of 108 nm from \citet{Chan1993n2}, who measure the extinction cross-section from 6.2-113 nm ($\leq$ 5 nm resolution). No absorption is detected longward of 108 nm \citep{Huffman1969, Chan1993n2}, hence we take Rayleigh scattering to account for the extinction for $\lambda>108$ nm. Figure~\ref{fig:n2xc} presents the total and Rayleigh scattering cross-sections for N$_2$ from 100-900 nm. 

\subsection{CO$_2$}
We compute the Rayleigh scattering cross-section of CO$_2$ using the formalism of \citet{Vardavas1984}: $\sigma=4.577\times10^{-21} \times KCF\times [A(1+B/\lambda^2)]/\lambda^4$, where $\lambda$ is in $
\mu$ m and $KCF$ is the King correction factor, where $KCF=(6+3\delta)/(6-7\delta)$, where $\delta$ is the depolarization factor. This approach accounts for the wavelength dependence of the index of refraction but assumes a constant depolarization factor. We take the values of the coefficients $A$ and $B$ from \citet{Cox2000}, and the depolarization factor of $\delta=0.0774$ from \citet{Shemansky1972}. 

\begin{figure}[H]
\centering
\includegraphics[width=16.5 cm, angle=0]{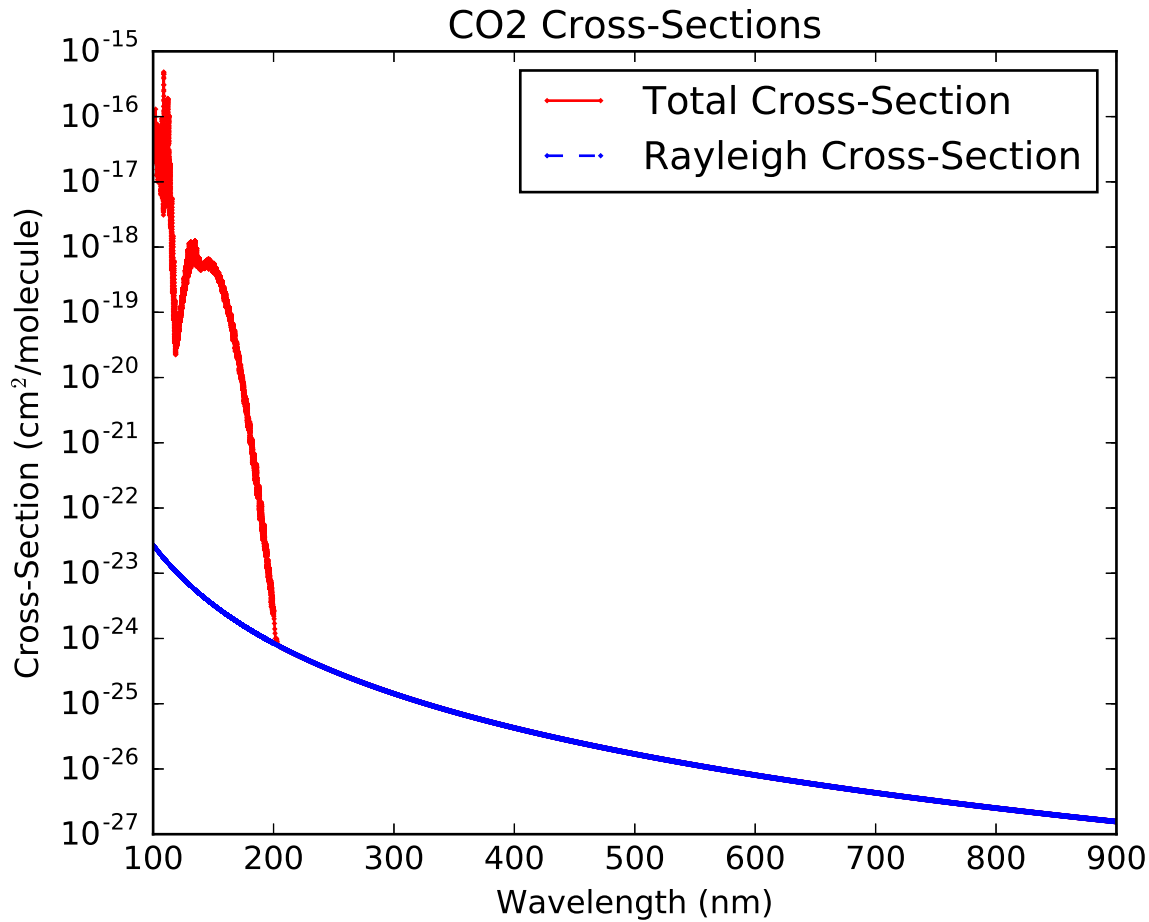}
\caption{Total extinction and Rayleigh scattering cross-sections for CO$_2$. \label{fig:co2xc}}
\end{figure}

We take empirically measured cross-sections shortward of 201.6 nm from \citet{Huestis2010}. \citet{Huestis2010} review existing measurements of extinction cross-sections for CO$_2$, and aggregate the most reliable ones into a single spectrum ($<1$ nm resolution). They test their composite spectrum with an electron-sum rule, and find it to agree with the theoretical expectation to 0.33\%. From 201.75-300 nm, the measurements of \citet{Shemansky1972} provide coverage. However, the resolution of these data ranges from 0.25 nm from 201.75-203.75 nm, to 12-25 nm from 210-300 nm. Further, \citet{Shemansky1972} finds that essentially all extinction at wavelengths longer than 203.5 nm is due to Rayleigh scattering (\citealt{Ityaksov2008co2} derive similar results) . Therefore, we adopt the measurements of 
\citet{Shemansky1972} from 201.75-203.75 nm, and take Rayleigh scattering to describe CO$_2$ extinction at longer wavelengths.
Figure~\ref{fig:co2xc} presents the total and Rayleigh scattering cross-sections for CO$_2$ from 100-900 nm. 

\subsection{H$_2$O}
\begin{figure}[H]
\centering
\includegraphics[width=16.5 cm, angle=0]{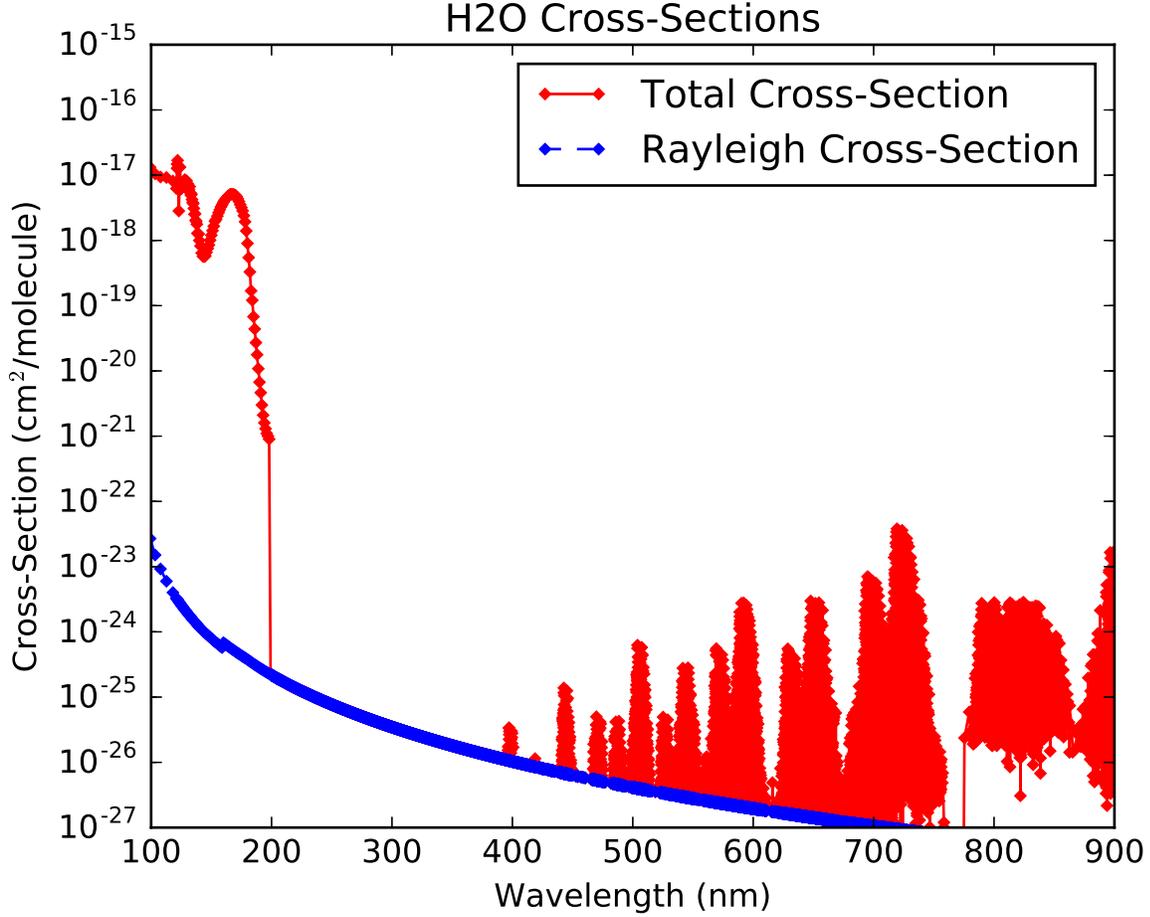}
\caption{Total extinction and Rayleigh scattering cross-sections for H$_2$O. \label{fig:h2oxc}}
\end{figure}

We compute the Rayleigh scattering cross-section of H$_2$O following the methodology of \citet{vonParis2010} and \citet{Kopparapu2013}. We compute the wavelength-dependent index of refraction of water vapor using the observation by \citet{Edlen1966} that the refractivity of water vapor is 15\% less than that of air (itself 1-2\% water vapor by volume). We compute the refractivity of standard air from the formulae of \citet{Bucholtz1995}. \citet{vonParis2010} and \citet{Kopparapu2013} used a value for the depolarization factor of 0.17, based on the work of \citet{Marshall1990}; however, this value was measured for liquid water. We instead adopt $\delta=0.000299$ from the work of \citet{Murphy1977}. We note that the equations of \citet{Bucholtz1995} have a singularity at 0.15946 $\mu$m (159.46 nm) due to a $39.32957-1/\lambda^2$ ($\lambda$ in $\mu$m) term in the denominator, which causes the Rayleigh scattering cross-section to go to infinity at that value. To circumvent this problem, in this term only we adopt $\lambda=0.140 \mu$m for $0.140 \mu$m$<\lambda<0.15946 \mu$m, and $\lambda=0.180 \mu$m for $0.15946 \mu$m$<\lambda< 0.180 \mu$m. This removes the singularity from the Rayleigh scattering curve (see Figure~\ref{fig:h2oxc}).

We take empirically measured cross-sections shortward of 121 nm from the dipole (e,e) spectroscopy measurements of \citet{Chan1993h2o} ($\leq$ 6 nm resolution). From 121-198 nm, we use the compilation of \citet{Sander2011}, who surveyed the measurements of H$_2$O vapor cross-section data to arrived at a recommended tabulation of cross-sections for planetary science studies ($<1$ nm resolution). From 396-755 nm we use the gas-cell absorption results of \citet{Coheur2002} and \citet{Fally2003}, as combined by the MPI Atlas (mode of 0.0073 nm resolution). From 775-1081 nm, we use the measurements of \citet{Merienne2003} (mode of 0.0077 nm resolution). At all wavelengths not covered by these datasets, we take the extinction to be due to Rayleigh scattering. Figure~\ref{fig:h2oxc} presents the total and Rayleigh scattering cross-sections for H$_2$O from 100-500 nm.

\subsection{CH$_4$}
We compute the Rayleigh scattering cross-section of CH$_4$ using the formalism of \citet{Sneep2004}. This method accounts for the wavelength dependence of the index of refraction, and assumes a constant depolarization factor of 0.0002 which is equal to the depolarization factor of CCl$_4$, which has a similar structure. Comparing their computations to data, \citet{Sneep2004} find their formalism overestimates the absorption cross-section of CH$_4$ at 532.5 nm by 15\%; therefore, we follow \citet{Kopparapu2013} in scaling down the \citet{Sneep2004} estimate by 15\% at all wavelengths. 

\begin{figure}[H]
\centering
\includegraphics[width=16.5 cm, angle=0]{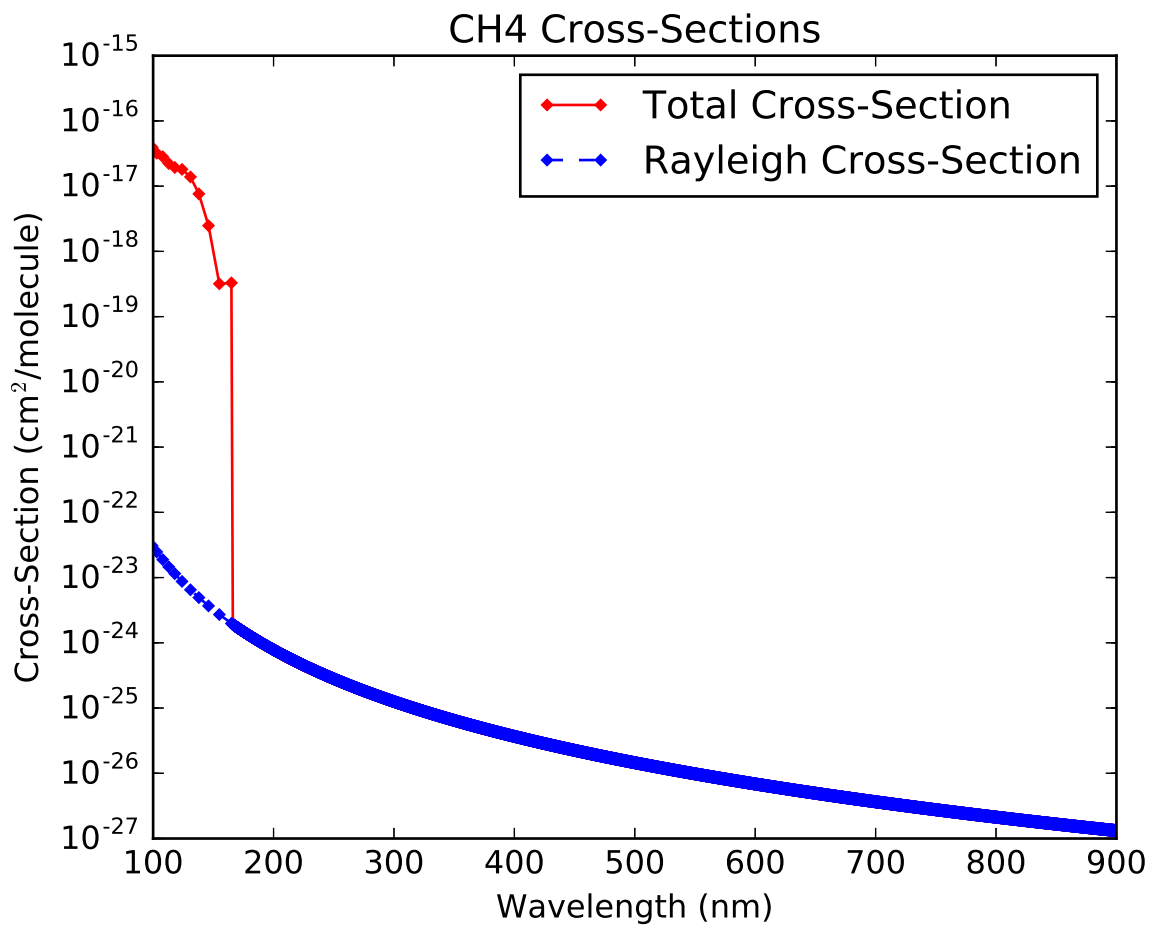}
\caption{Total extinction and Rayleigh scattering cross-sections for CH$_4$. \label{fig:ch4xc}}
\end{figure}

We take empirically measured cross-sections shortward of 165 nm from the dipole (e,e) spectroscopy measurements of \citet{Au1993} ($<5$ nm resolution for $\lambda<113$ nm, 6-10 nm resolution thereafter). Absorption due to CH$_4$has not been detected from 165-400 nm \citep{Mount1977, Chen2004}. Rayleigh scattering is taken to account for extinction at wavelengths longer than 165 nm. Figure~\ref{fig:ch4xc} presents the total and Rayleigh scattering cross-sections for CH$_4$ from 100-900 nm. 

\subsection{SO$_2$}
We compute the Rayleigh scattering cross-section of O$_3$ using the formalism of \citet{Cox2000}: $\sigma=1.306\times10^{20}\times KCF\times \alpha^2/\lambda^4$, where $\sigma$ is in cm$^2$ and $\lambda$ is in $\mu$m, and where $KCF$ refers to the King correction factor, $KCF=(6+3.\delta)/(6-7\delta)$ \citep{Sneep2004}, and $\alpha$ refers to the polarizability of the molecule and $\delta$ is the depolarization factor. \citet{Bogaard1978} list the polarizability $\alpha$ and depolarization ratio $\delta$ for SO$_2$ at 488, 514.5, and 632.8 nm. We use $\alpha_{488 nm}$ and $\delta_{488 nm}$  for $\lambda<501.25$ nm, $\alpha_{514.5 nm}$ and $\delta_{514.5 nm}$ for $501.25<\lambda<573.65$ nm, and $\alpha_{632.8 nm}$ and $\delta_{632.8 nm}$ for $573.65$nm$<\lambda$ nm.

\begin{figure}[H]
\centering
\includegraphics[width=16.5 cm, angle=0]{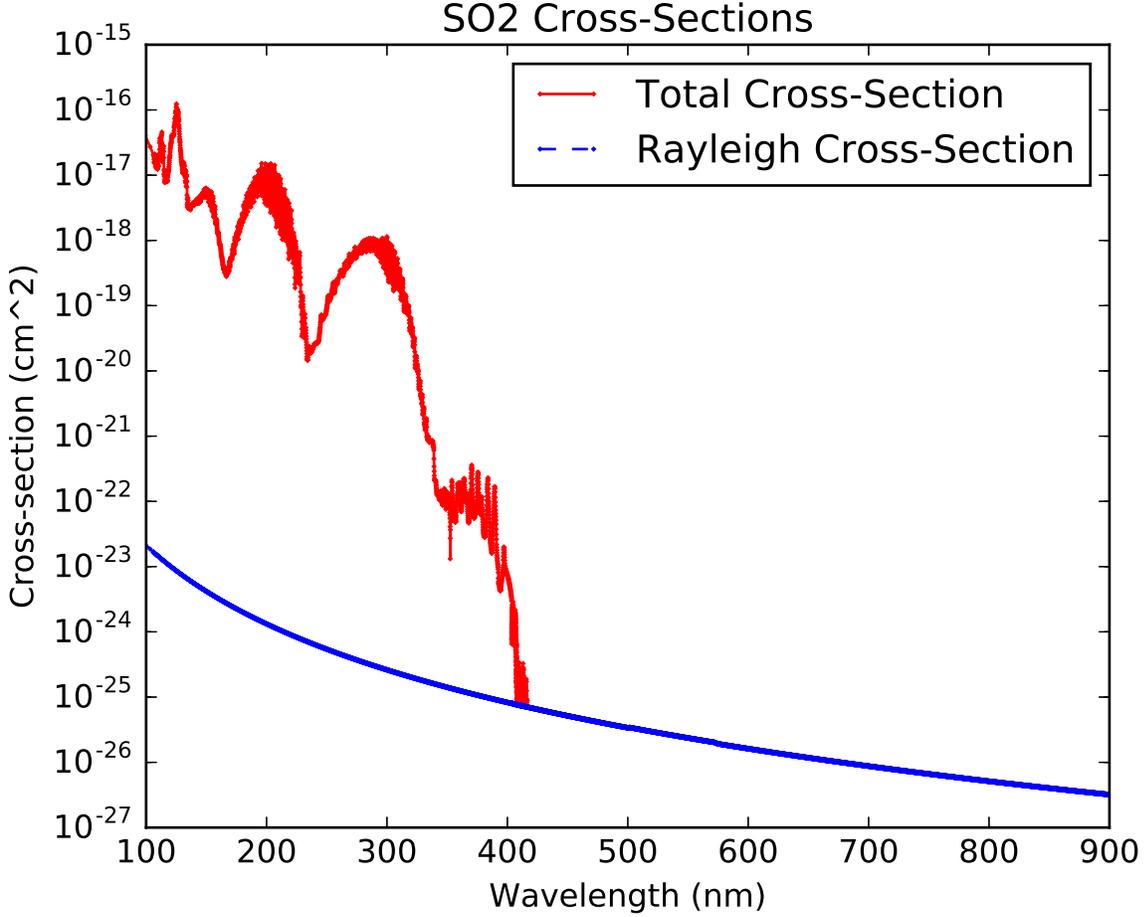}
\caption{Total extinction and Rayleigh scattering cross-sections for SO$_2$. \label{fig:so2xc}}
\end{figure}

We take empirically measured cross-sections of SO$_2$ shortward of 106.1 nm from the dipole (e,e) spectroscopy measurements of \citet{Feng1999} ($<5$ nm resolution). From 106.1-403.7 nm, we take the cross-sections for SO$_2$ extinction from the compendium of SO$_2$ cross-sections of \citet{Manatt1993} (0.1 nm resolution). \citet{Manatt1993} evaluate extant UV cross-sections for SO$_2$ extinction, and aggregate the most reliable into a single compendium covering this wavelength range at 293 $\pm$ 10 K. From 403.7-416.7 nm, we take the Fourier transform spectrometer measurements of \citet{Vandaele2009} ($<1$ nm resolution). Many of the cross-sections reported in this dataset are negative, corresponding to an increase in flux from traversing an gas-filled cell. These cross-sections are deemed unphysical and removed from our model. Figure~\ref{fig:so2xc} presents the total and Rayleigh scattering cross-sections for SO$_2$ from 100-900 nm. 

\subsection{H$_2$S}
We compute the Rayleigh scattering cross-section of H$_2$S using the formalism of \citet{Cox2000}: $\sigma=1.306\times10^{20}\times KCF\times \alpha^2/\lambda^4$, where $\sigma$ is in cm$^2$ and $\lambda$ is in $\mu$m, and where $KCF$ refers to the King correction factor, $KCF=(6+3.\delta)/(6-7\delta)$ \citep{Sneep2004}, and $\alpha$ refers to the polarizability of the molecule and $\delta$ is the depolarization factor. \citet{Bogaard1978} list the polarizability $\alpha$ and depolarization ratio $\delta$ for H$_2$S at 488, 514.5, and 632.8 nm. We use $\alpha_{488 nm}$ and $\delta_{488 nm}$  for $\lambda<501.25$ nm, $\alpha_{514.5 nm}$ and $\delta_{514.5 nm}$ for $501.25<\lambda<573.65$ nm, and $\alpha_{632.8 nm}$ and $\delta_{632.8 nm}$ for $573.65$nm$<\lambda$ nm.

\begin{figure}[H]
\centering
\includegraphics[width=16.5 cm, angle=0]{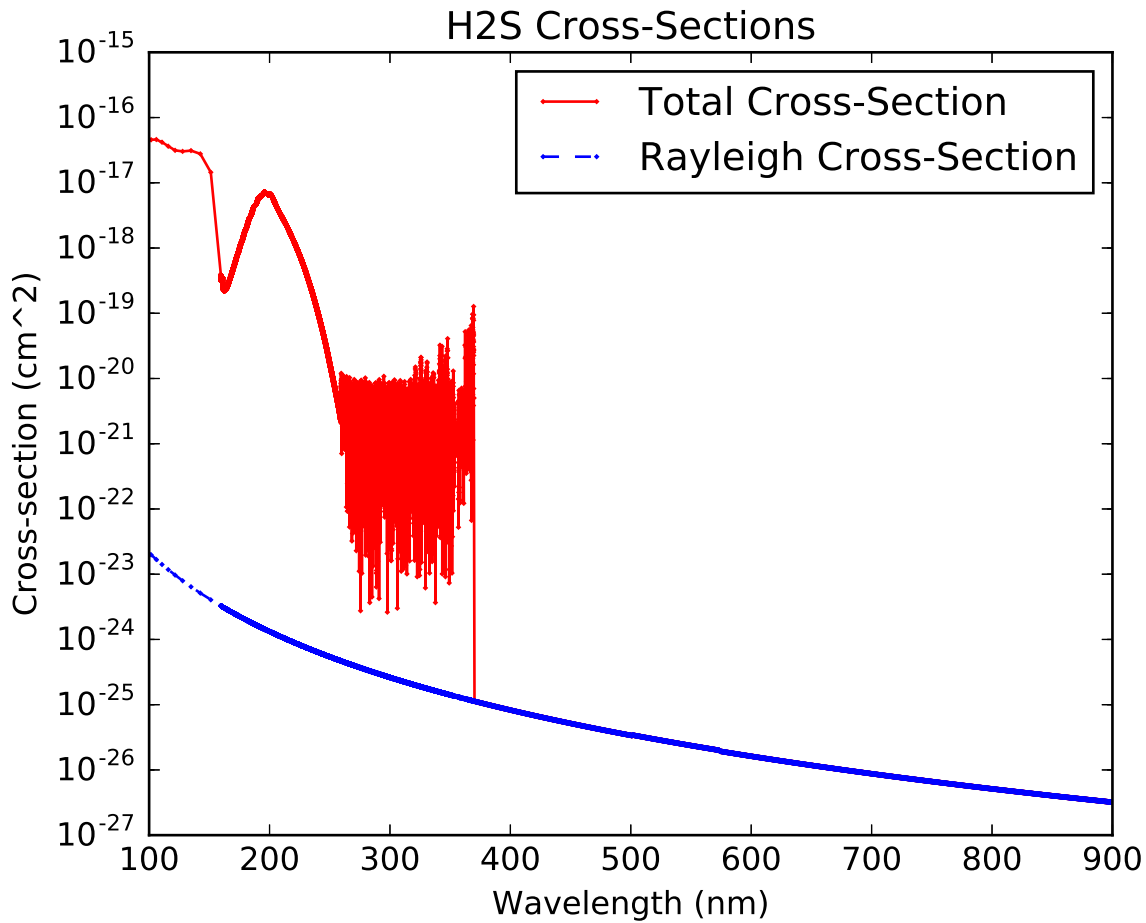}
\caption{Total extinction and Rayleigh scattering cross-sections for H$_2$S. \label{fig:h2sxc}}
\end{figure}

We take empirically measured cross-sections of H$_2$S shortward of 159.465 nm from the dipole (e,e) spectroscopy measurements of \citet{Feng1999h2s} ($<10$ nm resolution). From 159.465-259.460 nm, we take the cross-sections for H$_2$S extinction from the gas cell absorption measurements of \citet{Wu1998} (0.06 nm resolution), as recommended by \citet{Sander2011}. From 259.460-370.007 nm, we take the gas cell absorption measurements of \citet{Grosch2015} (0.018 nm resolution). Many of the cross-sections reported in this dataset are negative, corresponding to an increase in flux from traversing an gas-filled cell. These cross-sections are deemed unphysical and removed from our model. Figure~\ref{fig:h2sxc} presents the total and Rayleigh scattering cross-sections for H$_2$S from 100-900 nm. 

\subsection{O$_2$}
We compute the Rayleigh scattering cross-section of O$_2$ using the formalism of \citet{Vardavas1984}: $\sigma=4.577\times10^{-21} \times KCF\times [A(1+B/\lambda^2)]/\lambda^4$, where $\lambda$ is in $
\mu$ m and $KCF$ is the King correction factor, where $KCF=(6+3\delta)/(6-7\delta)$, where $\delta$ is the depolarization factor. This approach accounts for the wavelength dependence of the index of refraction but assumes a constant depolarization factor. We take the values of the coefficients $A$ and $B$ from \citet{Cox2000}, and the depolarization factor of $\delta=0.054$ from \citet{Penndorf1957}. 

\begin{figure}[H]
\centering
\includegraphics[width=16.5 cm, angle=0]{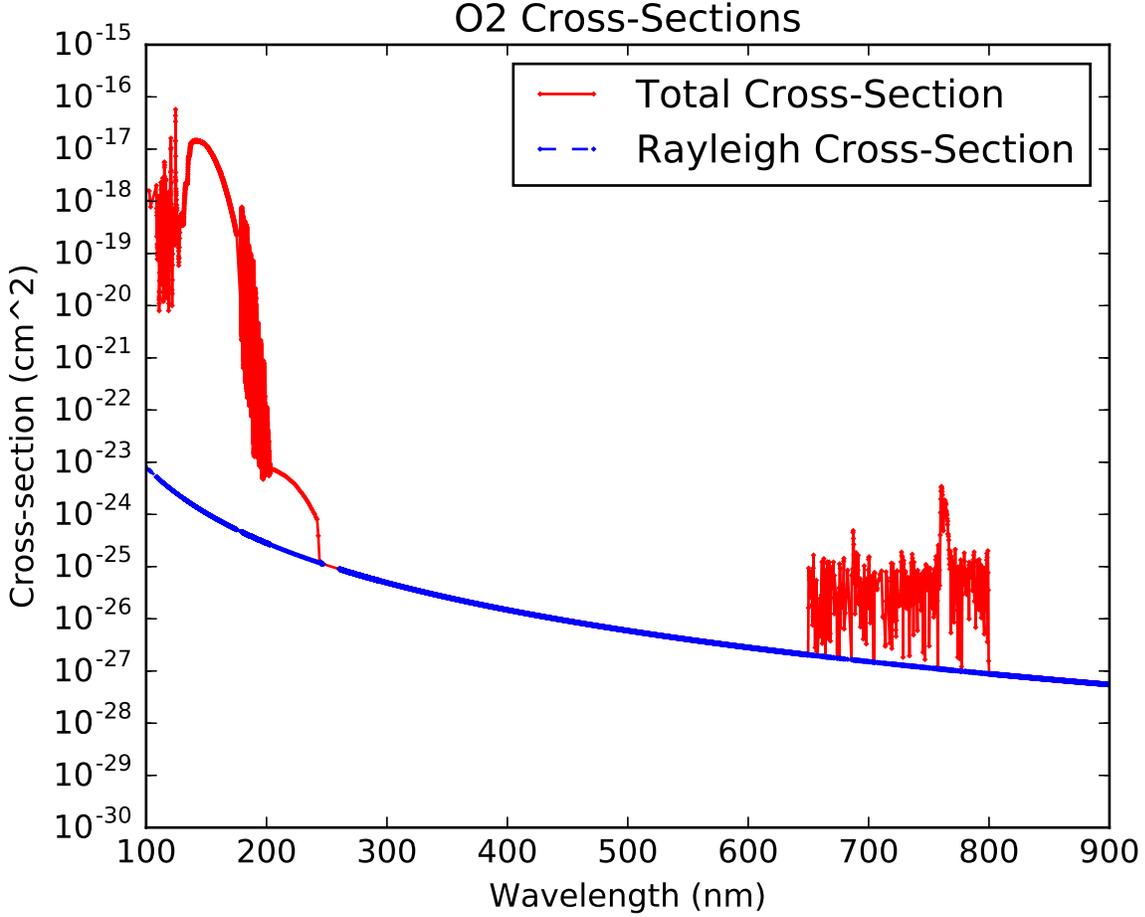}
\caption{Total extinction and Rayleigh scattering cross-sections for O$_2$. \label{fig:o2xc}}
\end{figure}

We take extinction cross-sections of O$_2$ shortward of 108.75 nm from the work of \citet{Huffman1969} ($<12.6$ nm resolution). \citet{Huffman1969} reviews previous literature measurements of VUV extinction cross-sections of O$_2$ and provides recommended values for aeronomic studies. From 108.75-130.0 nm, we use the gas absorption cell measurements for ground-state O$_2$ of \citet{Ogawa1975} ($<0.2$ nm resolution). From 130.04-175.24 nm, we use the absorption cell measurements of \citet{Yoshino2005} ($<0.42$ nm resolution). From 179.2-202.6 nm, we use the gas absorption cell measurements of \citet{Yoshino1992} (300 K, 0.01 nm resolution). Duplicate values in this database (presumably due to rounding error) were removed.  From 205-245 nm, we use the compilation of \citet{Sander2011}, recommended by JPL for use in planetary atmospheres studies ($<1$ nm resolution). From 245-294 nm, we use the gas absorption cell extinction cross-sections measured by \citet{Fally2000} ($<0.008$ nm resolution.) From 650-799.6, we use the gas-cell absorption measurements of the SCIAMACHY calibration data from \citet{Bogumil2003} ($<0.21$ nm resolution,). As in the case of SO$_2$, several of the cross-sections reported for this dataset are negative; these cross-sections are rejected as unphysical and removed from the model. Figure~\ref{fig:o2xc} presents the total and Rayleigh scattering cross-sections for O$_2$ from 100-500 nm. We note that the cross-sections presented here are for O$_2$ extinction only. Extinction due to molecular complexes, e.g. the O$_2$-O$_2$ complexes observed by \citet{Greenblatt1990}, are not included in our parametrization. This does not materially impact the fidelity of our model since the cross-sections associated with these complexes are small at relevant partial pressures of O$_2$ ($<3\times10^{-26}$ cm$^2$ for 1 atm of O$_2$).

\subsection{O$_3$}
We take the Rayleigh scattering cross-section of O$_3$ using the formalism of \citet{Cox2000}: $\sigma=1.306\times10^{20}\times KCF\times \alpha^2/\lambda^4$, where $\sigma$ is in cm$^2$ and $\lambda$ is in $\mu$m, and where $KCF$ refers to the King correction factor, $KCF=(6+3.\delta)/(6-7\delta)$ \citep{Sneep2004}, and $\alpha$ refers to the polarizability of the molecule. From \citet{Brasseur1986}, $KCF=1.06$ for ozone. We adopt $\alpha=3.21\times10^{-24}$ cm$^{3}$ based on the average electric dipole polarizability listed for ground state O$_3$ in \citet{CRC90}. This formulation assumes constant polarizability (index of refraction) and depolarization factor.

\begin{figure}[H]
\centering
\includegraphics[width=16.5 cm, angle=0]{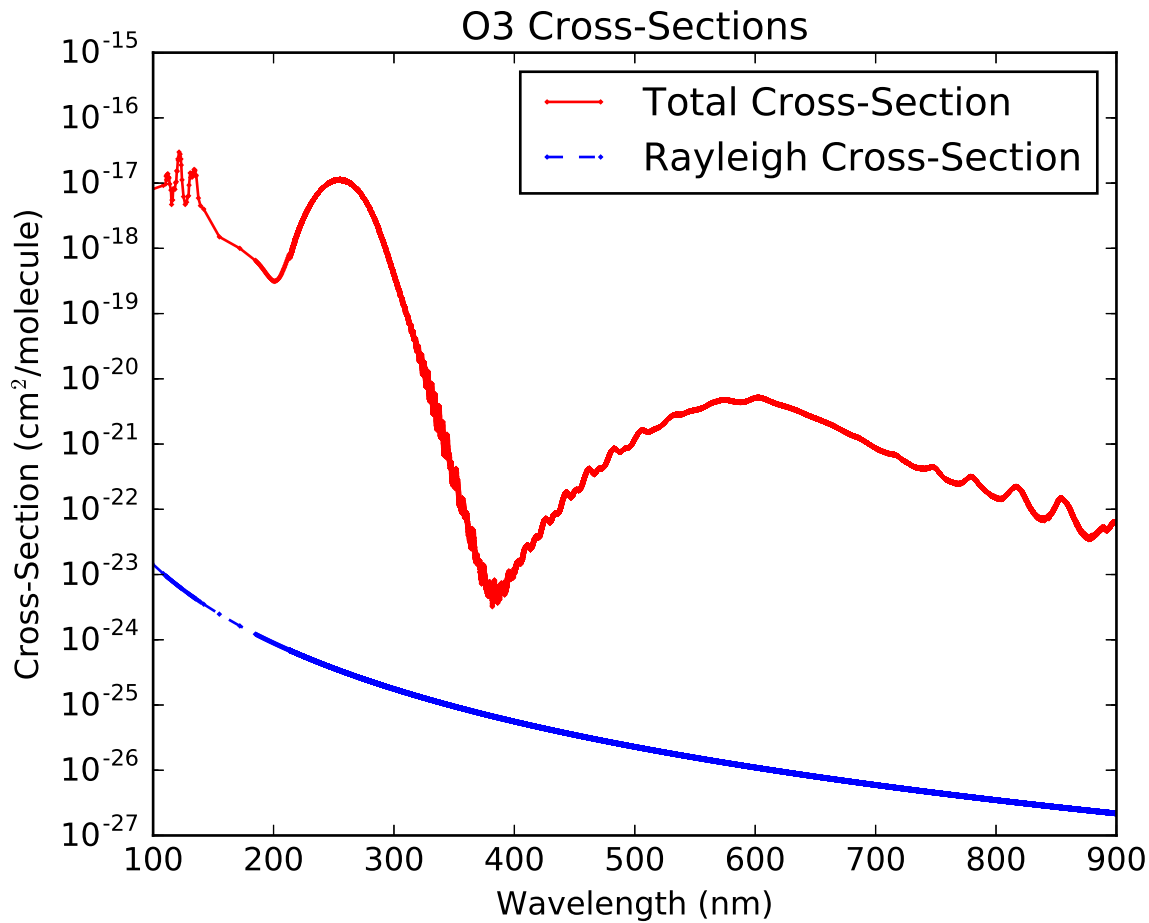}
\caption{Total extinction and Rayleigh scattering cross-sections for O$_3$. \label{fig:o3xc}}
\end{figure}

We take empirically measured cross-sections of O$_3$ shortward of 110 nm from the gas cell absorption measurements of \citet{Ogawa1958} ($<9.5$ nm resolution). We take cross-sections from 110-172 nm from the gas cell absorption measurements of \citet{Mason1996} ($<2$ nm resolution for $\lambda\leq139.31$, 3-17 nm resolution for $\lambda=139.31-172$ nm). Following the recommendations of \citet{Sander2011}, we take cross-sections from 185-213 nm from the gas cell absorption measurements of \citet{Molina1986} (0.5 nm resolution). Finally, we take the cross-sections for 213-1100 nm from the gas cell absorption measurements recently published in joint papers by \citet{Serdyuchenko2014}  and \citet{Gorshelev2014} (0.02-0.06 nm resolution, interpolated to 0.01 nm; 293 K). \citet{Gorshelev2014} compares this dataset to previous measurements and finds good agreement. Figure~\ref{fig:o3xc} presents the total and Rayleigh scattering cross-sections for O$_3$ from 100-500 nm. 

\section{Spectral Albedos\label{sec:Albedos}}
This section describes the sources for the direct and diffuse spectral albedos, $\alpha_{dir}$ and $\alpha_{dif}$, corresponding to different surface environments used in this study. In this section, $\mu=\cos(\theta_0)$ refers to the cosine of the solar zenith angle. For all albedos, we enforce a physical albedo range of 0-1 by setting any negative albedos to 0 and any albedos greater than 1 to 1. 

\subsection{Ocean}
We approximate the albedo of pure (ice- and land-free) ocean via the methodology of \citet{Briegleb1986}, who in turn rely upon \citet{Payne1972}.  \citet{Briegleb1986} take $\alpha_{dif}=0.06$ and $\alpha_{dir}=2.6/(\mu^{1.7}+0.065)+15(\mu-0.1)(\mu-0.5)(\mu-1.0)$. 

\citet{Payne1972} measured the reflectance of the ocean surface under a variety of conditions using a spectrometer with uniform sensitivity coverage from 280-2800 nm. Hence, albedos calculated from this work represent a fit to the mean albedo integrated 280-2800 nm, which includes 98\% of solar flux. \citet{Briegleb1986} argue that the variation in ocean albedo as a function of wavelength is modest, due to the modesty of variations of the index of refraction of water across this wavelength range. \citet{Briegleb1986} modeled this variation and found the spectral corrections to the oceanic albedo due to the wavelength dependence of the optical properties of water to be +0.02 for 200-500 nm, -0.003 for 500-700 nm, -0.007 for 700-850 nm, and -0.007 for 850 nm-4 $\mu$m.  They calculated these corrections to have minimal impact on the TOA broadband albedo, and hence ignored them. 

However, for our applications (computing the spectra radiance over different surface environments), these corrections can be significant, especially in the case of diffuse radiation, where the shortwave correction is 1/3 of the diffuse albedo. Hence, we include these corrections by adding a correction term $\alpha_{corr}$ to $\alpha_{dir}$ and $\alpha_{dif}$. Lacking any better assumption, we extend the 200-500 nm value to all wavelengths $<500$ nm. Hence $\alpha_{corr}=0.02$ for $\lambda<500$ nm, $\alpha_{corr}=-0.003$ for $\lambda=500-700$ nm,  and $\alpha_{corr}=-0.007$ for $\lambda>700$ nm.

\subsection{Snow}
We approximate the albedo of pure snow using the methodology outlined by \citet{Briegleb1982} and reviewed by \citet{Coakley2003}. This methodology treats new-fallen and old snow separately (new-fallen snow is brighter).

From Table 1 of \citet{Briegleb1982}, for new fallen snow: 
$$\alpha_{dif}(\lambda=200-500 \texttt{nm})=0.95$$
$$\alpha_{dif}(\lambda=500-700 \texttt{nm}))=0.95$$
$$\alpha_{dif}(\lambda=700-4000 \texttt{nm}))=0.65$$
  
Whereas for old snow:
 $$\alpha_{dif}(\lambda=200-500 \texttt{nm}))=0.76$$
 $$\alpha_{dif}(\lambda=500-700 \texttt{nm}))=0.76$$
 $$\alpha_{dif}(\lambda=700-4000 \texttt{nm}))=0.325$$ 
 
 As before, we extend the 200-500 nm value to all $\lambda<200$. 

To compute the direct albedo, we follow \citet{Briegleb1982} in using the formalism of \citet{Dickinson1981} to account for zenith angle dependence: $\alpha_{dir}=\alpha_{dif}+(1-\alpha_{dif})\times0.5\times[3/(1+4\mu)-1)$ for $\mu \leq 0.5$, and $\alpha_{dir}=\alpha_{dif}$ for $\mu>0.5$.

\subsection{Desert}
We approximate the albedo of desert environments following the methods of \citet{Briegleb1986}, as reviewed by \citet{Coakley2003}. Following Tables 1 and 2 of \citet{Briegleb1986},

$$\alpha_{dif}(\lambda<500 \texttt{nm}))=0.5\times0.28+0.5\times0.15=0.22$$
$$\alpha_{dif}(\lambda=500-700 \texttt{nm}))=0.5\times0.42+0.5\times0.25=0.34$$
$$\alpha_{dif}(\lambda=700-850 \texttt{nm}))=0.5\times0.50+0.5\times0.35=0.43$$
$$\alpha_{dif}(\lambda=850-4000 \texttt{nm}))=0.5\times0.50+0.5\times0.40=0.45$$

Where we have again extended the \citet{Briegleb1986} 200-500 nm albedos to all wavelengths $<200$ nm.

To compute the direct albedo, we follow \citet{Briegleb1986} in including zenith angle dependence by writing $\alpha_{dir}=\alpha_{dif}\times(1+d)/(1+2d\mu)$, where $d$ is a parameter derived from a fit to data for different terrain types. For desert, $d=0.4$.

\subsection{Tundra}
We approximate the albedo of tundra environments following the methods of \citet{Briegleb1986}, as reviewed by \citet{Coakley2003}. 

$$\alpha_{dif}(\lambda<500 \texttt{nm}))=0.5\times0.04+0.5\times0.07=0.06$$
$$\alpha_{dif}(\lambda=500-700 \texttt{nm}))=0.5\times0.10+0.5\times0.13=0.12$$
$$\alpha_{dif}(\lambda=700-850 \texttt{nm}))=0.5\times0.25+0.5\times0.19=0.22$$
$$\alpha_{dif}(\lambda=850-4000 \texttt{nm}))=0.5\times0.25+0.5\times0.28=0.27$$

Where we have again extended the \citet{Briegleb1986} 200-500 nm albedos to all wavelengths $<200$ nm.

To compute the direct albedo, we follow \citet{Briegleb1986} in including zenith angle dependence by writing $\alpha_{dir}=\alpha_{dif}\times(1+d)/(1+2d\mu)$, where $d$ is a parameter derived from a fit to data for different terrain types. For tundra, $d=0.1$.

\section{Enhancement of Upwelling Intensity in the Discrete Ordinates \& Two Stream Approximations\label{sec:enhancedupwellingradiance}}
In this section, we construct a detailed example to demonstrate that under the discrete ordinates approximation to atmospheric radiative transfer, of which the two-stream approximation is the $n=2$ special case, the total upwelling intensity through the atmosphere can exceed the total downwelling intensity if the planet surface is an Lambertian (isotropic) reflector. In outlining this example, we follow closely the DISORT User's Guide (DUG, \citealt{DUG}). The DUG outlines the formalism behind the DISORT code \citep{Stamnes1988}, one of the best-known implementations of plane-parallel discrete-ordinates radiative transfer.

Consider a planet with a homogenous atmosphere illuminated from above by the Sun. Let the Sun be located at direction $(\mu_0,\phi_0)$, where $\mu$ is the cosine of the solar zenith angle and $\phi$ is the azimuth, with incident intensity $I(\mu,\phi)=I_0\delta(\mu-\mu_0)\delta(\phi-\phi_0)$. Suppose thermal emission from the atmosphere and planet is negligible (i.e. solar forcing is the only source of photons to the system). Further, let the atmosphere have negligible optical depth $\tau_0/\mu_0<<1$ (transparent atmosphere approximation). Then atmospheric scattering and absorption have negligible effect on the planetary radiation field, the upwelling and downwelling intensity fields are essentially uncoupled, and the upward and downward fluxes and the mean intensity are constant throughout the atmosphere. Let the surface be a Lambertian reflector, with albedo $A=F^{+}/F^{-}$, where $F^{+}$ is the upwelling flux and $F^{-}$ is the downwelling flux.

The hemispherically-integrated upwelling and downwelling intensities $I^+$ and $I^-$ may be defined, following DUG equation 9c, by:

$$I^+=2\pi\overline{I^+}=2\pi\int^{1}_{0}I^0(\tau, +\mu)d\mu$$
$$I^-=2\pi\overline{I^-}=I_0\exp(-\tau_0/\mu_0)+2\pi\int^{1}_{0}I^0(\tau, -\mu)d\mu$$

Where $I^0(\tau, \mu)$ is the intensity at a depth $\tau$ in the atmosphere arriving from a direction $\mu$. Since $\tau_0/\mu_0<<1$, $I_0\exp(-\tau_0/\mu_0)\approx I_0(1-\tau_0/\mu_0)$. Further, since the atmosphere is optically very thin and noninteracting, essentially no flux is scattered out of the direct stream into the downward diffuse intensity, nor out of the upward diffuse intensity into the downward diffuse intensity (or vice versa). Therefore, we can further approximate that $I^0(\tau, -\mu)\approx0$. We can then simplify

$$I^-\approx I_0(1-\tau_0/\mu_0)$$

Next, from DUG equation 39, we can write the boundary condition at the planetary surface (i.e. the reflection condition):

$$I^0(\tau, +\mu)=\epsilon(\mu)B(T_g)+\frac{1}{\pi}\mu_0I_0\exp(-\tau_0/\mu_0)\rho_0(\mu, \mu_0)+2\int^{1}_{0}\mu' d\mu' I^0(\tau_0, -\mu) \rho_0(\mu, -\mu') $$

where $\epsilon(\mu)$ is the surface emissivity as a function of angle, $B(T_g)$ is the blackbody emission of the planet at temperature $T_g$, and. $\rho_0(\mu, -\mu')$ is the bidirectional reflection function. $B(T_g)\approx0$ by assumption of negligible blackbody emission at UV wavelengths. $\rho_0(\mu_i, \mu_j)=A$ for a Lambertian surface (see, e.g., \citealt{SpurrThesis}, equation 2.19). $I^0(\tau, -\mu)\approx0$ as above. We can therefore simplify:

$$I^0(\tau, +\mu)\approx\frac{1}{\pi}\mu_0I_0(1-\tau_0/\mu_0)A$$

Substituting for $I^0(\tau, +\mu)$ in our expression for $I^+$, we can then conclude

$$I^+\approx2\pi\int^{1}_{0}(\frac{1}{\pi}\mu_0I_0(1-\tau_0/\mu_0)A)d\mu$$
$$     =2\int^{1}_{0}(\mu_0I_0(1-\tau_0/\mu_0)A)d\mu$$
$$     =2A\mu_0I_0(1-\tau_0/\mu_0)$$

Therefore,

$$I^+/I^-=2A\mu_0$$.

For $A\mu_0>0.5$, $I^+>I^-$, with a maximum value of $2I^-$. This condition is physically plausible. One example satisfying this condition is the equator of a snowball Earth at noon on the equinox. At noon on the equinox, $\mu_0=1$ (SZA$=0$), and snow approximates a Lambertian scatterer\citep{Coakley2003} with a UV albedo of 0.95 \citep{Briegleb1982}. Under these conditions, $A\mu_0=0.95>0.5$, and $I^+=1.9I^-$. 

This example demonstrates that it is possible for the upwelling intensity to exceed the downwelling intensity. This result may be understood intuitively as a consequence of requiring flux conservation during reflection from a Lambertian surface. Consider the $\mu_0=1$, $A=1$ case in the transparent atmosphere approximation. In this case, all downward photons are arriving from the direction $\mu_0$, and the downwelling intensity and flux are both $I_0$. Since $A=1$, all the flux must be reflected, so the upwelling flux $F^+=I_0$ as well. But, since the surface is a Lambertian scatterer, the downward intensity was scattered uniformly in all directions. $I^+(\tau)=\int^1_0d\mu I^0(\tau, +\mu)>\int^1_0d\mu I^0(\tau, +\mu)=F^+(\tau)$ since $I^0(\tau, +\mu)>0$, so $I^+>I_0=I^-$. 

For another discussion of this phenomenon, see \citet{Madronich1987}, their Section 2.3.

\end{document}